\documentclass [prl,aps,letterpaper,twocolumn,superscriptaddress,amsmath,amssymb,floatfix] {revtex4-1}

\usepackage{graphicx}
\usepackage{textcomp}
\usepackage{epstopdf}
\usepackage{mathptmx}

\usepackage[latin9]{inputenc}
\usepackage{amsmath}
\usepackage{amssymb}
\usepackage{graphicx}
\usepackage{esint}

\makeatletter

\pdfpageheight\paperheight
\pdfpagewidth\paperwidth

\@ifundefined{textcolor}{}
{%
 \definecolor{BLACK}{gray}{0}
 \definecolor{WHITE}{gray}{1}
 \definecolor{RED}{rgb}{1,0,0}
 \definecolor{GREEN}{rgb}{0,1,0}
 \definecolor{BLUE}{rgb}{0,0,1}
 \definecolor{CYAN}{cmyk}{1,0,0,0}
 \definecolor{MAGENTA}{cmyk}{0,1,0,0}
 \definecolor{YELLOW}{cmyk}{0,0,1,0}
}

\usepackage{xcolor}\usepackage{soul}
\newcommand{\ket}[1]{\ensuremath{\left|#1\right\rangle}}

\definecolor{blue}{rgb}{0,0,1}
\definecolor{red}{rgb}{1,0,0}
\definecolor{green}{rgb}{0,1,0}

\begin{document}

\title{Demonstration of quantum error correction and universal gate set on a binomial bosonic logical qubit}

\author{L.~Hu}

\thanks{These two authors contributed equally to this work.}

\affiliation{Center for Quantum Information, Institute for Interdisciplinary Information
Sciences, Tsinghua University, Beijing 100084, China}

\author{Y.~Ma}

\thanks{These two authors contributed equally to this work.}

\affiliation{Center for Quantum Information, Institute for Interdisciplinary Information
Sciences, Tsinghua University, Beijing 100084, China}

\author{W.~Cai}

\affiliation{Center for Quantum Information, Institute for Interdisciplinary Information
Sciences, Tsinghua University, Beijing 100084, China}

\author{X.~Mu}

\affiliation{Center for Quantum Information, Institute for Interdisciplinary Information
Sciences, Tsinghua University, Beijing 100084, China}

\author{Y.~Xu}

\affiliation{Center for Quantum Information, Institute for Interdisciplinary Information
Sciences, Tsinghua University, Beijing 100084, China}

\author{W.~Wang}

\affiliation{Center for Quantum Information, Institute for Interdisciplinary Information
Sciences, Tsinghua University, Beijing 100084, China}

\author{Y.~Wu}
\affiliation{Department of Physics, University of Michigan, Ann Arbor, Michigan
48109, USA}

\author{H.~Wang}

\affiliation{Center for Quantum Information, Institute for Interdisciplinary Information
Sciences, Tsinghua University, Beijing 100084, China}

\author{Y.~P.~Song}

\affiliation{Center for Quantum Information, Institute for Interdisciplinary Information
Sciences, Tsinghua University, Beijing 100084, China}

\author{C.-L.~Zou}
\email{clzou321@ustc.edu.cn}
\affiliation{Key Laboratory of Quantum Information, CAS, University of Science
and Technology of China, Hefei, Anhui 230026, P. R. China}

\author{S.~M.~Girvin}
\affiliation{Departments of Applied Physics and Physics, Yale University, New Haven, CT 06511, USA}

\author{L-M.~Duan}
\affiliation{Center for Quantum Information, Institute for Interdisciplinary Information
Sciences, Tsinghua University, Beijing 100084, China}
\affiliation{Department of Physics, University of Michigan, Ann Arbor, Michigan
48109, USA}

\author{L.~Sun}
\email{luyansun@tsinghua.edu.cn}
\affiliation{Center for Quantum Information, Institute for Interdisciplinary Information
Sciences, Tsinghua University, Beijing 100084, China}


\begin{abstract}
Logical qubit encoding and quantum error correction (QEC) have been experimentally demonstrated in various physical systems with multiple physical qubits, however, logical operations are challenging due to the necessary nonlocal operations. Alternatively, logical qubits with bosonic-mode-encoding are of particular interest because their QEC protection is hardware efficient, but gate operations on QEC protected logical qubits remain elusive. Here, we experimentally demonstrate full control on a single logical qubit with a binomial bosonic code, including encoding, decoding, repetitive QEC, and high-fidelity (97.0\% process fidelity on average) universal quantum gate set on the logical qubit. The protected logical qubit has shown 2.8 times longer lifetime than the uncorrected one. A Ramsey experiment on a protected logical qubit is demonstrated for the first time with two times longer coherence than the unprotected one. Our experiment represents an important step towards fault-tolerant quantum computation based on bosonic encoding.
\end{abstract}

\maketitle


Quantum states are fragile and can be easily destroyed by their inevitable
coupling to the uncontrolled environment, which presents a major obstacle
to quantum computation~\cite{Nielsen}. A practical quantum computer ultimately calls for operations on logical qubits protected by quantum error correction (QEC)~\cite{Shor1995,Steane1996,Gottsman2010,Fowler2012} against unwanted
errors with hopefully long enough coherence time. Therefore, to realize such a logical qubit with a longer coherence time than its individual physical components is considered as one of the most challenging goals
for current quantum information processing~\cite{Devoret2013}. For
QEC, fragile quantum information needs first to be redundantly encoded
in a logical subspace of a large Hilbert space~\cite{Gottsman2010}.
Error syndromes that can distinguish the code space from other orthogonal
error subspaces need to be monitored repetitively without perturbing
the encoded information. Based on quantum non-demolition (QND) syndrome measurements, unitary recovery
gates are adaptively applied to restore the original encoded information.
To protect with QEC and to realize universal unitary manipulations
of the logical qubit within the code space are necessary steps towards
a practical quantum computer. The next stage is to demonstrate gate
operations on the logical qubit under continuous QEC protection.

\begin{figure}
\includegraphics{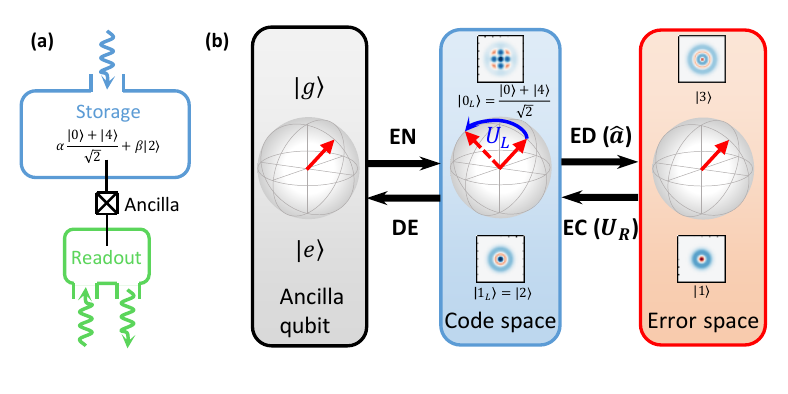} \caption{\textbf{Schematic of the experiments on binomial quantum code.} (a) The experimental device consists of a storage cavity
as an oscillator for logical quantum state encoding, an ancillary
transmon qubit facilitating all quantum operations, and a readout
cavity for the ancilla measurement. The quantum states are encoded
in the oscillator with the binomial code bases $\left|0_{\mathrm{L}}\right\rangle =\left(\left|0\right\rangle +\left|4\right\rangle \right)/\sqrt{2}$
and $\left|1_{\mathrm{L}}\right\rangle =\left|2\right\rangle $. (b) Operations on a single logical qubit. The quantum state
of the ancilla qubit $\left\{ \left|g\right\rangle ,\left|e\right\rangle \right\} $
can be encoded (EN) to and decoded (DE) from the code space of the
oscillator. A single photon loss changes the code space with even
parity to the error space $\{\left|3\right\rangle ,\left|1\right\rangle \}$
with odd parity. A high-fidelity and QND parity
measurement can detect this error event, while leaving encoded quantum
information untouched. Once a loss error is detected (ED), a unitary
error correction (EC) recovery gate $U_{\mathrm{R}}$ can convert
the error space back to the original code space. If there is no parity
jump detected, the deterministic evolution of the code word can be
corrected by another unitary recovery gate (not shown), completing
a closed-loop QEC on the logical qubit. With the assistance of the
ancilla, high-fidelity gate operations $U_{\mathrm{L}}$ on the logic
qubit can also be implemented within the code space.}
\label{fig:Fig1} \vspace{-6pt} 
\end{figure}



For previous logical qubit schemes based on multiple physical qubits
\cite{Cory1998,Chiaverini2004,Schindler2011,Reed2012,Waldherr2014,Taminiau2014,Cramer2016,Kelly2015,Corcoles2015,Riste2015},
it is technically challenging to realize QEC and logical operations,
because the number of error channels increases with the number of qubits
and it also requires non-local gates on a collection of physical qubits
for logical operations. Recently, a different encoding architecture
based on a single bosonic oscillator (e.g. a microwave cavity mode)
was proposed~\cite{Leghtas2013,Mirrahimi2014} and has attracted a lot of interest~\cite{Vlastakis2013,SunNature,Leghtas2015,Ofek2016,Wang2016,Heeres2017,Chou2018,Rosenblum2018}. In particular, QEC protection exceeding the break-even point has been demonstrated with a cat-code encoding~\cite{Ofek2016} and gate operations on such logical qubits alone have been demonstrated without any QEC protection~\cite{Heeres2017}. Taking advantage of the infinite dimensional Hilbert space of a harmonic
oscillator, quantum information can be encoded with a single degree of freedom.
Most importantly, photon loss remains the dominant error channel,
therefore there is still only one error syndrome that needs to be monitored.
In addition, universal operations on the oscillator can be realized
by dispersively coupling to a single ancilla transmon qubit, which
allows fast and high-fidelity operations~\cite{Krastanov2015}. As
a result, the requirements on hardware are greatly reduced~\cite{Leghtas2013,Mirrahimi2014}.

Besides the above advantages, a binomial code in a microwave cavity~\cite{Michael2016} is preferable because it has exact orthonormality of the code words
and an explicit unitary operation for repumping photons into the photonic
mode. The binomial code can exactly correct errors that are polynomial
up to a specific order in photon creation and annihilation operators,
including amplitude damping and displacement noise. Recently a controlled-NOT gate on a target qubit based on the lowest-order binomial code without QEC protection has been realized~\cite{Rosenblum2018}. Here, we also choose
the lowest-order binomial code that can be protected against a single
photon loss error, and demonstrate full control on a single bosonic logical qubit with a repetitive QEC and a high-fidelity ($97.0\%$ process fidelity on average) universal gate set. The protected logical qubit
has shown 2.8 times longer lifetime than the uncorrected one, and approaches the break-even point. The demonstrated complete logical qubit operations could be the foundations for developing fault-tolerant quantum computing~\cite{Devoret2013,Roseblum2018b} and one-way quantum repeaters~\cite{Li2017}. Additionally, we demonstrate for the first time a Ramsey experiment on the QEC protected logical qubit that shows two times longer coherence than the unprotected one, paving the way towards QEC enhanced quantum metrology~\cite{Zhou2018}.

\begin{figure*}
\includegraphics{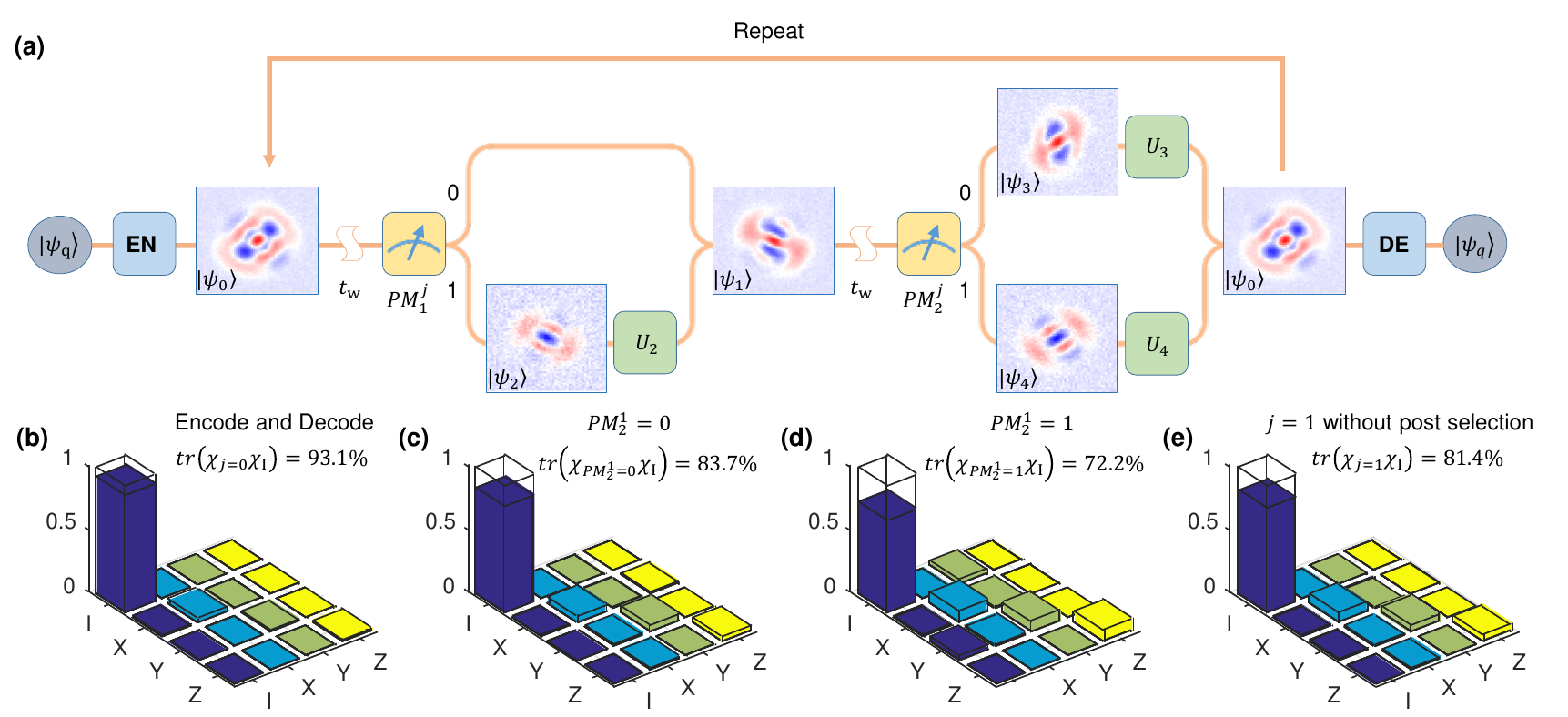} \caption{\textbf{Protocol of the repetitive QEC and process tomography.} \textbf{(a)}
The experimental procedure. The experiment begins with encoding an
arbitrary superposition state $\left|\psi_{q}\right\rangle =\alpha\left|g\right\rangle +\beta\left|e\right\rangle $
of the ancilla qubit onto the oscillator state $\left|\psi_{0}\right\rangle =\alpha\left(\left|0\right\rangle +\left|4\right\rangle \right)/\sqrt{2}+\beta\left|2\right\rangle $
(process EN). After a waiting time $t_{\mathrm{w}}\approx17.9~\mu$s,
a parity measurement is performed. $PM_{1}^{j}=0$ (the $1^{\mathrm{st}}$
parity measurement of the $j^{\mathrm{th}}$ round) indicates that
no photon loss error occurs (two or more photon loss errors can not
be distinguished but have a small probability), and the state of the
oscillator undergoes a deterministic evolution to a deformed code
space as $\left|\psi_{1}\right\rangle =\alpha(\mathrm{cos}{\theta_{1}}\left|0\right\rangle +\mathrm{sin}{\theta_{1}}e^{i\varphi_{4}}\left|4\right\rangle )/\sqrt{2}+\beta e^{i\varphi_{2}}\left|2\right\rangle $,
and no correction action is applied. $PM_{1}^{j}=1$ indicates that
one photon loss error occurs, and the oscillator state becomes $\left|\psi_{2}\right\rangle =\alpha e^{i\varphi_{3}}\left|3\right\rangle +\beta\left|1\right\rangle $.
A $\pi$ pulse is first applied to flip the ancilla qubit to the $\left|g\right\rangle $
state to minimize the detrimental effect from the ancilla qubit decoherence,
and then a unitary recovery gate $U_{2}$ is applied immediately to
convert $\left|\psi_{2}\right\rangle $ to $\left|\psi_{1}\right\rangle $.
Note that after another waiting time $t_{\mathrm{w}}$, a second parity
measurement is performed. Similarly, $PM_{2}^{j}=0$ or $1$ indicate
no or one photon loss occurs, and the oscillator state becomes $\left|\psi_{3}\right\rangle =\alpha(\mathrm{cos}{\theta_{2}}\left|0\right\rangle +\mathrm{sin}{\theta_{2}}e^{i\varphi_{4}'}\left|4\right\rangle )/\sqrt{2}+\beta e^{i\varphi_{2}'}\left|2\right\rangle $
and $\left|\psi_{4}\right\rangle =\alpha e^{i\varphi_{3}'}\left|3\right\rangle +\beta\left|1\right\rangle $,
respectively. Unitary gates $U_{3}$ and $U_{4}$ are then applied
correspondingly to restore the original state $\left|\psi_{0}\right\rangle $.
This error correction process is repeated $1-19$ times followed
by a decoding process (DE) to the ancilla qubit. $\left|\psi_{0-4}\right\rangle $
are all measured Wigner functions for $\alpha=1/\sqrt{2}$ and $\beta=-i/\sqrt{2}$.
\textbf{(b-e)} $\chi$ matrices of the process tomography, with (b)
for encoding followed immediately by the decoding process, (c)\&(d)
for one round of error correction for $PM_{2}^{1}=0$ (with probability 79.3\%) and $PM_{2}^{1}=1$ (with probability 20.7\%),
respectively, (e) for one round of correction without post-selecting $PM_{2}^{1}=0$ or 1.
Here, only the real parts are shown while the imaginary parts are
nearly zero.}
\label{fig:Fig2} \vspace{-6pt} 
\end{figure*}

\begin{figure}
\includegraphics{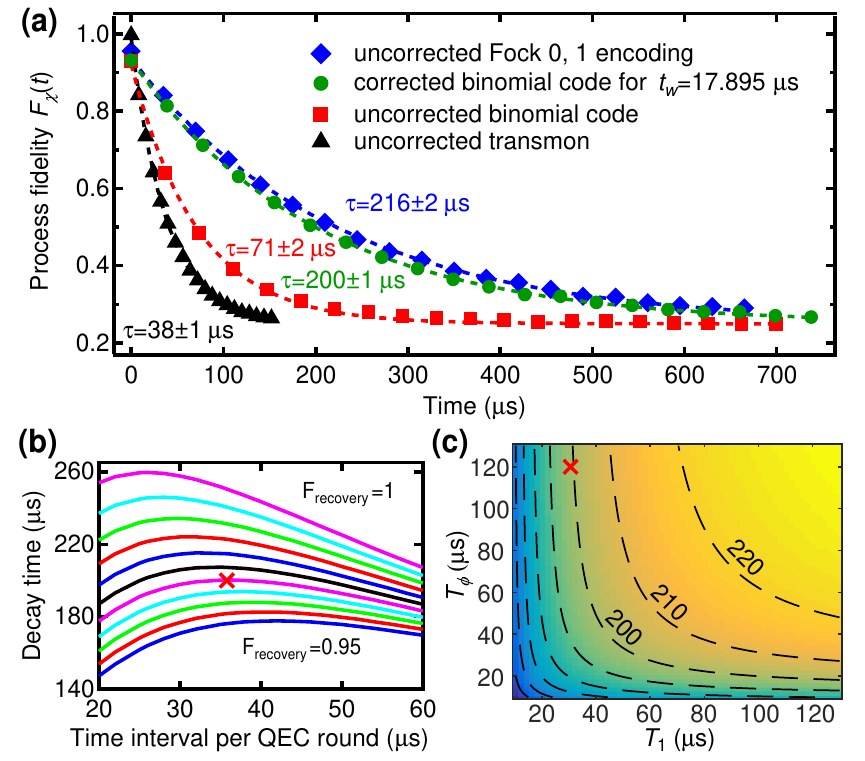} \caption{\textbf{QEC performance.} \textbf{(a)}
Green: Process fidelity $F_{\chi}(t)$ with the repetitive
QEC decays exponentially as a function of time. Each point is obtained
by 40,000 averages and the error bars are smaller than marker sizes.
For comparison, the process fidelity decay curves of the uncorrected
binomial quantum code (red), the uncorrected transmon qubit (black),
and the uncorrected Fock $\left\{ \left|0\right\rangle ,\left|1\right\rangle \right\} $
encoding (blue) are also plotted. All curves are fit using $F_{\chi}(t)=0.25+Ae^{-t/\tau}$
(dotted lines). \textbf{(b)} Theoretically expected process fidelity decay time
for the protocol in Fig.~\ref{fig:Fig2}a with the measured parameters but with different recovery gate fidelities (the colored lines correspond to fidelities from 0.95 to 1 with a step of 0.005). The cross indicates our experimental result, implying a recovery gate
fidelity of about 97.0\%. \textbf{(c)} Expected process fidelity decay
time as a function of $T_{1}$ and $T_{\phi}$. Dashed lines represent
isolines from 150~$\mu$s to 220~$\mu$s with a step of 10~$\mu$s.
The cross indicates our experimental result. In order to extend the
logical qubit lifetime beyond that of Fock state $\{\ket{0}$,$\ket{1}\}$
encoding, $T_{1}$ needs to be doubled while $T_{\phi}$ remains the
same.}
\label{fig:Fig3} \vspace{-6pt}
 
\end{figure}

\begin{figure}
\includegraphics{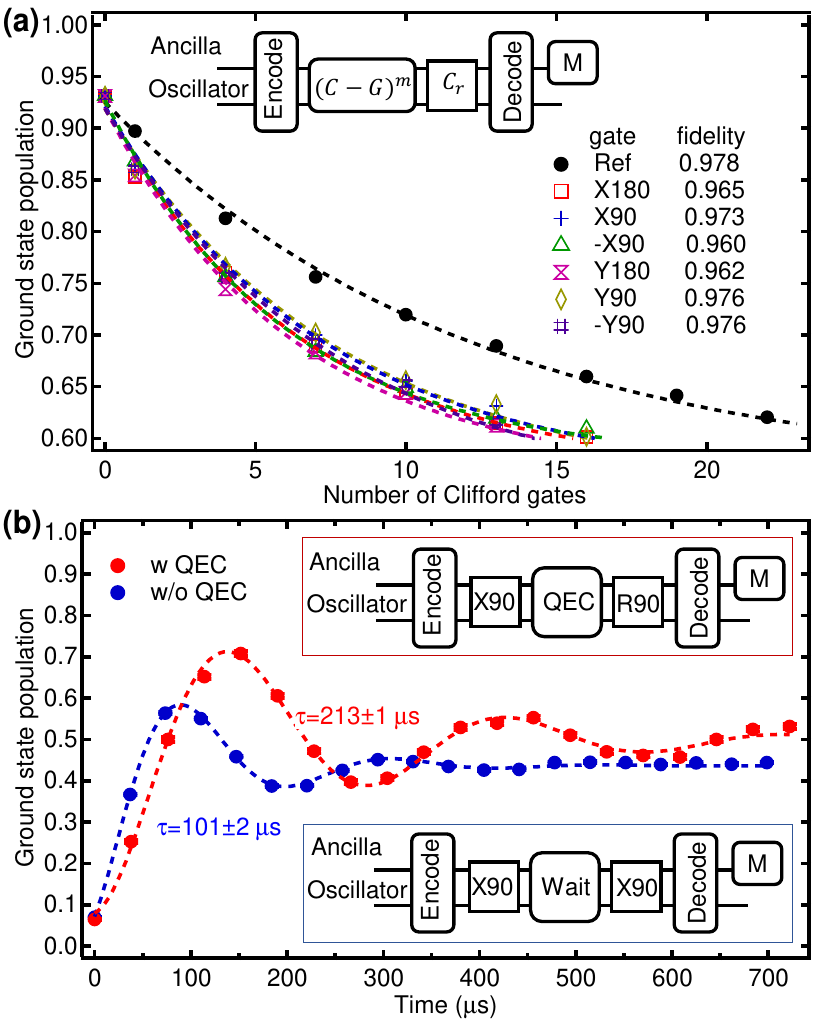} \caption{\textbf{Gate operations on the logical qubit and Ramsey interferometry
on the QEC protected logical qubit.} \textbf{(a)} Randomized benchmarking
is used to quantify the gate performance on the logical qubit, with
the protocol shown in the inset. All experiments start with an encoding
and end with a decoding followed by ancilla qubit measurement. The
reference curve is measured after applying sequences of $m$ random
Clifford gates, while the $\pm X90,\pm Y90,X180,Y180$ ($\pm\pi/2$
rotation along $x$- or $y$-axis, and $\pi$ rotation along $x$-
or $y$-axis, respectively) curves are realized after applying sequences
that interleave the corresponding gate with $m$ random Clifford gates.
Each sequence is followed by a recovery Clifford gate in the end right
before the final measurement. The number of random sequences for each
length $m$ in our experiment is $k=100$ and each random sequence
is repeated over 5,000 times. All curves are fitted by $F_{\mathrm{RB}}=Ap^{m}+B$
with different sequence decay $p$. The reference decay indicates
the average error of the single-qubit gates, while the ratio of the
interleaved and reference decay gives the specific gate fidelity.
The averaged single-qubit gate error $r_{\mathrm{gate}}\sim0.031$.
\textbf{(b)} Ramsey interferometry of logical qubit, with insets showing
the experimental sequences. The Ramsey oscillation on the QEC protected
logical qubit (red) gives a coherence time twice longer than that
without QEC protection (blue), demonstrating a real gain of the coherence
from QEC. For the process without QEC, the intervals between
gates are chosen such that the Kerr rotations of $\left|4\right\rangle $
are integer multiples of $2\pi$. The rotation axis of the second $\pi/2$
gate is fixed and the oscillation comes from the Kerr rotation of
$\left|2\right\rangle $. The population converges to a value slightly lower than 0.5, which comes from the leakage out of the computational space during
the long process without protection. For the QEC protected case, the
rotation axis of the second $\pi/2$ gate (R90) varies in the $x$-$y$
plane with an interval of $\pi/4$.}
\label{fig:Fig4} \vspace{-6pt}
 
\end{figure}

The bosonic logical qubit experiment is implemented on a circuit quantum electrodynamics
(cQED) architecture~\cite{Wallraff2004,Paik2011} with a transmon
qubit dispersively coupled to two three-dimensional (3D) cavities~\cite{Paik2011,Vlastakis2013,SunNature,Krastanov2015,Ofek2016,Liu2017},
which is illustrated schematically in Fig.~\ref{fig:Fig1}a. The
ancilla qubit has an energy relaxation time $T_{1}=30$~$\mu$s and
a pure dephasing time $T_{\phi}=120$~$\mu$s. The storage cavity
serving as an oscillator (henceforth referred as the ``oscillator\char`\"{})
for encoding logical quantum states has a single-photon lifetime $\tau_{\mathrm{s}}=143$~$\mu$s
and a coherence time $252~\mu$s. Utilizing the large Hilbert space
of the harmonic oscillator, the binomial code is constructed by superposing
Fock states with binomial coefficients~\cite{Michael2016}. In our
experiment, we choose the lowest-order binomial code with the code
words 
\begin{align}
\left|0_{\mathrm{L}}\right\rangle  & =\left(\left|0\right\rangle +\left|4\right\rangle \right)/\sqrt{2},\\
\left|1_{\mathrm{L}}\right\rangle  & =\left|2\right\rangle ,
\end{align}
as shown in Fig.~\ref{fig:Fig1}b. Both bases have the same average
photon number of two as required by the QEC criteria~\cite{Nielsen}.
This code can protect quantum information from one photon loss error,
which is the dominant error in the microwave oscillator. A single
photon loss delivers the logical states from the code space with even
parity to an error space with odd parity, i.e. $\hat{a}\left|0_{\mathrm{L}}\right\rangle =\left|3\right\rangle $
and $\hat{a}\left|1_{\mathrm{L}}\right\rangle =\left|1\right\rangle $,
where $\hat{a}$ is the photon annihilation operator of the oscillator.

A complete set of operations on the bosonic logical qubit is illustrated in
Fig.~\ref{fig:Fig1}b. Besides serving as an ancilla for the error
syndrome detection of the logical qubit, the transmon qubit provides
the necessary non-linearity for implementing the quantum encoding,
decoding, and error correcting recovery operation ($U_{\mathrm{R}}$),
as well as the universal single logical qubit operations in the binomial
code space (logical gates $U_{\mathrm{L}}$). Through a sequence of
control pulses on the system, all these operations on the logical
qubit are realized based on the dispersive interaction between the
ancilla and the oscillator~\cite{Ofek2016,Kirchmair}
\begin{equation}
H_{\mathrm{int}}=-\chi_{\mathrm{qs}}\hat{a}^{\dagger}\hat{a}\left|e\right\rangle \left\langle e\right|-\frac{K}{2}\hat{a}^{\dagger2}\hat{a}^{2},
\end{equation}
where $\left|e\right\rangle $ is the excited state of the ancilla
qubit ($\ket{g}$ is the ground state), $\chi_{\mathrm{qs}}/2\pi=1.90$~MHz
is the dispersive interaction strength, and $K/2\pi=4.2$~kHz is
the self-Kerr coefficient of the oscillator. The control pulses are
synchronized and generated by field programmable gate arrays (FPGA)
with home-made logics, and allow for fast real-time feedback control
of the logical qubit (see Supplementary Materials for the experimental
apparatus). Specially, the pulse shapes are numerically optimized
with the gradient ascent pulse engineering (GRAPE) method~\cite{Khaneja2005,DeFouquieres2011}
based on carefully calibrated experimental parameters.

First, we demonstrate the encoding and decoding process, where quantum
information is transferred between the ancilla qubit $\{\left|g\right\rangle ,\left|e\right\rangle \}$
and the binomial code space $\{\left|0_{\mathrm{L}}\right\rangle ,\left|1_{\mathrm{L}}\right\rangle \}$
of the oscillator. Process tomography~\cite{Nielsen} is used to
benchmark our encoding and decoding performance, and the fidelity
is defined as $F_{\chi}=\mathrm{tr}(\chi_{\mathrm{M}}\chi_{\mathrm{ideal}})$,
with $\chi_{\mathrm{M}}\left(\chi_{\mathrm{ideal}}\right)$ being the
derived $4\times4$ process matrix for experimental (ideal) operation
(see Methods). Figure~\ref{fig:Fig2}a shows the $\chi_{\mathrm{M}}$
for a sequential encoding and decoding process ($\chi_{\mathrm{ideal}}=\mathbb{I}$)
and indicates a fidelity of $93.1\%$.


The quantum information in the binomial code space can be protected
from a single photon loss by the QEC process (including both error
detection and correction). A photon number parity measurement can
distinguish the code and error spaces (Fig.~\ref{fig:Fig1}b) without
perturbing the encoded information~\cite{SunNature}, and thus serves
as the error syndrome for error detection. Once a parity change is
detected, the state in the error space can be converted back to the
original code space by a unitary recovery gate $U_{\mathrm{R}}$.
As opposed to the cat code, where corrections can be performed at
the end of error-syndrome tracking~\cite{Ofek2016}, the photon loss
error in our experiment needs to be corrected immediately since this
particular binomial code does not tolerate two or more photon losses.
Repeated such processes can therefore protect the information stored
in the logical qubit. However, a trade-off needs to be considered
between more frequent parity measurement to avoid missing photon loss
errors and finite detection and recovery fidelities causing extra
information loss.

The experimental protocol of repetitive QEC is shown in Fig.~\ref{fig:Fig2}a,
where a two-layer QEC procedure is adapted to balance the operation
and photon loss errors. The top-layer QEC consists of several bottom-layer
QEC steps, and each step corrects photon loss error but tolerates the backaction from the detection of no parity jump
until the last step. Therefore, the top-layer QEC recovers the quantum
information in code space and is repeated many times, while the bottom-layer
QEC conserves parity in a deformed code space. The optimal scheme
for current experiments consists of two bottom-layer QEC steps with
a waiting time $t_{\mathrm{w}}=17.9~\mu$s. The outcome of first error
detection $PM=0$ indicates that no photon loss error occurs and the
state of the oscillator undergoes an evolution from $\left|\psi_{0}\right\rangle =\alpha\left|0_{\mathrm{L}}\right\rangle +\beta\left|0_{\mathrm{L}}\right\rangle $
to $\left|\psi_{1}\right\rangle =\alpha\left|0_{\mathrm{L}}^{\prime}\right\rangle +\beta\left|1_{\mathrm{L}}^{\prime}\right\rangle $,
with a deformation
of code space bases $\left|0_{\mathrm{L}}^{\prime}\right\rangle =(\mathrm{cos}{\theta_{1}}\left|0\right\rangle +\mathrm{sin}{\theta_{1}}e^{i\varphi_{4}}\left|4\right\rangle )/\sqrt{2}$
and $\left|1_{\mathrm{L}}^{\prime}\right\rangle =e^{i\varphi_{2}}\left|2\right\rangle$. We note that the above unitary evolution is only an approximation to the non-unitary backaction associated with the no-parity-jump evolution $e^{-(\kappa/2) \hat{a}^\dagger\hat{a} t}$ valid to first order in $\kappa t_{\mathrm{w}}$~\cite{Michael2016}. It is also worth noting that no detected parity change can not rule
out the possibility of having two-photon losses ($\sim2.1\%$), which
causes complete quantum information loss. For $PM=1$, one photon
loss error occurs (the probability of having a three-photon loss $\sim0.14\%$
can be neglected), and the oscillator state becomes $\left|\psi_{2}\right\rangle =\alpha e^{i\varphi_{3}}\left|3\right\rangle +\beta\left|1\right\rangle $
and a unitary recovery gate $U_{2}$ has to be applied immediately
to convert $\left|\psi_{2}\right\rangle $ to $\left|\psi_{1}\right\rangle $
in the deformed code space. After an additional waiting time $t_{\mathrm{w}}$, another
error detection is performed for the second bottom-layer QEC step.
Similarly, $PM=0$ and $1$ indicate a further deformation of code space to
$\left|\psi_{3}\right\rangle $ and a jump to error space with $\left|\psi_{4}\right\rangle =\alpha e^{i\varphi_{3}'}\left|3\right\rangle +\beta\left|1\right\rangle $,
respectively. Then, unitary gates $U_{3}$ and $U_{4}$ are applied
correspondingly to restore the original state $\left|\psi_{0}\right\rangle $,
completing one round of top-layer error correction. In Figs.$\,$\ref{fig:Fig2}c-e,
the process matrices are presented for the process after the second
error detection in the first round of top-layer QEC, showing a good
fidelity of $81.4\%$.

Figure~\ref{fig:Fig3}a shows the process fidelity $F_{\chi}(t)$
with the repetitive QEC (green) decaying exponentially as a function
of time, where the two-layer error correction process is repeated
$j=1-19$ times. For comparison, the process fidelities of the uncorrected
binomial quantum code (red), the uncorrected transmon qubit (black),
and the uncorrected Fock $\{\left|0\right\rangle $,$\left|1\right\rangle \}$
encoding of the oscillator (blue) are also plotted. All curves are
fitted using $F_{\chi}(t)=0.25+Ae^{-t/\tau}$, where $\tau$
is the lifetime. $\tau$ of the corrected binomial code is 5.3 times
longer than the uncorrected transmon qubit, 2.8 times longer than
the uncorrected binomial code, and only less than $8\%$ lower than
the uncorrected Fock encoding that defines the break-even point for the QEC~\cite{Ofek2016}. These results demonstrate that the experimental
system and scheme are robust and can indeed protect the encoded bosonic
code from the photon loss error.

The experiment is mainly limited by the decoherence of the ancilla qubit
that induces imperfections and deserves further investigation. Since
the ancilla facilitates both error detection and gate operations,
the ancilla decoherence induces errors for these processes and prevents
more frequent QEC. On the other hand, with a larger interval between
QECs, there is a higher probability of having undetectable two-photon
losses and a larger dephasing effect induced by photon losses due
to the non-commutativity of the annihilation operation $\hat{a}$
and the self-Kerr term $\frac{K}{2}\hat{a}^{\dagger2}\hat{a}^{2}$
(see Supplementary Materials). To illustrate this trade-off relation,
Fig.~\ref{fig:Fig3}b shows the numerically predicted decay time as
a function of the tracking interval per two error detections following
the protocol in Fig.~\ref{fig:Fig2}a. It predicts a shorter optimal
$t_{\mathrm{w}}$ and a longer decay time for a higher recovery gate fidelity.
Our experimental result is indicated by the cross, implying a recovery
gate fidelity of about $97.0\%$ that is mainly limited by the qubit
decoherence during the recovery gates. In order to extend the logical
qubit life time beyond that of Fock state $\{\ket{0}$,$\ket{1}\}$
encoding, either a better strategy with at least four bottom-layer
QEC steps needs to be implemented with current qubit coherence (but
demanding more adaptive gate calibrations, see Supplementary Materials),
or the coherence time of the ancilla qubit needs to be improved. As
depicted in Fig.~\ref{fig:Fig3}c, this goal can be achieved if we
can double $T_{1}$ while keep the same $T_{\phi}$. By comparison,
a cat code does not have this recovery gate error issue, and instead
only requires tracking the number of errors and a correction at the
end of the whole QEC process~\cite{Ofek2016}. If the recovery gate
itself is perfect and we only have detection errors, we in principle
can achieve a decay time after QEC over 260~$\mu$s, more than $20\%$
longer than for Fock state $\{\left|0\right\rangle $,$\left|1\right\rangle \}$
encoding.

To fully exploit the binomial encoding for future quantum information
processing, gate operations on the logical qubits are indispensable.
Thanks to the single bosonic oscillator encoding, we are able to implement
the logical qubit gates by a universal control of the state of the oscillator, instead of encoding/decoding to the ancilla qubit.
A randomized benchmarking experiment~\cite{Barends2014,Knill2008,Ryan2009,Magesan2012}
is performed to determine the fidelity of the Clifford gates on the
logical qubit. Figure~\ref{fig:Fig4}a shows the results with an
averaged gate error $r_{\mathrm{gate}}\sim0.031$. These gates fidelities
could be improved further by a more careful calibration of the GRAPE
pulses. We note that the $T$ gate does not belong to the Clifford group, so it can not be characterized by randomized benchmarking.
We instead perform repeated gates to extract the gate fidelity of
about $98.7\%$ (data shown in the Supplementary Materials).

With the QEC and the high-fidelity universal operation of the logical
qubit in hand, we demonstrate a Ramsey experiment on the QEC protected
logical qubit that indeed shows longer coherence. Figure~\ref{fig:Fig4}b
insets show the experimental sequences for the logical qubit Ramsey
experiments with and without QEC, and both experiments also combine
an encoding and decoding process. The results are fitted with an exponentially damped sinusoidal function. The Ramsey experiment performed
on the logical qubit without QEC gives a coherence time of $101~\mu$s.
The interference fringes against the evolution time correspond to
a Kerr nonlinearity induced phase change of the logical qubit. It is worth noting that the self-Kerr does not cause errors in the lowest-order binomial code in the absence of cavity photon decay. A particular frame can be chosen to match the rotation of $\ket{4}$ relative to $\ket{0}$ so that $\ket{0}$ and $\ket{4}$ are degenerate, while $\ket{2}$ has a slightly different frequency. However, the resulting phase rotation of the logical qubit is purely deterministic and thus there is no decoherence associated with the self-Kerr effect. This is true even if the higher-order corrections to Kerr nonlinearity are taken into account, which is one of the nice features of the lowest-order binomial code. 

For comparison, the experiment on the QEC protected logical qubit gives a twice longer
coherence time of $213~\mu$s, demonstrating a real gain of the coherence
from QEC. The Ramsey oscillation in this QEC protected case comes
from the variable rotation axis of the second $\pi/2$ gate. Since
the Ramsey interferometry has been widely used for precision measurements,
our results represent an important advance towards QEC enhanced
metrology based on logical qubits.


In summary, we have demonstrated a real-time repetitive QEC and universal
gate set on a bosonic logical qubit in a superconducting circuit. Our experiment
shows that the binomial encoding in a 3D cQED architecture is hardware-efficient
for QEC and logical operations and represents a promising platform
towards future quantum algorithms based on multiple logical qubits
with QEC protection. Our work provides a starting point for future
works: realize logical qudit with binomial code, generalize the tools
realized in this work to multiple oscillators (logical qubits), and
realize fault-tolerant error detection, correction, as well as logical
gates. Current experimental techniques are feasible to realize more
logical qubits by integrating more cavities or modes in a compact
3D structure~\cite{Axline2016}. Our results motivate further investigations
of the superconducting quantum processor and envision the route towards
fault-tolerant implementation of QEC and gates based on bosonic encoding~\cite{Devoret2013,Roseblum2018b}.



\begin{thebibliography}{47}%
\makeatletter
\providecommand \@ifxundefined [1]{%
 \@ifx{#1\undefined}
}%
\providecommand \@ifnum [1]{%
 \ifnum #1\expandafter \@firstoftwo
 \else \expandafter \@secondoftwo
 \fi
}%
\providecommand \@ifx [1]{%
 \ifx #1\expandafter \@firstoftwo
 \else \expandafter \@secondoftwo
 \fi
}%
\providecommand \natexlab [1]{#1}%
\providecommand \enquote  [1]{``#1''}%
\providecommand \bibnamefont  [1]{#1}%
\providecommand \bibfnamefont [1]{#1}%
\providecommand \citenamefont [1]{#1}%
\providecommand \href@noop [0]{\@secondoftwo}%
\providecommand \href [0]{\begingroup \@sanitize@url \@href}%
\providecommand \@href[1]{\@@startlink{#1}\@@href}%
\providecommand \@@href[1]{\endgroup#1\@@endlink}%
\providecommand \@sanitize@url [0]{\catcode `\\12\catcode `\$12\catcode
  `\&12\catcode `\#12\catcode `\^12\catcode `\_12\catcode `\%12\relax}%
\providecommand \@@startlink[1]{}%
\providecommand \@@endlink[0]{}%
\providecommand \url  [0]{\begingroup\@sanitize@url \@url }%
\providecommand \@url [1]{\endgroup\@href {#1}{\urlprefix }}%
\providecommand \urlprefix  [0]{URL }%
\providecommand \Eprint [0]{\href }%
\providecommand \doibase [0]{http://dx.doi.org/}%
\providecommand \selectlanguage [0]{\@gobble}%
\providecommand \bibinfo  [0]{\@secondoftwo}%
\providecommand \bibfield  [0]{\@secondoftwo}%
\providecommand \translation [1]{[#1]}%
\providecommand \BibitemOpen [0]{}%
\providecommand \bibitemStop [0]{}%
\providecommand \bibitemNoStop [0]{.\EOS\space}%
\providecommand \EOS [0]{\spacefactor3000\relax}%
\providecommand \BibitemShut  [1]{\csname bibitem#1\endcsname}%
\let\auto@bib@innerbib\@empty
\bibitem [{\citenamefont {Nielsen}\ and\ \citenamefont
  {Chuang}(2000)}]{Nielsen}%
  \BibitemOpen
  \bibfield  {author} {\bibinfo {author} {\bibfnamefont {M.~A.}\ \bibnamefont
  {Nielsen}}\ and\ \bibinfo {author} {\bibfnamefont {I.~L.}\ \bibnamefont
  {Chuang}},\ }\href@noop {} {\emph {\bibinfo {title} {Quantum Computation and
  Quantum Information}}}\ (\bibinfo  {publisher} {Cambridge Univ. Press},\
  \bibinfo {year} {2000})\BibitemShut {NoStop}%
\bibitem [{\citenamefont {Shor}(1995)}]{Shor1995}%
  \BibitemOpen
  \bibfield  {author} {\bibinfo {author} {\bibfnamefont {P.~W.}\ \bibnamefont
  {Shor}},\ }\bibfield  {title} {\enquote {\bibinfo {title} {Scheme for
  reducing decoherence in quantum computer memory},}\ }\href {\doibase
  10.1103/PhysRevA.52.R2493} {\bibfield  {journal} {\bibinfo  {journal} {Phys.
  Rev. A}\ }\textbf {\bibinfo {volume} {52}},\ \bibinfo {pages} {2493}
  (\bibinfo {year} {1995})}\BibitemShut {NoStop}%
\bibitem [{\citenamefont {Steane}(1996)}]{Steane1996}%
  \BibitemOpen
  \bibfield  {author} {\bibinfo {author} {\bibfnamefont {A.}~\bibnamefont
  {Steane}},\ }\bibfield  {title} {\enquote {\bibinfo {title} {Multiple
  particle interference and quantum error correction},}\ }\href {\doibase
  10.1098/rspa.1996.0136} {\bibfield  {journal} {\bibinfo  {journal} {Proc.
  Roy. Soc. Lond. A}\ }\textbf {\bibinfo {volume} {452}},\ \bibinfo {pages}
  {2551} (\bibinfo {year} {1996})}\BibitemShut {NoStop}%
\bibitem [{\citenamefont {Gottsman}(2010)}]{Gottsman2010}%
  \BibitemOpen
  \bibfield  {author} {\bibinfo {author} {\bibfnamefont {D.}~\bibnamefont
  {Gottsman}},\ }\bibfield  {title} {\enquote {\bibinfo {title} {An
  introduction to quantum error correction and fault-tolerant quantum
  computation},}\ }\href@noop {} {\bibfield  {journal} {\bibinfo  {journal}
  {Proceeding of Symposia in Applied Mathematics}\ }\textbf {\bibinfo {volume}
  {68}},\ \bibinfo {pages} {13} (\bibinfo {year} {2010})}\BibitemShut {NoStop}%
\bibitem [{\citenamefont {Fowler}\ \emph {et~al.}(2012)\citenamefont {Fowler},
  \citenamefont {Mariantoni}, \citenamefont {Martinis},\ and\ \citenamefont
  {Cleland}}]{Fowler2012}%
  \BibitemOpen
  \bibfield  {author} {\bibinfo {author} {\bibfnamefont {A.~G.}\ \bibnamefont
  {Fowler}}, \bibinfo {author} {\bibfnamefont {M.}~\bibnamefont {Mariantoni}},
  \bibinfo {author} {\bibfnamefont {J.~M.}\ \bibnamefont {Martinis}}, \ and\
  \bibinfo {author} {\bibfnamefont {A.~N.}\ \bibnamefont {Cleland}},\
  }\bibfield  {title} {\enquote {\bibinfo {title} {Surface codes: Towards
  practical large-scale quantum computation},}\ }\href {\doibase
  10.1103/PhysRevA.86.032324} {\bibfield  {journal} {\bibinfo  {journal} {Phys.
  Rev. A}\ }\textbf {\bibinfo {volume} {86}},\ \bibinfo {pages} {032324}
  (\bibinfo {year} {2012})}\BibitemShut {NoStop}%
\bibitem [{\citenamefont {Devoret}\ and\ \citenamefont
  {Schoelkopf}(2013)}]{Devoret2013}%
  \BibitemOpen
  \bibfield  {author} {\bibinfo {author} {\bibfnamefont {M.~H.}\ \bibnamefont
  {Devoret}}\ and\ \bibinfo {author} {\bibfnamefont {R.~J.}\ \bibnamefont
  {Schoelkopf}},\ }\bibfield  {title} {\enquote {\bibinfo {title}
  {{Superconducting circuits for quantum information: an outlook.}}}\ }\href
  {\doibase 10.1126/science.1231930} {\bibfield  {journal} {\bibinfo  {journal}
  {Science}\ }\textbf {\bibinfo {volume} {339}},\ \bibinfo {pages} {1169}
  (\bibinfo {year} {2013})}\BibitemShut {NoStop}%
\bibitem [{\citenamefont {Cory}\ \emph {et~al.}(1998)\citenamefont {Cory},
  \citenamefont {Price}, \citenamefont {Maas}, \citenamefont {Knill},
  \citenamefont {Laflamme}, \citenamefont {Zurek}, \citenamefont {Havel},\ and\
  \citenamefont {Somaroo}}]{Cory1998}%
  \BibitemOpen
  \bibfield  {author} {\bibinfo {author} {\bibfnamefont {D.~G.}\ \bibnamefont
  {Cory}}, \bibinfo {author} {\bibfnamefont {M.}~\bibnamefont {Price}},
  \bibinfo {author} {\bibfnamefont {W.}~\bibnamefont {Maas}}, \bibinfo {author}
  {\bibfnamefont {E.}~\bibnamefont {Knill}}, \bibinfo {author} {\bibfnamefont
  {R.}~\bibnamefont {Laflamme}}, \bibinfo {author} {\bibfnamefont {W.~H.}\
  \bibnamefont {Zurek}}, \bibinfo {author} {\bibfnamefont {T.~F.}\ \bibnamefont
  {Havel}}, \ and\ \bibinfo {author} {\bibfnamefont {S.}~\bibnamefont
  {Somaroo}},\ }\bibfield  {title} {\enquote {\bibinfo {title} {Experimental
  quantum error correction},}\ }\href {\doibase 10.1103/PhysRevLett.81.2152}
  {\bibfield  {journal} {\bibinfo  {journal} {Phys. Rev. Lett.}\ }\textbf
  {\bibinfo {volume} {81}},\ \bibinfo {pages} {2152} (\bibinfo {year}
  {1998})}\BibitemShut {NoStop}%
\bibitem [{\citenamefont {Chiaverini}\ \emph {et~al.}(2004)\citenamefont
  {Chiaverini}, \citenamefont {Leibfried}, \citenamefont {Schaetz},
  \citenamefont {Barrett}, \citenamefont {Blakestad}, \citenamefont {Britton},
  \citenamefont {Itano}, \citenamefont {Jost}, \citenamefont {Knill},
  \citenamefont {Langer}, \citenamefont {Ozeri},\ and\ \citenamefont
  {Wineland}}]{Chiaverini2004}%
  \BibitemOpen
  \bibfield  {author} {\bibinfo {author} {\bibfnamefont {J.}~\bibnamefont
  {Chiaverini}}, \bibinfo {author} {\bibfnamefont {D.}~\bibnamefont
  {Leibfried}}, \bibinfo {author} {\bibfnamefont {T.}~\bibnamefont {Schaetz}},
  \bibinfo {author} {\bibfnamefont {M.~D.}\ \bibnamefont {Barrett}}, \bibinfo
  {author} {\bibfnamefont {R.~B.}\ \bibnamefont {Blakestad}}, \bibinfo {author}
  {\bibfnamefont {J.}~\bibnamefont {Britton}}, \bibinfo {author} {\bibfnamefont
  {W.~M.}\ \bibnamefont {Itano}}, \bibinfo {author} {\bibfnamefont {J.~D.}\
  \bibnamefont {Jost}}, \bibinfo {author} {\bibfnamefont {E.}~\bibnamefont
  {Knill}}, \bibinfo {author} {\bibfnamefont {C.}~\bibnamefont {Langer}},
  \bibinfo {author} {\bibfnamefont {R.}~\bibnamefont {Ozeri}}, \ and\ \bibinfo
  {author} {\bibfnamefont {D.~J.}\ \bibnamefont {Wineland}},\ }\bibfield
  {title} {\enquote {\bibinfo {title} {Realization of quantum error
  correction},}\ }\href {\doibase 10.1038/nature03074} {\bibfield  {journal}
  {\bibinfo  {journal} {Nature}\ }\textbf {\bibinfo {volume} {432}},\ \bibinfo
  {pages} {602} (\bibinfo {year} {2004})}\BibitemShut {NoStop}%
\bibitem [{\citenamefont {Schindler}\ \emph {et~al.}(2011)\citenamefont
  {Schindler}, \citenamefont {Barreiro}, \citenamefont {Monz}, \citenamefont
  {Nebendahl}, \citenamefont {Nigg}, \citenamefont {Chwalla}, \citenamefont
  {Hennrich},\ and\ \citenamefont {Blatt}}]{Schindler2011}%
  \BibitemOpen
  \bibfield  {author} {\bibinfo {author} {\bibfnamefont {P.}~\bibnamefont
  {Schindler}}, \bibinfo {author} {\bibfnamefont {J.~T.}\ \bibnamefont
  {Barreiro}}, \bibinfo {author} {\bibfnamefont {T.}~\bibnamefont {Monz}},
  \bibinfo {author} {\bibfnamefont {V.}~\bibnamefont {Nebendahl}}, \bibinfo
  {author} {\bibfnamefont {D.}~\bibnamefont {Nigg}}, \bibinfo {author}
  {\bibfnamefont {M.}~\bibnamefont {Chwalla}}, \bibinfo {author} {\bibfnamefont
  {M.}~\bibnamefont {Hennrich}}, \ and\ \bibinfo {author} {\bibfnamefont
  {R.}~\bibnamefont {Blatt}},\ }\bibfield  {title} {\enquote {\bibinfo {title}
  {{Experimental Repetitive Quantum Error Correction}},}\ }\href {\doibase
  10.1126/science.1203329} {\bibfield  {journal} {\bibinfo  {journal}
  {Science}\ }\textbf {\bibinfo {volume} {332}},\ \bibinfo {pages} {1059}
  (\bibinfo {year} {2011})}\BibitemShut {NoStop}%
\bibitem [{\citenamefont {Reed}\ \emph {et~al.}(2012)\citenamefont {Reed},
  \citenamefont {DiCarlo}, \citenamefont {Nigg}, \citenamefont {Sun},
  \citenamefont {Frunzio}, \citenamefont {Girvin},\ and\ \citenamefont
  {Schoelkopf}}]{Reed2012}%
  \BibitemOpen
  \bibfield  {author} {\bibinfo {author} {\bibfnamefont {M.~D.}\ \bibnamefont
  {Reed}}, \bibinfo {author} {\bibfnamefont {L.}~\bibnamefont {DiCarlo}},
  \bibinfo {author} {\bibfnamefont {S.~E.}\ \bibnamefont {Nigg}}, \bibinfo
  {author} {\bibfnamefont {L.}~\bibnamefont {Sun}}, \bibinfo {author}
  {\bibfnamefont {L.}~\bibnamefont {Frunzio}}, \bibinfo {author} {\bibfnamefont
  {S.~M.}\ \bibnamefont {Girvin}}, \ and\ \bibinfo {author} {\bibfnamefont
  {R.~J.}\ \bibnamefont {Schoelkopf}},\ }\bibfield  {title} {\enquote {\bibinfo
  {title} {{Realization of three-qubit quantum error correction with
  superconducting circuits}},}\ }\href {\doibase 10.1038/nature10786}
  {\bibfield  {journal} {\bibinfo  {journal} {Nature}\ }\textbf {\bibinfo
  {volume} {482}},\ \bibinfo {pages} {382} (\bibinfo {year}
  {2012})}\BibitemShut {NoStop}%
\bibitem [{\citenamefont {Waldherr}\ \emph {et~al.}(2014)\citenamefont
  {Waldherr}, \citenamefont {Wang}, \citenamefont {Zaiser}, \citenamefont
  {Jamali}, \citenamefont {Schulte-Herbr{\"{u}}ggen}, \citenamefont {Abe},
  \citenamefont {Ohshima}, \citenamefont {Isoya}, \citenamefont {Du},
  \citenamefont {Neumann},\ and\ \citenamefont {Wrachtrup}}]{Waldherr2014}%
  \BibitemOpen
  \bibfield  {author} {\bibinfo {author} {\bibfnamefont {G.}~\bibnamefont
  {Waldherr}}, \bibinfo {author} {\bibfnamefont {Y.}~\bibnamefont {Wang}},
  \bibinfo {author} {\bibfnamefont {S.}~\bibnamefont {Zaiser}}, \bibinfo
  {author} {\bibfnamefont {M.}~\bibnamefont {Jamali}}, \bibinfo {author}
  {\bibfnamefont {T.}~\bibnamefont {Schulte-Herbr{\"{u}}ggen}}, \bibinfo
  {author} {\bibfnamefont {H.}~\bibnamefont {Abe}}, \bibinfo {author}
  {\bibfnamefont {T.}~\bibnamefont {Ohshima}}, \bibinfo {author} {\bibfnamefont
  {J.}~\bibnamefont {Isoya}}, \bibinfo {author} {\bibfnamefont {J.~F.}\
  \bibnamefont {Du}}, \bibinfo {author} {\bibfnamefont {P.}~\bibnamefont
  {Neumann}}, \ and\ \bibinfo {author} {\bibfnamefont {J.}~\bibnamefont
  {Wrachtrup}},\ }\bibfield  {title} {\enquote {\bibinfo {title} {{Quantum
  error correction in a solid-state hybrid spin register.}}}\ }\href {\doibase
  10.1038/nature12919} {\bibfield  {journal} {\bibinfo  {journal} {Nature}\
  }\textbf {\bibinfo {volume} {506}},\ \bibinfo {pages} {204} (\bibinfo {year}
  {2014})}\BibitemShut {NoStop}%
\bibitem [{\citenamefont {Taminiau}\ \emph {et~al.}(2014)\citenamefont
  {Taminiau}, \citenamefont {Cramer}, \citenamefont {van~der Sar},
  \citenamefont {Dobrovitski},\ and\ \citenamefont {Hanson}}]{Taminiau2014}%
  \BibitemOpen
  \bibfield  {author} {\bibinfo {author} {\bibfnamefont {T.~H.}\ \bibnamefont
  {Taminiau}}, \bibinfo {author} {\bibfnamefont {J.}~\bibnamefont {Cramer}},
  \bibinfo {author} {\bibfnamefont {T.}~\bibnamefont {van~der Sar}}, \bibinfo
  {author} {\bibfnamefont {V.~V.}\ \bibnamefont {Dobrovitski}}, \ and\ \bibinfo
  {author} {\bibfnamefont {R.}~\bibnamefont {Hanson}},\ }\bibfield  {title}
  {\enquote {\bibinfo {title} {{Universal control and error correction in
  multi-qubit spin registers in diamond.}}}\ }\href {\doibase
  10.1038/nnano.2014.2} {\bibfield  {journal} {\bibinfo  {journal} {Nat.
  Nanotechnol.}\ }\textbf {\bibinfo {volume} {9}},\ \bibinfo {pages} {171}
  (\bibinfo {year} {2014})}\BibitemShut {NoStop}%
\bibitem [{\citenamefont {Cramer}\ \emph {et~al.}(2016)\citenamefont {Cramer},
  \citenamefont {Kalb}, \citenamefont {Rol}, \citenamefont {Hensen},
  \citenamefont {Blok}, \citenamefont {Markham}, \citenamefont {Twitchen},
  \citenamefont {Hanson},\ and\ \citenamefont {Taminiau}}]{Cramer2016}%
  \BibitemOpen
  \bibfield  {author} {\bibinfo {author} {\bibfnamefont {J.}~\bibnamefont
  {Cramer}}, \bibinfo {author} {\bibfnamefont {N.}~\bibnamefont {Kalb}},
  \bibinfo {author} {\bibfnamefont {M.~A.}\ \bibnamefont {Rol}}, \bibinfo
  {author} {\bibfnamefont {B.}~\bibnamefont {Hensen}}, \bibinfo {author}
  {\bibfnamefont {M.~S.}\ \bibnamefont {Blok}}, \bibinfo {author}
  {\bibfnamefont {M.}~\bibnamefont {Markham}}, \bibinfo {author} {\bibfnamefont
  {D.~J.}\ \bibnamefont {Twitchen}}, \bibinfo {author} {\bibfnamefont
  {R.}~\bibnamefont {Hanson}}, \ and\ \bibinfo {author} {\bibfnamefont {T.~H.}\
  \bibnamefont {Taminiau}},\ }\bibfield  {title} {\enquote {\bibinfo {title}
  {{Repeated quantum error correction on a continuously encoded qubit by
  real-time feedback}},}\ }\href {\doibase 10.1038/ncomms11526} {\bibfield
  {journal} {\bibinfo  {journal} {Nat. Commun.}\ }\textbf {\bibinfo {volume}
  {7}},\ \bibinfo {pages} {11526} (\bibinfo {year} {2016})}\BibitemShut
  {NoStop}%
\bibitem [{\citenamefont {Kelly}\ \emph {et~al.}(2015)\citenamefont {Kelly},
  \citenamefont {Barends}, \citenamefont {Fowler}, \citenamefont {Megrant},
  \citenamefont {Jeffrey}, \citenamefont {White}, \citenamefont {Sank},
  \citenamefont {Mutus}, \citenamefont {Campbell}, \citenamefont {Chen},
  \citenamefont {Chen}, \citenamefont {Chiaro}, \citenamefont {Dunsworth},
  \citenamefont {Hoi}, \citenamefont {Neill}, \citenamefont {O'Malley},
  \citenamefont {Quintana}, \citenamefont {Roushan}, \citenamefont
  {Vainsencher}, \citenamefont {Wenner}, \citenamefont {Cleland},\ and\
  \citenamefont {Martinis}}]{Kelly2015}%
  \BibitemOpen
  \bibfield  {author} {\bibinfo {author} {\bibfnamefont {J.}~\bibnamefont
  {Kelly}}, \bibinfo {author} {\bibfnamefont {R.}~\bibnamefont {Barends}},
  \bibinfo {author} {\bibfnamefont {A.~G.}\ \bibnamefont {Fowler}}, \bibinfo
  {author} {\bibfnamefont {A.}~\bibnamefont {Megrant}}, \bibinfo {author}
  {\bibfnamefont {E.}~\bibnamefont {Jeffrey}}, \bibinfo {author} {\bibfnamefont
  {T.~C.}\ \bibnamefont {White}}, \bibinfo {author} {\bibfnamefont
  {D.}~\bibnamefont {Sank}}, \bibinfo {author} {\bibfnamefont {J.~Y.}\
  \bibnamefont {Mutus}}, \bibinfo {author} {\bibfnamefont {B.}~\bibnamefont
  {Campbell}}, \bibinfo {author} {\bibfnamefont {Y.}~\bibnamefont {Chen}},
  \bibinfo {author} {\bibfnamefont {Z.}~\bibnamefont {Chen}}, \bibinfo {author}
  {\bibfnamefont {B.}~\bibnamefont {Chiaro}}, \bibinfo {author} {\bibfnamefont
  {A.}~\bibnamefont {Dunsworth}}, \bibinfo {author} {\bibfnamefont {I.-C.}\
  \bibnamefont {Hoi}}, \bibinfo {author} {\bibfnamefont {C.}~\bibnamefont
  {Neill}}, \bibinfo {author} {\bibfnamefont {P.~J.~J.}\ \bibnamefont
  {O'Malley}}, \bibinfo {author} {\bibfnamefont {C.}~\bibnamefont {Quintana}},
  \bibinfo {author} {\bibfnamefont {P.}~\bibnamefont {Roushan}}, \bibinfo
  {author} {\bibfnamefont {A.}~\bibnamefont {Vainsencher}}, \bibinfo {author}
  {\bibfnamefont {J.}~\bibnamefont {Wenner}}, \bibinfo {author} {\bibfnamefont
  {A.~N.}\ \bibnamefont {Cleland}}, \ and\ \bibinfo {author} {\bibfnamefont
  {J.~M.}\ \bibnamefont {Martinis}},\ }\bibfield  {title} {\enquote {\bibinfo
  {title} {{State preservation by repetitive error detection in a
  superconducting quantum circuit}},}\ }\href {\doibase 10.1038/nature14270}
  {\bibfield  {journal} {\bibinfo  {journal} {Nature}\ }\textbf {\bibinfo
  {volume} {519}},\ \bibinfo {pages} {66} (\bibinfo {year} {2015})}\BibitemShut
  {NoStop}%
\bibitem [{\citenamefont {C\'{o}rcoles}\ \emph {et~al.}(2015)\citenamefont
  {C\'{o}rcoles}, \citenamefont {Magesan}, \citenamefont {Srinivasan},
  \citenamefont {Cross}, \citenamefont {Steffen}, \citenamefont {Gambetta},\
  and\ \citenamefont {Chow}}]{Corcoles2015}%
  \BibitemOpen
  \bibfield  {author} {\bibinfo {author} {\bibfnamefont {A.~D.}\ \bibnamefont
  {C\'{o}rcoles}}, \bibinfo {author} {\bibfnamefont {E.}~\bibnamefont
  {Magesan}}, \bibinfo {author} {\bibfnamefont {S.~J.}\ \bibnamefont
  {Srinivasan}}, \bibinfo {author} {\bibfnamefont {A.~W.}\ \bibnamefont
  {Cross}}, \bibinfo {author} {\bibfnamefont {M.}~\bibnamefont {Steffen}},
  \bibinfo {author} {\bibfnamefont {J.~M.}\ \bibnamefont {Gambetta}}, \ and\
  \bibinfo {author} {\bibfnamefont {J.~M.}\ \bibnamefont {Chow}},\ }\bibfield
  {title} {\enquote {\bibinfo {title} {Demonstration of a quantum error
  detection code using a square lattice of four superconducting qubits},}\
  }\href@noop {} {\bibfield  {journal} {\bibinfo  {journal} {Nature
  Communications}\ }\textbf {\bibinfo {volume} {6}},\ \bibinfo {pages} {6979}
  (\bibinfo {year} {2015})}\BibitemShut {NoStop}%
\bibitem [{\citenamefont {Rist{\`{e}}}\ \emph {et~al.}(2015)\citenamefont
  {Rist{\`{e}}}, \citenamefont {Poletto}, \citenamefont {Huang}, \citenamefont
  {Bruno}, \citenamefont {Vesterinen}, \citenamefont {Saira},\ and\
  \citenamefont {DiCarlo}}]{Riste2015}%
  \BibitemOpen
  \bibfield  {author} {\bibinfo {author} {\bibfnamefont {D.}~\bibnamefont
  {Rist{\`{e}}}}, \bibinfo {author} {\bibfnamefont {S.}~\bibnamefont
  {Poletto}}, \bibinfo {author} {\bibfnamefont {M.-Z.}\ \bibnamefont {Huang}},
  \bibinfo {author} {\bibfnamefont {A.}~\bibnamefont {Bruno}}, \bibinfo
  {author} {\bibfnamefont {V.}~\bibnamefont {Vesterinen}}, \bibinfo {author}
  {\bibfnamefont {O.-P.}\ \bibnamefont {Saira}}, \ and\ \bibinfo {author}
  {\bibfnamefont {L.}~\bibnamefont {DiCarlo}},\ }\bibfield  {title} {\enquote
  {\bibinfo {title} {{Detecting bit-flip errors in a logical qubit using
  stabilizer measurements}},}\ }\href {\doibase 10.1038/ncomms7983} {\bibfield
  {journal} {\bibinfo  {journal} {Nat. Commun.}\ }\textbf {\bibinfo {volume}
  {6}},\ \bibinfo {pages} {6983} (\bibinfo {year} {2015})}\BibitemShut
  {NoStop}%
\bibitem [{\citenamefont {Leghtas}\ \emph {et~al.}(2013)\citenamefont
  {Leghtas}, \citenamefont {Kirchmair}, \citenamefont {Vlastakis},
  \citenamefont {Schoelkopf}, \citenamefont {Devoret},\ and\ \citenamefont
  {Mirrahimi}}]{Leghtas2013}%
  \BibitemOpen
  \bibfield  {author} {\bibinfo {author} {\bibfnamefont {Z.}~\bibnamefont
  {Leghtas}}, \bibinfo {author} {\bibfnamefont {G.}~\bibnamefont {Kirchmair}},
  \bibinfo {author} {\bibfnamefont {B.}~\bibnamefont {Vlastakis}}, \bibinfo
  {author} {\bibfnamefont {R.~J.}\ \bibnamefont {Schoelkopf}}, \bibinfo
  {author} {\bibfnamefont {M.~H.}\ \bibnamefont {Devoret}}, \ and\ \bibinfo
  {author} {\bibfnamefont {M.}~\bibnamefont {Mirrahimi}},\ }\bibfield  {title}
  {\enquote {\bibinfo {title} {{Hardware-Efficient Autonomous Quantum Memory
  Protection}},}\ }\href {\doibase 10.1103/PhysRevLett.111.120501} {\bibfield
  {journal} {\bibinfo  {journal} {Phys. Rev. Lett.}\ }\textbf {\bibinfo
  {volume} {111}},\ \bibinfo {pages} {120501} (\bibinfo {year}
  {2013})}\BibitemShut {NoStop}%
\bibitem [{\citenamefont {Mirrahimi}\ \emph {et~al.}(2014)\citenamefont
  {Mirrahimi}, \citenamefont {Leghtas}, \citenamefont {Albert}, \citenamefont
  {Touzard}, \citenamefont {Schoelkopf}, \citenamefont {Jiang},\ and\
  \citenamefont {Devoret}}]{Mirrahimi2014}%
  \BibitemOpen
  \bibfield  {author} {\bibinfo {author} {\bibfnamefont {M.}~\bibnamefont
  {Mirrahimi}}, \bibinfo {author} {\bibfnamefont {Z.}~\bibnamefont {Leghtas}},
  \bibinfo {author} {\bibfnamefont {V.~V.}\ \bibnamefont {Albert}}, \bibinfo
  {author} {\bibfnamefont {S.}~\bibnamefont {Touzard}}, \bibinfo {author}
  {\bibfnamefont {R.~J.}\ \bibnamefont {Schoelkopf}}, \bibinfo {author}
  {\bibfnamefont {L.}~\bibnamefont {Jiang}}, \ and\ \bibinfo {author}
  {\bibfnamefont {M.~H.}\ \bibnamefont {Devoret}},\ }\bibfield  {title}
  {\enquote {\bibinfo {title} {{Dynamically protected cat-qubits: a new
  paradigm for universal quantum computation}},}\ }\href {\doibase
  10.1088/1367-2630/16/4/045014} {\bibfield  {journal} {\bibinfo  {journal}
  {New J. Phys.}\ }\textbf {\bibinfo {volume} {16}},\ \bibinfo {pages} {045014}
  (\bibinfo {year} {2014})}\BibitemShut {NoStop}%
\bibitem [{\citenamefont {Vlastakis}\ \emph {et~al.}(2013)\citenamefont
  {Vlastakis}, \citenamefont {Kirchmair}, \citenamefont {Leghtas},
  \citenamefont {Nigg}, \citenamefont {Frunzio}, \citenamefont {Girvin},
  \citenamefont {Mirrahimi}, \citenamefont {Devoret},\ and\ \citenamefont
  {Schoelkopf}}]{Vlastakis2013}%
  \BibitemOpen
  \bibfield  {author} {\bibinfo {author} {\bibfnamefont {B.}~\bibnamefont
  {Vlastakis}}, \bibinfo {author} {\bibfnamefont {G.}~\bibnamefont
  {Kirchmair}}, \bibinfo {author} {\bibfnamefont {Z.}~\bibnamefont {Leghtas}},
  \bibinfo {author} {\bibfnamefont {S.~E.}\ \bibnamefont {Nigg}}, \bibinfo
  {author} {\bibfnamefont {L.}~\bibnamefont {Frunzio}}, \bibinfo {author}
  {\bibfnamefont {S.~M.}\ \bibnamefont {Girvin}}, \bibinfo {author}
  {\bibfnamefont {M.}~\bibnamefont {Mirrahimi}}, \bibinfo {author}
  {\bibfnamefont {M.~H.}\ \bibnamefont {Devoret}}, \ and\ \bibinfo {author}
  {\bibfnamefont {R.~J.}\ \bibnamefont {Schoelkopf}},\ }\bibfield  {title}
  {\enquote {\bibinfo {title} {{Deterministically encoding quantum information
  using 100-photon Schr{\"{o}}dinger cat states.}}}\ }\href {\doibase
  10.1126/science.1243289} {\bibfield  {journal} {\bibinfo  {journal}
  {Science}\ }\textbf {\bibinfo {volume} {342}},\ \bibinfo {pages} {607}
  (\bibinfo {year} {2013})}\BibitemShut {NoStop}%
\bibitem [{\citenamefont {Sun}\ \emph {et~al.}(2014)\citenamefont {Sun},
  \citenamefont {Petrenko}, \citenamefont {Leghtas}, \citenamefont {Vlastakis},
  \citenamefont {Kirchmair}, \citenamefont {Sliwa}, \citenamefont {Narla},
  \citenamefont {Hatridge}, \citenamefont {Shankar}, \citenamefont {Blumoff},
  \citenamefont {Frunzio}, \citenamefont {Mirrahimi}, \citenamefont {Devoret},\
  and\ \citenamefont {Schoelkopf}}]{SunNature}%
  \BibitemOpen
  \bibfield  {author} {\bibinfo {author} {\bibfnamefont {L.}~\bibnamefont
  {Sun}}, \bibinfo {author} {\bibfnamefont {A.}~\bibnamefont {Petrenko}},
  \bibinfo {author} {\bibfnamefont {Z.}~\bibnamefont {Leghtas}}, \bibinfo
  {author} {\bibfnamefont {B.}~\bibnamefont {Vlastakis}}, \bibinfo {author}
  {\bibfnamefont {G.}~\bibnamefont {Kirchmair}}, \bibinfo {author}
  {\bibfnamefont {K.~M.}\ \bibnamefont {Sliwa}}, \bibinfo {author}
  {\bibfnamefont {A.}~\bibnamefont {Narla}}, \bibinfo {author} {\bibfnamefont
  {M.}~\bibnamefont {Hatridge}}, \bibinfo {author} {\bibfnamefont
  {S.}~\bibnamefont {Shankar}}, \bibinfo {author} {\bibfnamefont
  {J.}~\bibnamefont {Blumoff}}, \bibinfo {author} {\bibfnamefont
  {L.}~\bibnamefont {Frunzio}}, \bibinfo {author} {\bibfnamefont
  {M.}~\bibnamefont {Mirrahimi}}, \bibinfo {author} {\bibfnamefont {M.~H.}\
  \bibnamefont {Devoret}}, \ and\ \bibinfo {author} {\bibfnamefont {R.~J.}\
  \bibnamefont {Schoelkopf}},\ }\bibfield  {title} {\enquote {\bibinfo {title}
  {Tracking photon jumps with repeated quantum non-demolition parity
  measurements},}\ }\href {\doibase 10.1038/nature13436} {\bibfield  {journal}
  {\bibinfo  {journal} {Nature}\ }\textbf {\bibinfo {volume} {511}},\ \bibinfo
  {pages} {444} (\bibinfo {year} {2014})}\BibitemShut {NoStop}%
\bibitem [{\citenamefont {Leghtas}\ \emph {et~al.}(2015)\citenamefont
  {Leghtas}, \citenamefont {Touzard}, \citenamefont {Pop}, \citenamefont {Kou},
  \citenamefont {Vlastakis}, \citenamefont {Petrenko}, \citenamefont {Sliwa},
  \citenamefont {Narla}, \citenamefont {Shankar}, \citenamefont {Hatridge},
  \citenamefont {Reagor}, \citenamefont {Frunzio}, \citenamefont {Schoelkopf},
  \citenamefont {Mirrahimi},\ and\ \citenamefont {Devoret}}]{Leghtas2015}%
  \BibitemOpen
  \bibfield  {author} {\bibinfo {author} {\bibfnamefont {Z.}~\bibnamefont
  {Leghtas}}, \bibinfo {author} {\bibfnamefont {S.}~\bibnamefont {Touzard}},
  \bibinfo {author} {\bibfnamefont {I.~M.}\ \bibnamefont {Pop}}, \bibinfo
  {author} {\bibfnamefont {A.}~\bibnamefont {Kou}}, \bibinfo {author}
  {\bibfnamefont {B.}~\bibnamefont {Vlastakis}}, \bibinfo {author}
  {\bibfnamefont {A.}~\bibnamefont {Petrenko}}, \bibinfo {author}
  {\bibfnamefont {K.~M.}\ \bibnamefont {Sliwa}}, \bibinfo {author}
  {\bibfnamefont {A.}~\bibnamefont {Narla}}, \bibinfo {author} {\bibfnamefont
  {S.}~\bibnamefont {Shankar}}, \bibinfo {author} {\bibfnamefont {M.~J.}\
  \bibnamefont {Hatridge}}, \bibinfo {author} {\bibfnamefont {M.}~\bibnamefont
  {Reagor}}, \bibinfo {author} {\bibfnamefont {L.}~\bibnamefont {Frunzio}},
  \bibinfo {author} {\bibfnamefont {R.~J.}\ \bibnamefont {Schoelkopf}},
  \bibinfo {author} {\bibfnamefont {M.}~\bibnamefont {Mirrahimi}}, \ and\
  \bibinfo {author} {\bibfnamefont {M.~H.}\ \bibnamefont {Devoret}},\
  }\bibfield  {title} {\enquote {\bibinfo {title} {{Confining the state of
  light to a quantum manifold by engineered two-photon loss}},}\ }\href
  {\doibase 10.1126/science.aaa2085} {\bibfield  {journal} {\bibinfo  {journal}
  {Science}\ }\textbf {\bibinfo {volume} {347}},\ \bibinfo {pages} {853}
  (\bibinfo {year} {2015})}\BibitemShut {NoStop}%
\bibitem [{\citenamefont {Ofek}\ \emph {et~al.}(2016)\citenamefont {Ofek},
  \citenamefont {Petrenko}, \citenamefont {Heeres}, \citenamefont {Reinhold},
  \citenamefont {Leghtas}, \citenamefont {Vlastakis}, \citenamefont {Liu},
  \citenamefont {Frunzio}, \citenamefont {Girvin}, \citenamefont {Jiang},
  \citenamefont {Mirrahimi}, \citenamefont {Devoret},\ and\ \citenamefont
  {Schoelkopf}}]{Ofek2016}%
  \BibitemOpen
  \bibfield  {author} {\bibinfo {author} {\bibfnamefont {N.}~\bibnamefont
  {Ofek}}, \bibinfo {author} {\bibfnamefont {A.}~\bibnamefont {Petrenko}},
  \bibinfo {author} {\bibfnamefont {R.}~\bibnamefont {Heeres}}, \bibinfo
  {author} {\bibfnamefont {P.}~\bibnamefont {Reinhold}}, \bibinfo {author}
  {\bibfnamefont {Z.}~\bibnamefont {Leghtas}}, \bibinfo {author} {\bibfnamefont
  {B.}~\bibnamefont {Vlastakis}}, \bibinfo {author} {\bibfnamefont
  {Y.}~\bibnamefont {Liu}}, \bibinfo {author} {\bibfnamefont {L.}~\bibnamefont
  {Frunzio}}, \bibinfo {author} {\bibfnamefont {S.~M.}\ \bibnamefont {Girvin}},
  \bibinfo {author} {\bibfnamefont {L.}~\bibnamefont {Jiang}}, \bibinfo
  {author} {\bibfnamefont {M.}~\bibnamefont {Mirrahimi}}, \bibinfo {author}
  {\bibfnamefont {M.~H.}\ \bibnamefont {Devoret}}, \ and\ \bibinfo {author}
  {\bibfnamefont {R.~J.}\ \bibnamefont {Schoelkopf}},\ }\bibfield  {title}
  {\enquote {\bibinfo {title} {{Extending the lifetime of a quantum bit with
  error correction in superconducting circuits}},}\ }\href {\doibase
  10.1038/nature18949} {\bibfield  {journal} {\bibinfo  {journal} {Nature}\
  }\textbf {\bibinfo {volume} {536}},\ \bibinfo {pages} {441} (\bibinfo {year}
  {2016})}\BibitemShut {NoStop}%
\bibitem [{\citenamefont {Wang}\ \emph {et~al.}(2016)\citenamefont {Wang},
  \citenamefont {Gao}, \citenamefont {Reinhold}, \citenamefont {Heeres},
  \citenamefont {Ofek}, \citenamefont {Chou}, \citenamefont {Axline},
  \citenamefont {Reagor}, \citenamefont {Blumoff}, \citenamefont {Sliwa},
  \citenamefont {Frunzio}, \citenamefont {Girvin}, \citenamefont {Jiang},
  \citenamefont {Mirrahimi}, \citenamefont {Devoret},\ and\ \citenamefont
  {Schoelkopf}}]{Wang2016}%
  \BibitemOpen
  \bibfield  {author} {\bibinfo {author} {\bibfnamefont {C.}~\bibnamefont
  {Wang}}, \bibinfo {author} {\bibfnamefont {Y.~Y.}\ \bibnamefont {Gao}},
  \bibinfo {author} {\bibfnamefont {P.}~\bibnamefont {Reinhold}}, \bibinfo
  {author} {\bibfnamefont {R.~W.}\ \bibnamefont {Heeres}}, \bibinfo {author}
  {\bibfnamefont {N.}~\bibnamefont {Ofek}}, \bibinfo {author} {\bibfnamefont
  {K.}~\bibnamefont {Chou}}, \bibinfo {author} {\bibfnamefont {C.}~\bibnamefont
  {Axline}}, \bibinfo {author} {\bibfnamefont {M.}~\bibnamefont {Reagor}},
  \bibinfo {author} {\bibfnamefont {J.}~\bibnamefont {Blumoff}}, \bibinfo
  {author} {\bibfnamefont {K.~M.}\ \bibnamefont {Sliwa}}, \bibinfo {author}
  {\bibfnamefont {L.}~\bibnamefont {Frunzio}}, \bibinfo {author} {\bibfnamefont
  {S.~M.}\ \bibnamefont {Girvin}}, \bibinfo {author} {\bibfnamefont
  {L.}~\bibnamefont {Jiang}}, \bibinfo {author} {\bibfnamefont
  {M.}~\bibnamefont {Mirrahimi}}, \bibinfo {author} {\bibfnamefont {M.~H.}\
  \bibnamefont {Devoret}}, \ and\ \bibinfo {author} {\bibfnamefont {R.~J.}\
  \bibnamefont {Schoelkopf}},\ }\bibfield  {title} {\enquote {\bibinfo {title}
  {{A Schrodinger cat living in two boxes}},}\ }\href {\doibase
  10.1126/science.aaf2941} {\bibfield  {journal} {\bibinfo  {journal}
  {Science}\ }\textbf {\bibinfo {volume} {352}},\ \bibinfo {pages} {1087}
  (\bibinfo {year} {2016})}\BibitemShut {NoStop}%
\bibitem [{\citenamefont {Heeres}\ \emph {et~al.}(2017)\citenamefont {Heeres},
  \citenamefont {Reinhold}, \citenamefont {Ofek}, \citenamefont {Frunzio},
  \citenamefont {Jiang}, \citenamefont {Devoret},\ and\ \citenamefont
  {Schoelkopf}}]{Heeres2017}%
  \BibitemOpen
  \bibfield  {author} {\bibinfo {author} {\bibfnamefont {R.~W.}\ \bibnamefont
  {Heeres}}, \bibinfo {author} {\bibfnamefont {P.}~\bibnamefont {Reinhold}},
  \bibinfo {author} {\bibfnamefont {N.}~\bibnamefont {Ofek}}, \bibinfo {author}
  {\bibfnamefont {L.}~\bibnamefont {Frunzio}}, \bibinfo {author} {\bibfnamefont
  {L.}~\bibnamefont {Jiang}}, \bibinfo {author} {\bibfnamefont {M.~H.}\
  \bibnamefont {Devoret}}, \ and\ \bibinfo {author} {\bibfnamefont {R.~J.}\
  \bibnamefont {Schoelkopf}},\ }\bibfield  {title} {\enquote {\bibinfo {title}
  {{Implementing a universal gate set on a logical qubit encoded in an
  oscillator}},}\ }\href {\doibase 10.1038/s41467-017-00045-1} {\bibfield
  {journal} {\bibinfo  {journal} {Nat. Commun.}\ }\textbf {\bibinfo {volume}
  {8}},\ \bibinfo {pages} {94} (\bibinfo {year} {2017})}\BibitemShut {NoStop}%
\bibitem [{\citenamefont {Chou}\ \emph {et~al.}(2018)\citenamefont {Chou},
  \citenamefont {Blumoff}, \citenamefont {Wang}, \citenamefont {Reinhold},
  \citenamefont {Axline}, \citenamefont {Gao}, \citenamefont {Frunzio},
  \citenamefont {Devoret}, \citenamefont {Jiang},\ and\ \citenamefont
  {Schoelkopf}}]{Chou2018}%
  \BibitemOpen
  \bibfield  {author} {\bibinfo {author} {\bibfnamefont {K.~S.}\ \bibnamefont
  {Chou}}, \bibinfo {author} {\bibfnamefont {J.~Z.}\ \bibnamefont {Blumoff}},
  \bibinfo {author} {\bibfnamefont {C.~S.}\ \bibnamefont {Wang}}, \bibinfo
  {author} {\bibfnamefont {P.~C.}\ \bibnamefont {Reinhold}}, \bibinfo {author}
  {\bibfnamefont {C.~J.}\ \bibnamefont {Axline}}, \bibinfo {author}
  {\bibfnamefont {Y.~Y.}\ \bibnamefont {Gao}}, \bibinfo {author} {\bibfnamefont
  {L.}~\bibnamefont {Frunzio}}, \bibinfo {author} {\bibfnamefont {M.~H.}\
  \bibnamefont {Devoret}}, \bibinfo {author} {\bibfnamefont {L.}~\bibnamefont
  {Jiang}}, \ and\ \bibinfo {author} {\bibfnamefont {R.~J.}\ \bibnamefont
  {Schoelkopf}},\ }\bibfield  {title} {\enquote {\bibinfo {title}
  {{Deterministic teleportation of a quantum gate between two logical
  qubits}},}\ }\href {http://arxiv.org/abs/1801.05283} {\bibfield  {journal}
  {\bibinfo  {journal} {arXiv}\ ,\ \bibinfo {pages} {1801.05283}} (\bibinfo
  {year} {2018})}\BibitemShut {NoStop}%
\bibitem [{\citenamefont {Rosenblum}\ \emph
  {et~al.}(2018{\natexlab{a}})\citenamefont {Rosenblum}, \citenamefont {Gao},
  \citenamefont {Reinhold}, \citenamefont {Wang}, \citenamefont {Axline},
  \citenamefont {Frunzio}, \citenamefont {Girvin}, \citenamefont {Jiang},
  \citenamefont {Mirrahimi}, \citenamefont {Devoret},\ and\ \citenamefont
  {Schoelkopf}}]{Rosenblum2018}%
  \BibitemOpen
  \bibfield  {author} {\bibinfo {author} {\bibfnamefont {S.}~\bibnamefont
  {Rosenblum}}, \bibinfo {author} {\bibfnamefont {Y.~Y.}\ \bibnamefont {Gao}},
  \bibinfo {author} {\bibfnamefont {P.}~\bibnamefont {Reinhold}}, \bibinfo
  {author} {\bibfnamefont {C.}~\bibnamefont {Wang}}, \bibinfo {author}
  {\bibfnamefont {C.~J.}\ \bibnamefont {Axline}}, \bibinfo {author}
  {\bibfnamefont {L.}~\bibnamefont {Frunzio}}, \bibinfo {author} {\bibfnamefont
  {S.~M.}\ \bibnamefont {Girvin}}, \bibinfo {author} {\bibfnamefont
  {L.}~\bibnamefont {Jiang}}, \bibinfo {author} {\bibfnamefont
  {M.}~\bibnamefont {Mirrahimi}}, \bibinfo {author} {\bibfnamefont {M.~H.}\
  \bibnamefont {Devoret}}, \ and\ \bibinfo {author} {\bibfnamefont {R.~J.}\
  \bibnamefont {Schoelkopf}},\ }\bibfield  {title} {\enquote {\bibinfo {title}
  {{A CNOT gate between multiphoton qubits encoded in two cavities}},}\ }\href
  {\doibase 10.1038/s41467-018-03059-5} {\bibfield  {journal} {\bibinfo
  {journal} {Nat. Commun.}\ }\textbf {\bibinfo {volume} {9}},\ \bibinfo {pages}
  {652} (\bibinfo {year} {2018}{\natexlab{a}})}\BibitemShut {NoStop}%
\bibitem [{\citenamefont {Krastanov}\ \emph {et~al.}(2015)\citenamefont
  {Krastanov}, \citenamefont {Albert}, \citenamefont {Shen}, \citenamefont
  {Zou}, \citenamefont {Heeres}, \citenamefont {Vlastakis}, \citenamefont
  {Schoelkopf},\ and\ \citenamefont {Jiang}}]{Krastanov2015}%
  \BibitemOpen
  \bibfield  {author} {\bibinfo {author} {\bibfnamefont {S.}~\bibnamefont
  {Krastanov}}, \bibinfo {author} {\bibfnamefont {V.~V.}\ \bibnamefont
  {Albert}}, \bibinfo {author} {\bibfnamefont {C.}~\bibnamefont {Shen}},
  \bibinfo {author} {\bibfnamefont {C.-L.}\ \bibnamefont {Zou}}, \bibinfo
  {author} {\bibfnamefont {R.~W.}\ \bibnamefont {Heeres}}, \bibinfo {author}
  {\bibfnamefont {B.}~\bibnamefont {Vlastakis}}, \bibinfo {author}
  {\bibfnamefont {R.~J.}\ \bibnamefont {Schoelkopf}}, \ and\ \bibinfo {author}
  {\bibfnamefont {L.}~\bibnamefont {Jiang}},\ }\bibfield  {title} {\enquote
  {\bibinfo {title} {{Universal control of an oscillator with dispersive
  coupling to a qubit}},}\ }\href {\doibase 10.1103/PhysRevA.92.040303}
  {\bibfield  {journal} {\bibinfo  {journal} {Phys. Rev. A}\ }\textbf {\bibinfo
  {volume} {92}},\ \bibinfo {pages} {040303} (\bibinfo {year}
  {2015})}\BibitemShut {NoStop}%
\bibitem [{\citenamefont {Michael}\ \emph {et~al.}(2016)\citenamefont
  {Michael}, \citenamefont {Silveri}, \citenamefont {Brierley}, \citenamefont
  {Albert}, \citenamefont {Salmilehto}, \citenamefont {Jiang},\ and\
  \citenamefont {Girvin}}]{Michael2016}%
  \BibitemOpen
  \bibfield  {author} {\bibinfo {author} {\bibfnamefont {M.~H.}\ \bibnamefont
  {Michael}}, \bibinfo {author} {\bibfnamefont {M.}~\bibnamefont {Silveri}},
  \bibinfo {author} {\bibfnamefont {R.~T.}\ \bibnamefont {Brierley}}, \bibinfo
  {author} {\bibfnamefont {V.~V.}\ \bibnamefont {Albert}}, \bibinfo {author}
  {\bibfnamefont {J.}~\bibnamefont {Salmilehto}}, \bibinfo {author}
  {\bibfnamefont {L.}~\bibnamefont {Jiang}}, \ and\ \bibinfo {author}
  {\bibfnamefont {S.~M.}\ \bibnamefont {Girvin}},\ }\bibfield  {title}
  {\enquote {\bibinfo {title} {{New Class of Quantum Error-Correcting Codes for
  a Bosonic Mode}},}\ }\href {\doibase 10.1103/PhysRevX.6.031006} {\bibfield
  {journal} {\bibinfo  {journal} {Phys. Rev. X}\ }\textbf {\bibinfo {volume}
  {6}},\ \bibinfo {pages} {031006} (\bibinfo {year} {2016})}\BibitemShut
  {NoStop}%
\bibitem [{\citenamefont {Rosenblum}\ \emph
  {et~al.}(2018{\natexlab{b}})\citenamefont {Rosenblum}, \citenamefont
  {Reinhold}, \citenamefont {Mirrahimi}, \citenamefont {Jiang}, \citenamefont
  {Frunzio},\ and\ \citenamefont {Schoelkopf}}]{Roseblum2018b}%
  \BibitemOpen
  \bibfield  {author} {\bibinfo {author} {\bibfnamefont {S.}~\bibnamefont
  {Rosenblum}}, \bibinfo {author} {\bibfnamefont {P.}~\bibnamefont {Reinhold}},
  \bibinfo {author} {\bibfnamefont {M.}~\bibnamefont {Mirrahimi}}, \bibinfo
  {author} {\bibfnamefont {L.}~\bibnamefont {Jiang}}, \bibinfo {author}
  {\bibfnamefont {L.}~\bibnamefont {Frunzio}}, \ and\ \bibinfo {author}
  {\bibfnamefont {R.}~\bibnamefont {Schoelkopf}},\ }\bibfield  {title}
  {\enquote {\bibinfo {title} {Fault-tolerant measurement of a quantum error
  syndrome},}\ }\href@noop {} {\bibfield  {journal} {\bibinfo  {journal}
  {arXiv:1803.00102}\ } (\bibinfo {year} {2018}{\natexlab{b}})}\BibitemShut
  {NoStop}%
\bibitem [{\citenamefont {Li}\ \emph {et~al.}(2017)\citenamefont {Li},
  \citenamefont {Zou}, \citenamefont {Albert}, \citenamefont {Muralidharan},
  \citenamefont {Girvin},\ and\ \citenamefont {Jiang}}]{Li2017}%
  \BibitemOpen
  \bibfield  {author} {\bibinfo {author} {\bibfnamefont {L.}~\bibnamefont
  {Li}}, \bibinfo {author} {\bibfnamefont {C.-L.}\ \bibnamefont {Zou}},
  \bibinfo {author} {\bibfnamefont {V.~V.}\ \bibnamefont {Albert}}, \bibinfo
  {author} {\bibfnamefont {S.}~\bibnamefont {Muralidharan}}, \bibinfo {author}
  {\bibfnamefont {S.~M.}\ \bibnamefont {Girvin}}, \ and\ \bibinfo {author}
  {\bibfnamefont {L.}~\bibnamefont {Jiang}},\ }\bibfield  {title} {\enquote
  {\bibinfo {title} {{Cat Codes with Optimal Decoherence Suppression for a
  Lossy Bosonic Channel}},}\ }\href {\doibase 10.1103/PhysRevLett.119.030502}
  {\bibfield  {journal} {\bibinfo  {journal} {Phys. Rev. Lett.}\ }\textbf
  {\bibinfo {volume} {119}},\ \bibinfo {pages} {030502} (\bibinfo {year}
  {2017})}\BibitemShut {NoStop}%
\bibitem [{\citenamefont {Zhou}\ \emph {et~al.}(2018)\citenamefont {Zhou},
  \citenamefont {Zhang}, \citenamefont {Preskill},\ and\ \citenamefont
  {Jiang}}]{Zhou2018}%
  \BibitemOpen
  \bibfield  {author} {\bibinfo {author} {\bibfnamefont {S.}~\bibnamefont
  {Zhou}}, \bibinfo {author} {\bibfnamefont {M.}~\bibnamefont {Zhang}},
  \bibinfo {author} {\bibfnamefont {J.}~\bibnamefont {Preskill}}, \ and\
  \bibinfo {author} {\bibfnamefont {L.}~\bibnamefont {Jiang}},\ }\bibfield
  {title} {\enquote {\bibinfo {title} {{Achieving the Heisenberg limit in
  quantum metrology using quantum error correction}},}\ }\href {\doibase
  10.1038/s41467-017-02510-3} {\bibfield  {journal} {\bibinfo  {journal} {Nat.
  Commun.}\ }\textbf {\bibinfo {volume} {9}},\ \bibinfo {pages} {78} (\bibinfo
  {year} {2018})}\BibitemShut {NoStop}%
\bibitem [{\citenamefont {Wallraff}\ \emph {et~al.}(2004)\citenamefont
  {Wallraff}, \citenamefont {Schuster}, \citenamefont {Blais}, \citenamefont
  {Frunzio}, \citenamefont {Huang}, \citenamefont {Majer}, \citenamefont
  {Kumar}, \citenamefont {Girvin},\ and\ \citenamefont
  {Schoelkopf}}]{Wallraff2004}%
  \BibitemOpen
  \bibfield  {author} {\bibinfo {author} {\bibfnamefont {A.}~\bibnamefont
  {Wallraff}}, \bibinfo {author} {\bibfnamefont {D.~I.}\ \bibnamefont
  {Schuster}}, \bibinfo {author} {\bibfnamefont {A.}~\bibnamefont {Blais}},
  \bibinfo {author} {\bibfnamefont {L.}~\bibnamefont {Frunzio}}, \bibinfo
  {author} {\bibfnamefont {R.-S.}\ \bibnamefont {Huang}}, \bibinfo {author}
  {\bibfnamefont {J.}~\bibnamefont {Majer}}, \bibinfo {author} {\bibfnamefont
  {S.}~\bibnamefont {Kumar}}, \bibinfo {author} {\bibfnamefont {S.~M.}\
  \bibnamefont {Girvin}}, \ and\ \bibinfo {author} {\bibfnamefont {R.~J.}\
  \bibnamefont {Schoelkopf}},\ }\bibfield  {title} {\enquote {\bibinfo {title}
  {{Strong coupling of a single photon to a superconducting qubit using circuit
  quantum electrodynamics}},}\ }\href {\doibase 10.1038/nature02851} {\bibfield
   {journal} {\bibinfo  {journal} {Nature}\ }\textbf {\bibinfo {volume}
  {431}},\ \bibinfo {pages} {162} (\bibinfo {year} {2004})}\BibitemShut
  {NoStop}%
\bibitem [{\citenamefont {Paik}\ \emph {et~al.}(2011)\citenamefont {Paik},
  \citenamefont {Schuster}, \citenamefont {Bishop}, \citenamefont {Kirchmair},
  \citenamefont {Catelani}, \citenamefont {Sears}, \citenamefont {Johnson},
  \citenamefont {Reagor}, \citenamefont {Frunzio}, \citenamefont {Glazman},
  \citenamefont {Girvin}, \citenamefont {Devoret},\ and\ \citenamefont
  {Schoelkopf}}]{Paik2011}%
  \BibitemOpen
  \bibfield  {author} {\bibinfo {author} {\bibfnamefont {H.}~\bibnamefont
  {Paik}}, \bibinfo {author} {\bibfnamefont {D.~I.}\ \bibnamefont {Schuster}},
  \bibinfo {author} {\bibfnamefont {L.~S.}\ \bibnamefont {Bishop}}, \bibinfo
  {author} {\bibfnamefont {G.}~\bibnamefont {Kirchmair}}, \bibinfo {author}
  {\bibfnamefont {G.}~\bibnamefont {Catelani}}, \bibinfo {author}
  {\bibfnamefont {a.~P.}\ \bibnamefont {Sears}}, \bibinfo {author}
  {\bibfnamefont {B.~R.}\ \bibnamefont {Johnson}}, \bibinfo {author}
  {\bibfnamefont {M.~J.}\ \bibnamefont {Reagor}}, \bibinfo {author}
  {\bibfnamefont {L.}~\bibnamefont {Frunzio}}, \bibinfo {author} {\bibfnamefont
  {L.~I.}\ \bibnamefont {Glazman}}, \bibinfo {author} {\bibfnamefont {S.~M.}\
  \bibnamefont {Girvin}}, \bibinfo {author} {\bibfnamefont {M.~H.}\
  \bibnamefont {Devoret}}, \ and\ \bibinfo {author} {\bibfnamefont {R.~J.}\
  \bibnamefont {Schoelkopf}},\ }\bibfield  {title} {\enquote {\bibinfo {title}
  {{Observation of High Coherence in Josephson Junction Qubits Measured in a
  Three-Dimensional Circuit QED Architecture}},}\ }\href {\doibase
  10.1103/PhysRevLett.107.240501} {\bibfield  {journal} {\bibinfo  {journal}
  {Phys. Rev. Lett.}\ }\textbf {\bibinfo {volume} {107}},\ \bibinfo {pages}
  {240501} (\bibinfo {year} {2011})}\BibitemShut {NoStop}%
\bibitem [{\citenamefont {Liu}\ \emph {et~al.}(2017)\citenamefont {Liu},
  \citenamefont {Xu}, \citenamefont {Wang}, \citenamefont {Zheng},
  \citenamefont {Roy}, \citenamefont {Kundu}, \citenamefont {Chand},
  \citenamefont {Ranadive}, \citenamefont {Vijay}, \citenamefont {Song},
  \citenamefont {Duan},\ and\ \citenamefont {Sun}}]{Liu2017}%
  \BibitemOpen
  \bibfield  {author} {\bibinfo {author} {\bibfnamefont {K.}~\bibnamefont
  {Liu}}, \bibinfo {author} {\bibfnamefont {Y.}~\bibnamefont {Xu}}, \bibinfo
  {author} {\bibfnamefont {W.}~\bibnamefont {Wang}}, \bibinfo {author}
  {\bibfnamefont {S.-B.}\ \bibnamefont {Zheng}}, \bibinfo {author}
  {\bibfnamefont {T.}~\bibnamefont {Roy}}, \bibinfo {author} {\bibfnamefont
  {S.}~\bibnamefont {Kundu}}, \bibinfo {author} {\bibfnamefont
  {M.}~\bibnamefont {Chand}}, \bibinfo {author} {\bibfnamefont
  {A.}~\bibnamefont {Ranadive}}, \bibinfo {author} {\bibfnamefont
  {R.}~\bibnamefont {Vijay}}, \bibinfo {author} {\bibfnamefont
  {Y.}~\bibnamefont {Song}}, \bibinfo {author} {\bibfnamefont {L.}~\bibnamefont
  {Duan}}, \ and\ \bibinfo {author} {\bibfnamefont {L.}~\bibnamefont {Sun}},\
  }\bibfield  {title} {\enquote {\bibinfo {title} {{A twofold quantum
  delayed-choice experiment in a superconducting circuit}},}\ }\href {\doibase
  10.1126/sciadv.1603159} {\bibfield  {journal} {\bibinfo  {journal} {Sci.
  Adv.}\ }\textbf {\bibinfo {volume} {3}},\ \bibinfo {pages} {e1603159}
  (\bibinfo {year} {2017})}\BibitemShut {NoStop}%
\bibitem [{\citenamefont {Kirchmair}\ \emph {et~al.}(2013)\citenamefont
  {Kirchmair}, \citenamefont {Vlastakis}, \citenamefont {Leghtas},
  \citenamefont {Nigg}, \citenamefont {Paik}, \citenamefont {Ginossar},
  \citenamefont {Mirrahimi}, \citenamefont {Frunzio}, \citenamefont {Girvin},\
  and\ \citenamefont {Schoelkopf}}]{Kirchmair}%
  \BibitemOpen
  \bibfield  {author} {\bibinfo {author} {\bibfnamefont {G.}~\bibnamefont
  {Kirchmair}}, \bibinfo {author} {\bibfnamefont {B.}~\bibnamefont
  {Vlastakis}}, \bibinfo {author} {\bibfnamefont {Z.}~\bibnamefont {Leghtas}},
  \bibinfo {author} {\bibfnamefont {S.~E.}\ \bibnamefont {Nigg}}, \bibinfo
  {author} {\bibfnamefont {H.}~\bibnamefont {Paik}}, \bibinfo {author}
  {\bibfnamefont {E.}~\bibnamefont {Ginossar}}, \bibinfo {author}
  {\bibfnamefont {M.}~\bibnamefont {Mirrahimi}}, \bibinfo {author}
  {\bibfnamefont {L.}~\bibnamefont {Frunzio}}, \bibinfo {author} {\bibfnamefont
  {S.~M.}\ \bibnamefont {Girvin}}, \ and\ \bibinfo {author} {\bibfnamefont
  {R.~J.}\ \bibnamefont {Schoelkopf}},\ }\bibfield  {title} {\enquote {\bibinfo
  {title} {Observation of quantum state collapse and revival due to the
  single-photon {K}err effect},}\ }\href@noop {} {\bibfield  {journal}
  {\bibinfo  {journal} {Nature}\ }\textbf {\bibinfo {volume} {495}},\ \bibinfo
  {pages} {205} (\bibinfo {year} {2013})}\BibitemShut {NoStop}%
\bibitem [{\citenamefont {Khaneja}\ \emph {et~al.}(2005)\citenamefont
  {Khaneja}, \citenamefont {Reiss}, \citenamefont {Kehlet}, \citenamefont
  {Schulte-Herbr{\"u}ggen},\ and\ \citenamefont {Glaser}}]{Khaneja2005}%
  \BibitemOpen
  \bibfield  {author} {\bibinfo {author} {\bibfnamefont {N.}~\bibnamefont
  {Khaneja}}, \bibinfo {author} {\bibfnamefont {T.}~\bibnamefont {Reiss}},
  \bibinfo {author} {\bibfnamefont {C.}~\bibnamefont {Kehlet}}, \bibinfo
  {author} {\bibfnamefont {T.}~\bibnamefont {Schulte-Herbr{\"u}ggen}}, \ and\
  \bibinfo {author} {\bibfnamefont {S.~J.}\ \bibnamefont {Glaser}},\ }\bibfield
   {title} {\enquote {\bibinfo {title} {Optimal control of coupled spin
  dynamics: design of nmr pulse sequences by gradient ascent algorithms},}\
  }\href {\doibase 10.1016/j.jmr.2004.11.004} {\bibfield  {journal} {\bibinfo
  {journal} {Journal of magnetic resonance}\ }\textbf {\bibinfo {volume}
  {172}},\ \bibinfo {pages} {296} (\bibinfo {year} {2005})}\BibitemShut
  {NoStop}%
\bibitem [{\citenamefont {De~Fouquieres}\ \emph {et~al.}(2011)\citenamefont
  {De~Fouquieres}, \citenamefont {Schirmer}, \citenamefont {Glaser},\ and\
  \citenamefont {Kuprov}}]{DeFouquieres2011}%
  \BibitemOpen
  \bibfield  {author} {\bibinfo {author} {\bibfnamefont {P.}~\bibnamefont
  {De~Fouquieres}}, \bibinfo {author} {\bibfnamefont {S.}~\bibnamefont
  {Schirmer}}, \bibinfo {author} {\bibfnamefont {S.}~\bibnamefont {Glaser}}, \
  and\ \bibinfo {author} {\bibfnamefont {I.}~\bibnamefont {Kuprov}},\
  }\bibfield  {title} {\enquote {\bibinfo {title} {Second order gradient ascent
  pulse engineering},}\ }\href {\doibase 10.1016/j.jmr.2011.07.023} {\bibfield
  {journal} {\bibinfo  {journal} {Journal of Magnetic Resonance}\ }\textbf
  {\bibinfo {volume} {212}},\ \bibinfo {pages} {412} (\bibinfo {year}
  {2011})}\BibitemShut {NoStop}%
\bibitem [{\citenamefont {Barends}\ \emph {et~al.}(2014)\citenamefont
  {Barends}, \citenamefont {Kelly}, \citenamefont {Megrant}, \citenamefont
  {Veitia}, \citenamefont {Sank}, \citenamefont {Jeffrey}, \citenamefont
  {White}, \citenamefont {Mutus}, \citenamefont {Fowler}, \citenamefont
  {Campbell}, \citenamefont {Chen}, \citenamefont {Chen}, \citenamefont
  {Chiaro}, \citenamefont {Dunsworth}, \citenamefont {Neill}, \citenamefont
  {O'Malley}, \citenamefont {Roushan}, \citenamefont {Vainsencher},
  \citenamefont {Wenner}, \citenamefont {Korotkov}, \citenamefont {Cleland},\
  and\ \citenamefont {Martinis}}]{Barends2014}%
  \BibitemOpen
  \bibfield  {author} {\bibinfo {author} {\bibfnamefont {R.}~\bibnamefont
  {Barends}}, \bibinfo {author} {\bibfnamefont {J.}~\bibnamefont {Kelly}},
  \bibinfo {author} {\bibfnamefont {A.}~\bibnamefont {Megrant}}, \bibinfo
  {author} {\bibfnamefont {A.}~\bibnamefont {Veitia}}, \bibinfo {author}
  {\bibfnamefont {D.}~\bibnamefont {Sank}}, \bibinfo {author} {\bibfnamefont
  {E.}~\bibnamefont {Jeffrey}}, \bibinfo {author} {\bibfnamefont {T.~C.}\
  \bibnamefont {White}}, \bibinfo {author} {\bibfnamefont {J.}~\bibnamefont
  {Mutus}}, \bibinfo {author} {\bibfnamefont {a.~G.}\ \bibnamefont {Fowler}},
  \bibinfo {author} {\bibfnamefont {B.}~\bibnamefont {Campbell}}, \bibinfo
  {author} {\bibfnamefont {Y.}~\bibnamefont {Chen}}, \bibinfo {author}
  {\bibfnamefont {Z.}~\bibnamefont {Chen}}, \bibinfo {author} {\bibfnamefont
  {B.}~\bibnamefont {Chiaro}}, \bibinfo {author} {\bibfnamefont
  {A.}~\bibnamefont {Dunsworth}}, \bibinfo {author} {\bibfnamefont
  {C.}~\bibnamefont {Neill}}, \bibinfo {author} {\bibfnamefont
  {P.}~\bibnamefont {O'Malley}}, \bibinfo {author} {\bibfnamefont
  {P.}~\bibnamefont {Roushan}}, \bibinfo {author} {\bibfnamefont
  {A.}~\bibnamefont {Vainsencher}}, \bibinfo {author} {\bibfnamefont
  {J.}~\bibnamefont {Wenner}}, \bibinfo {author} {\bibfnamefont {a.~N.}\
  \bibnamefont {Korotkov}}, \bibinfo {author} {\bibfnamefont {a.~N.}\
  \bibnamefont {Cleland}}, \ and\ \bibinfo {author} {\bibfnamefont {J.~M.}\
  \bibnamefont {Martinis}},\ }\bibfield  {title} {\enquote {\bibinfo {title}
  {{Superconducting quantum circuits at the surface code threshold for fault
  tolerance}},}\ }\href {\doibase 10.1038/nature13171} {\bibfield  {journal}
  {\bibinfo  {journal} {Nature}\ }\textbf {\bibinfo {volume} {508}},\ \bibinfo
  {pages} {500} (\bibinfo {year} {2014})}\BibitemShut {NoStop}%
\bibitem [{\citenamefont {Knill}\ \emph {et~al.}(2008)\citenamefont {Knill},
  \citenamefont {Leibfried}, \citenamefont {Reichle}, \citenamefont {Britton},
  \citenamefont {Blakestad}, \citenamefont {Jost}, \citenamefont {Langer},
  \citenamefont {Ozeri}, \citenamefont {Seidelin},\ and\ \citenamefont
  {Wineland}}]{Knill2008}%
  \BibitemOpen
  \bibfield  {author} {\bibinfo {author} {\bibfnamefont {E.}~\bibnamefont
  {Knill}}, \bibinfo {author} {\bibfnamefont {D.}~\bibnamefont {Leibfried}},
  \bibinfo {author} {\bibfnamefont {R.}~\bibnamefont {Reichle}}, \bibinfo
  {author} {\bibfnamefont {J.}~\bibnamefont {Britton}}, \bibinfo {author}
  {\bibfnamefont {R.~B.}\ \bibnamefont {Blakestad}}, \bibinfo {author}
  {\bibfnamefont {J.~D.}\ \bibnamefont {Jost}}, \bibinfo {author}
  {\bibfnamefont {C.}~\bibnamefont {Langer}}, \bibinfo {author} {\bibfnamefont
  {R.}~\bibnamefont {Ozeri}}, \bibinfo {author} {\bibfnamefont
  {S.}~\bibnamefont {Seidelin}}, \ and\ \bibinfo {author} {\bibfnamefont
  {D.~J.}\ \bibnamefont {Wineland}},\ }\bibfield  {title} {\enquote {\bibinfo
  {title} {Randomized benchmarking of quantum gates},}\ }\href {\doibase
  10.1103/PhysRevA.77.012307} {\bibfield  {journal} {\bibinfo  {journal} {Phys.
  Rev. A}\ }\textbf {\bibinfo {volume} {77}},\ \bibinfo {pages} {012307}
  (\bibinfo {year} {2008})}\BibitemShut {NoStop}%
\bibitem [{\citenamefont {Ryan}\ \emph {et~al.}(2009)\citenamefont {Ryan},
  \citenamefont {Laforest},\ and\ \citenamefont {Laflamme}}]{Ryan2009}%
  \BibitemOpen
  \bibfield  {author} {\bibinfo {author} {\bibfnamefont {C.~A.}\ \bibnamefont
  {Ryan}}, \bibinfo {author} {\bibfnamefont {M.}~\bibnamefont {Laforest}}, \
  and\ \bibinfo {author} {\bibfnamefont {R.}~\bibnamefont {Laflamme}},\
  }\bibfield  {title} {\enquote {\bibinfo {title} {Randomized benchmarking of
  single- and multi-qubit control in liquid-state nmr quantum information
  processing},}\ }\href {\doibase 10.1088/1367-2630/11/1/013034} {\bibfield
  {journal} {\bibinfo  {journal} {New J. Phys.}\ }\textbf {\bibinfo {volume}
  {11}},\ \bibinfo {pages} {013034} (\bibinfo {year} {2009})}\BibitemShut
  {NoStop}%
\bibitem [{\citenamefont {Magesan}\ \emph {et~al.}(2012)\citenamefont
  {Magesan}, \citenamefont {Gambetta}, \citenamefont {Johnson}, \citenamefont
  {Ryan}, \citenamefont {Chow}, \citenamefont {Merkel}, \citenamefont
  {da~Silva}, \citenamefont {Keefe}, \citenamefont {Rothwell}, \citenamefont
  {Ohki}, \citenamefont {Ketchen},\ and\ \citenamefont
  {Steffen}}]{Magesan2012}%
  \BibitemOpen
  \bibfield  {author} {\bibinfo {author} {\bibfnamefont {E.}~\bibnamefont
  {Magesan}}, \bibinfo {author} {\bibfnamefont {J.~M.}\ \bibnamefont
  {Gambetta}}, \bibinfo {author} {\bibfnamefont {B.~R.}\ \bibnamefont
  {Johnson}}, \bibinfo {author} {\bibfnamefont {C.~A.}\ \bibnamefont {Ryan}},
  \bibinfo {author} {\bibfnamefont {J.~M.}\ \bibnamefont {Chow}}, \bibinfo
  {author} {\bibfnamefont {S.~T.}\ \bibnamefont {Merkel}}, \bibinfo {author}
  {\bibfnamefont {M.~P.}\ \bibnamefont {da~Silva}}, \bibinfo {author}
  {\bibfnamefont {G.~A.}\ \bibnamefont {Keefe}}, \bibinfo {author}
  {\bibfnamefont {M.~B.}\ \bibnamefont {Rothwell}}, \bibinfo {author}
  {\bibfnamefont {T.~A.}\ \bibnamefont {Ohki}}, \bibinfo {author}
  {\bibfnamefont {M.~B.}\ \bibnamefont {Ketchen}}, \ and\ \bibinfo {author}
  {\bibfnamefont {M.}~\bibnamefont {Steffen}},\ }\bibfield  {title} {\enquote
  {\bibinfo {title} {Efficient measurement of quantum gate error by interleaved
  randomized benchmarking},}\ }\href {\doibase 10.1103/PhysRevLett.109.080505}
  {\bibfield  {journal} {\bibinfo  {journal} {Phys. Rev. Lett.}\ }\textbf
  {\bibinfo {volume} {109}},\ \bibinfo {pages} {080505} (\bibinfo {year}
  {2012})}\BibitemShut {NoStop}%
\bibitem [{\citenamefont {Axline}\ \emph {et~al.}(2016)\citenamefont {Axline},
  \citenamefont {Reagor}, \citenamefont {Heeres}, \citenamefont {Reinhold},
  \citenamefont {Wang}, \citenamefont {Shain}, \citenamefont {Pfaff},
  \citenamefont {Chu}, \citenamefont {Frunzio},\ and\ \citenamefont
  {Schoelkopf}}]{Axline2016}%
  \BibitemOpen
  \bibfield  {author} {\bibinfo {author} {\bibfnamefont {C.}~\bibnamefont
  {Axline}}, \bibinfo {author} {\bibfnamefont {M.}~\bibnamefont {Reagor}},
  \bibinfo {author} {\bibfnamefont {R.}~\bibnamefont {Heeres}}, \bibinfo
  {author} {\bibfnamefont {P.}~\bibnamefont {Reinhold}}, \bibinfo {author}
  {\bibfnamefont {C.}~\bibnamefont {Wang}}, \bibinfo {author} {\bibfnamefont
  {K.}~\bibnamefont {Shain}}, \bibinfo {author} {\bibfnamefont
  {W.}~\bibnamefont {Pfaff}}, \bibinfo {author} {\bibfnamefont
  {Y.}~\bibnamefont {Chu}}, \bibinfo {author} {\bibfnamefont {L.}~\bibnamefont
  {Frunzio}}, \ and\ \bibinfo {author} {\bibfnamefont {R.~J.}\ \bibnamefont
  {Schoelkopf}},\ }\bibfield  {title} {\enquote {\bibinfo {title} {{An
  architecture for integrating planar and 3D cQED devices}},}\ }\href {\doibase
  10.1063/1.4959241} {\bibfield  {journal} {\bibinfo  {journal} {Appl. Phys.
  Lett.}\ }\textbf {\bibinfo {volume} {109}},\ \bibinfo {pages} {042601}
  (\bibinfo {year} {2016})}\BibitemShut {NoStop}%
\bibitem [{\citenamefont {Hatridge}\ \emph {et~al.}(2011)\citenamefont
  {Hatridge}, \citenamefont {Vijay}, \citenamefont {Slichter}, \citenamefont
  {Clarke},\ and\ \citenamefont {Siddiqi}}]{Hatridge}%
  \BibitemOpen
  \bibfield  {author} {\bibinfo {author} {\bibfnamefont {M.}~\bibnamefont
  {Hatridge}}, \bibinfo {author} {\bibfnamefont {R.}~\bibnamefont {Vijay}},
  \bibinfo {author} {\bibfnamefont {D.~H.}\ \bibnamefont {Slichter}}, \bibinfo
  {author} {\bibfnamefont {J.}~\bibnamefont {Clarke}}, \ and\ \bibinfo {author}
  {\bibfnamefont {I.}~\bibnamefont {Siddiqi}},\ }\bibfield  {title} {\enquote
  {\bibinfo {title} {Dispersive magnetometry with a quantum limited {SQUID}
  parametric amplifier},}\ }\href {\doibase 10.1103/PhysRevB.83.134501}
  {\bibfield  {journal} {\bibinfo  {journal} {Phys. Rev. B}\ }\textbf {\bibinfo
  {volume} {83}},\ \bibinfo {pages} {134501} (\bibinfo {year}
  {2011})}\BibitemShut {NoStop}%
\bibitem [{\citenamefont {Roy}\ \emph {et~al.}(2015)\citenamefont {Roy},
  \citenamefont {Kundu}, \citenamefont {Chand}, \citenamefont {Vadiraj},
  \citenamefont {Ranadive}, \citenamefont {Nehra}, \citenamefont {Patankar},
  \citenamefont {Aumentado}, \citenamefont {Clerk},\ and\ \citenamefont
  {Vijay}}]{Roy2015}%
  \BibitemOpen
  \bibfield  {author} {\bibinfo {author} {\bibfnamefont {T.}~\bibnamefont
  {Roy}}, \bibinfo {author} {\bibfnamefont {S.}~\bibnamefont {Kundu}}, \bibinfo
  {author} {\bibfnamefont {M.}~\bibnamefont {Chand}}, \bibinfo {author}
  {\bibfnamefont {A.~M.}\ \bibnamefont {Vadiraj}}, \bibinfo {author}
  {\bibfnamefont {A.}~\bibnamefont {Ranadive}}, \bibinfo {author}
  {\bibfnamefont {N.}~\bibnamefont {Nehra}}, \bibinfo {author} {\bibfnamefont
  {M.~P.}\ \bibnamefont {Patankar}}, \bibinfo {author} {\bibfnamefont
  {J.}~\bibnamefont {Aumentado}}, \bibinfo {author} {\bibfnamefont {A.~A.}\
  \bibnamefont {Clerk}}, \ and\ \bibinfo {author} {\bibfnamefont
  {R.}~\bibnamefont {Vijay}},\ }\bibfield  {title} {\enquote {\bibinfo {title}
  {{Broadband parametric amplification with impedance engineering: Beyond the
  gain-bandwidth product}},}\ }\href {\doibase 10.1063/1.4939148} {\bibfield
  {journal} {\bibinfo  {journal} {Appl. Phys. Lett.}\ }\textbf {\bibinfo
  {volume} {107}},\ \bibinfo {pages} {262601} (\bibinfo {year}
  {2015})}\BibitemShut {NoStop}%
\bibitem [{\citenamefont {Kamal}\ \emph {et~al.}(2009)\citenamefont {Kamal},
  \citenamefont {Marblestone},\ and\ \citenamefont {Devoret}}]{Kamal}%
  \BibitemOpen
  \bibfield  {author} {\bibinfo {author} {\bibfnamefont {A.}~\bibnamefont
  {Kamal}}, \bibinfo {author} {\bibfnamefont {A.}~\bibnamefont {Marblestone}},
  \ and\ \bibinfo {author} {\bibfnamefont {M.~H.}\ \bibnamefont {Devoret}},\
  }\bibfield  {title} {\enquote {\bibinfo {title} {Signal-to-pump back action
  and self-oscillation in double-pump {J}osephson parametric amplifier},}\
  }\href {\doibase 10.1103/PhysRevB.79.184301} {\bibfield  {journal} {\bibinfo
  {journal} {Phys. Rev. B}\ }\textbf {\bibinfo {volume} {79}},\ \bibinfo
  {pages} {184301} (\bibinfo {year} {2009})}\BibitemShut {NoStop}%
\bibitem [{\citenamefont {Murch}\ \emph {et~al.}(2013)\citenamefont {Murch},
  \citenamefont {Weber}, \citenamefont {Macklin},\ and\ \citenamefont
  {Siddiqi}}]{Murch}%
  \BibitemOpen
  \bibfield  {author} {\bibinfo {author} {\bibfnamefont {K.~W.}\ \bibnamefont
  {Murch}}, \bibinfo {author} {\bibfnamefont {S.~J.}\ \bibnamefont {Weber}},
  \bibinfo {author} {\bibfnamefont {C.}~\bibnamefont {Macklin}}, \ and\
  \bibinfo {author} {\bibfnamefont {I.}~\bibnamefont {Siddiqi}},\ }\bibfield
  {title} {\enquote {\bibinfo {title} {Observing single quantum trajectories of
  a superconducting quantum bit},}\ }\href {\doibase 10.1038/nature12539}
  {\bibfield  {journal} {\bibinfo  {journal} {Nature}\ }\textbf {\bibinfo
  {volume} {502}},\ \bibinfo {pages} {211} (\bibinfo {year}
  {2013})}\BibitemShut {NoStop}%
\bibitem [{\citenamefont {Chow}\ \emph {et~al.}(2009)\citenamefont {Chow},
  \citenamefont {Gambetta}, \citenamefont {Tornberg}, \citenamefont {Koch},
  \citenamefont {Bishop}, \citenamefont {Houck}, \citenamefont {Johnson},
  \citenamefont {Frunzio}, \citenamefont {Girvin},\ and\ \citenamefont
  {Schoelkopf}}]{Chow2009}%
  \BibitemOpen
  \bibfield  {author} {\bibinfo {author} {\bibfnamefont {J.~M.}\ \bibnamefont
  {Chow}}, \bibinfo {author} {\bibfnamefont {J.~M.}\ \bibnamefont {Gambetta}},
  \bibinfo {author} {\bibfnamefont {L.}~\bibnamefont {Tornberg}}, \bibinfo
  {author} {\bibfnamefont {J.}~\bibnamefont {Koch}}, \bibinfo {author}
  {\bibfnamefont {L.~S.}\ \bibnamefont {Bishop}}, \bibinfo {author}
  {\bibfnamefont {A.~A.}\ \bibnamefont {Houck}}, \bibinfo {author}
  {\bibfnamefont {B.~R.}\ \bibnamefont {Johnson}}, \bibinfo {author}
  {\bibfnamefont {L.}~\bibnamefont {Frunzio}}, \bibinfo {author} {\bibfnamefont
  {S.~M.}\ \bibnamefont {Girvin}}, \ and\ \bibinfo {author} {\bibfnamefont
  {R.~J.}\ \bibnamefont {Schoelkopf}},\ }\bibfield  {title} {\enquote {\bibinfo
  {title} {{Randomized Benchmarking and Process Tomography for Gate Errors in a
  Solid-State Qubit}},}\ }\href {\doibase 10.1103/PhysRevLett.102.090502}
  {\bibfield  {journal} {\bibinfo  {journal} {Phys. Rev. Lett.}\ }\textbf
  {\bibinfo {volume} {102}},\ \bibinfo {pages} {090502} (\bibinfo {year}
  {2009})}\BibitemShut {NoStop}%
\end{thebibliography}

%

\vspace{0.2in}

\vbox{}

\noindent \textbf{Acknowledgments}\\
We thank N. Ofek and Y. Liu for valuable suggestions on FPGA
programming, L. Jiang, L. Li, and R. Schoelkopf for helpful discussions. LS acknowledges the
support from National Natural Science Foundation of China Grant No.11474177,
National Key Research and Development Program of China No.2017YFA0304303,
and the Thousand Youth Fellowship program in China. SMG acknowledges grants from ARO W911NF1410011 and NSF DMR-1609326. LS also thanks
R. Vijay and his group for help on the parametric amplifier measurements.

\vbox{}

\noindent \textbf{Author contributions}\\
LH and LS developed the FPGA logic. LH and YM
performed the experiment and analyzed the data with the assistance
of WC, XM, YX, and WW. LS directed the experiment. LMD, CLZ, and LS proposed the experiment. CLZ, YW, SMG, and LMD provided theoretical support. WC fabricated the JPA.
LH and XM fabricated the devices with the assistance of YX, HW, and YPS. CLZ, SMG, and LS wrote the manuscript with feedback from all authors.

\vbox{}

\noindent \textbf{Competing financial interests}\\
The authors declare no competing financial interests.

\clearpage{}

\newpage{}

\newpage{}

\onecolumngrid
\renewcommand{\thefigure}{S\arabic{figure}}
\setcounter{figure}{0} 
\renewcommand{\thepage}{S\arabic{page}}
\setcounter{page}{1} 
\renewcommand{\theequation}{S.\arabic{equation}}
\setcounter{equation}{0} 
\setcounter{section}{0}

\end{document}


\title{Supplementary Information ``Demonstration of quantum error correction and universal gate set
on a binomial bosonic logical qubit"}


\author{L.~Hu}

\thanks{These two authors contributed equally to this work.}

\affiliation{Center for Quantum Information, Institute for Interdisciplinary Information
Sciences, Tsinghua University, Beijing 100084, China}

\author{Y.~Ma}

\thanks{These two authors contributed equally to this work.}

\affiliation{Center for Quantum Information, Institute for Interdisciplinary Information
Sciences, Tsinghua University, Beijing 100084, China}

\author{W.~Cai}

\affiliation{Center for Quantum Information, Institute for Interdisciplinary Information
Sciences, Tsinghua University, Beijing 100084, China}

\author{X.~Mu}

\affiliation{Center for Quantum Information, Institute for Interdisciplinary Information
Sciences, Tsinghua University, Beijing 100084, China}

\author{Y.~Xu}

\affiliation{Center for Quantum Information, Institute for Interdisciplinary Information
Sciences, Tsinghua University, Beijing 100084, China}

\author{W.~Wang}

\affiliation{Center for Quantum Information, Institute for Interdisciplinary Information
Sciences, Tsinghua University, Beijing 100084, China}

\author{Y.~Wu}
\affiliation{Department of Physics, University of Michigan, Ann Arbor, Michigan
48109, USA}

\author{H.~Wang}

\affiliation{Center for Quantum Information, Institute for Interdisciplinary Information
Sciences, Tsinghua University, Beijing 100084, China}

\author{Y.~P.~Song}

\affiliation{Center for Quantum Information, Institute for Interdisciplinary Information
Sciences, Tsinghua University, Beijing 100084, China}

\author{C.-L.~Zou}
\email{clzou321@ustc.edu.cn}
\affiliation{Key Laboratory of Quantum Information, CAS, University of Science
and Technology of China, Hefei, Anhui 230026, P. R. China}

\author{S.~M.~Girvin}
\affiliation{Departments of Applied Physics and Physics, Yale University, New Haven, CT 06511, USA}

\author{L-M.~Duan}

\affiliation{Center for Quantum Information, Institute for Interdisciplinary Information
Sciences, Tsinghua University, Beijing 100084, China}
\affiliation{Department of Physics, University of Michigan, Ann Arbor, Michigan
48109, USA}

\author{L.~Sun}
\email{luyansun@tsinghua.edu.cn}
\affiliation{Center for Quantum Information, Institute for Interdisciplinary Information
Sciences, Tsinghua University, Beijing 100084, China}

\maketitle

\section{Experimental Device and Setup}
\begin{figure*}
\centering \includegraphics[scale=0.6]{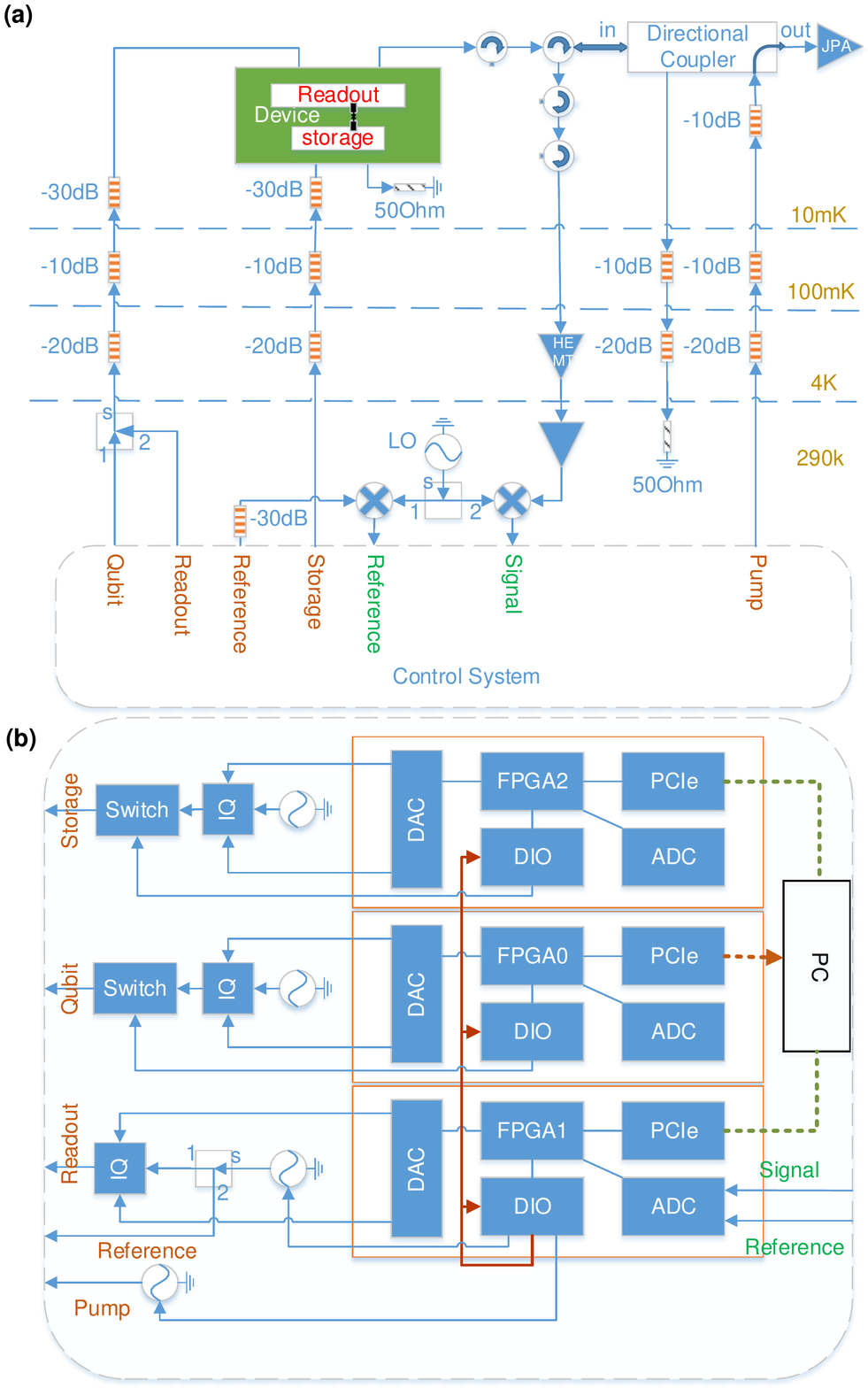} 
\caption{\textbf{Experimental setup.} \textbf{(a)} Details of wiring and circuit components.
The experimental device has a transmon qubit (the ancilla) dispersively
coupled to both a readout cavity and a storage cavity (the oscillator).
The cavities are made of high purity 5N5 aluminum and are chemically
etched to improve coherence time. A JPA is located at 10~mk, after
two circulators, allowing for a high-fidelity single-shot measurement
of the ancilla. A high-electron-mobility-transistor (HEMT) amplifier
at 4~K and an amplifier at room temperature are also used before
the down-conversion of the signal to 50~MHz with a local oscillator
(LO). Part of the readout signal does not go through the dilution
refrigerator and is used as a reference to lock the phase of the returning
readout signal from the device for a better measurement stability.
\textbf{(b)} The FPGA control system. Three X6-1000M boards and a
VPXI-ePC from Innovative Integration are used to control the qubit
(FPGA0), the readout cavity (FPGA1), and the oscillator (FPGA2). Each
board contains two DACs, two ADCs, and 64 independent DIO ports (as
input or output). The control pulse envelopes modulated at 100~MHz
are generated from DACs. Microwave switches are used for the qubit
and the oscillator control signals for better protection against
leakage and are controlled by the DIO port on the corresponding board.
Both microwave generators for the readout and the JPA pump work in
a pulsed mode controlled by the DIO ports as well. The down-converted
returning readout signal from the device is fed into the ADCs on the
same FPGA board (FPGA1) that also sends out the readout signal. The
demodulation and calculation are performed inside FPGA1 and the results
are discriminated into digitized signals. These final readout signals
are not only sent to the host PC, but also distributed among all FPGAs
through the DIO ports for inter-board communication. The feedback
operations coordinated among all three boards depend on the same digitized
readout result.}
\label{fig:experimentsetup} 
\end{figure*}

Our experimental device is measured in a dilution refrigerator with a base temperature
about 10~mK. Details of the experimental setup are shown in Fig.~\ref{fig:experimentsetup}a.
The device has a transmon qubit (the ancilla) dispersively
coupled to both a readout cavity and a storage cavity (the oscillator).
The ancilla qubit is fabricated on a $c$-plane sapphire (Al$_{2}$O$_{3}$)
substrate with a double-angle evaporation of aluminum after a single
electron-beam lithography step. The cavities are made of high purity
5N5 aluminum and are chemically etched for a better coherence time~\cite{Reagor2013,Reagor2016PRB}.
A Josephson parametric Amplifier (JPA)~\cite{Hatridge,Roy2015,Kamal,Murch}
is used as the first stage of amplification, allowing for a high-fidelity
single-shot measurement of the qubit (see the section on the Readout
Properties).

We follow a similar hardware setup and quantum control architecture
strategy based on field programmable gate array (FPGA) boards as Dr.
Schoelkopf's lab at Yale University for our experiment~\cite{Ofek2016}.
The ancilla qubit, the oscillator, and the readout cavity are each
controlled by a FPGA board. All three FPGA boards are X6-1000M from
Innovative Integration in a VPXI-ePC chassis as shown in Fig.~\ref{fig:experimentsetup}b.
Each board integrates two digital-to-analog converters (DACs), two
analog-to-digital converters (ADCs), 64 independent digital inputs/outputs
(DIO), and a Xilinx VIRTEX-6 FPGA with customized logic for control.
All FPGAs are loaded with the same logic, and we send different but
synchronized instructions to each FPGA to realize the desired operations.
A VPX-COMEX module is used to generate 1~GHz clocks (1~GHz sampling
rate for all ADCs and DACs) and synchronized triggers for all three
boards, so that all DIO, ADCs, and DACs on different boards are synchronized
into one entirety. The down-converted returning readout signal from
the device is fed into the ADCs of one FPGA board, within which the
demodulation and calculations are performed and the results are discriminated
into digitized signals. These final readout signals are not only sent
to the host PC, but also distributed among all FPGAs through the DIO
ports for inter-board communication. The latter is essential
because the feedback operations coordinated among all three boards depend
on the same digitized readout result. The total feedback latency, the time interval between
receiving the last point of the readout signal and sending out the
first point of the control signal, is 336~ns in
our experiment. This time includes the signal travel
time through the experimental circuitry.

\begin{table}
\centering %
\begin{tabular}{cccccc}
\hline 
Term  & Measured (Predicted) &  &  &  & \tabularnewline
\hline 
$\omega_{\mathrm{q}}/2\pi$ & 5.692 GHz &  &  &  & \tabularnewline
$\omega_{\mathrm{s}}/2\pi$ & 7.634 GHz &  &  &  & \tabularnewline
$\omega_{\mathrm{r}}/2\pi$ & 8.610 GHz &  &  &  & \tabularnewline
\hline 
$K_{\mathrm{q}}/2\pi$ & 232 MHz &  &  &  & \tabularnewline
$K_{\mathrm{r}}/2\pi$ & (14.4 kHz) &  &  &  & \tabularnewline
$K_{\mathrm{s}}/2\pi$ & 4.23 kHz &  &  &  & \tabularnewline
$K_{\mathrm{s}}^{'}/2\pi$ & 0.45 kHz &  &  &  & \tabularnewline
\hline 
$\chi_{\mathrm{qs}}/2\pi$ & 1.90 MHz &  &  &  & \tabularnewline
$\chi_{\mathrm{qr}}/2\pi$ & 3.65 MHz &  &  &  & \tabularnewline
$\chi_{\mathrm{sr}}/2\pi$ & (15.6 kHz) &  &  &  & \tabularnewline
\hline 
\end{tabular}\caption{Hamiltonian parameters. The predicted values in parentheses are calculated according to $\chi_{pp^{'}}=-2\sqrt{K_{p}K_{p}^{'}}$
based on the black-box quantization theory~\cite{Nigg2012}.}
\label{Table:DeviceParameters} 
\end{table}

\begin{table}
\centering %
\begin{tabular}{cccccc}
\hline 
$\quad$ & Ancilla & Oscillator & Readout &  & \tabularnewline
\hline 
$T_{1}$ & $30\ \mu$s & - & - &  & \tabularnewline
$T_{2}$ & $40\ \mu$s & - & - &  & \tabularnewline
$T_{\phi}$ & $120\ \mu$s & - & - &  & \tabularnewline
$\tau_{\mathrm{s,r}}(\kappa_{\mathrm{s,r}}/2\pi)$ & - & $143\ \mu$s~(1.1~kHz)  & 44 ns~(3.62~MHz) &  & \tabularnewline
$T_{2}^{\mathrm{s}}$ & - & $252\ \mu$s & - &  & \tabularnewline
\hline 
ground state & $99.2\%$ & $99.4\%$ & $>99.9\%$ &  & \tabularnewline
\hline 
\end{tabular}\caption{Coherent times and thermal populations of the ancilla qubit, the oscillator,
and the readout cavity.}
\label{Table:coherenttime} 
\end{table}

\begin{table}
\centering %
\begin{tabular}{cccccc}
\hline 
$\bar{n}$  & Parity fidelity &  &  &  & \tabularnewline
\hline 
0 & $98.8\%$ &  &  &  & \tabularnewline
1 & $97.8\%$ &  &  &  & \tabularnewline
2 & $97.2\%$ &  &  &  & \tabularnewline
3 & $97.1\%$ &  &  &  & \tabularnewline
\hline 
\end{tabular}\caption{Average parity measurement fidelity vs $\bar{n}$ based on the protocol
($R_{\pi/2}^{Y}$, $\pi/\chi_{\mathrm{qs}}$, $R_{-\pi/2}^{Y}$).
The single parity measurement is 99.92\% QND (see Fig.~\ref{fig:QNDP}
below).}
\label{Table:parity_fidelity} 
\end{table}


\section{Measured System Parameters, Coherence Times, and Measurement Properties}

\begin{figure*}[hbt]
\centering \includegraphics{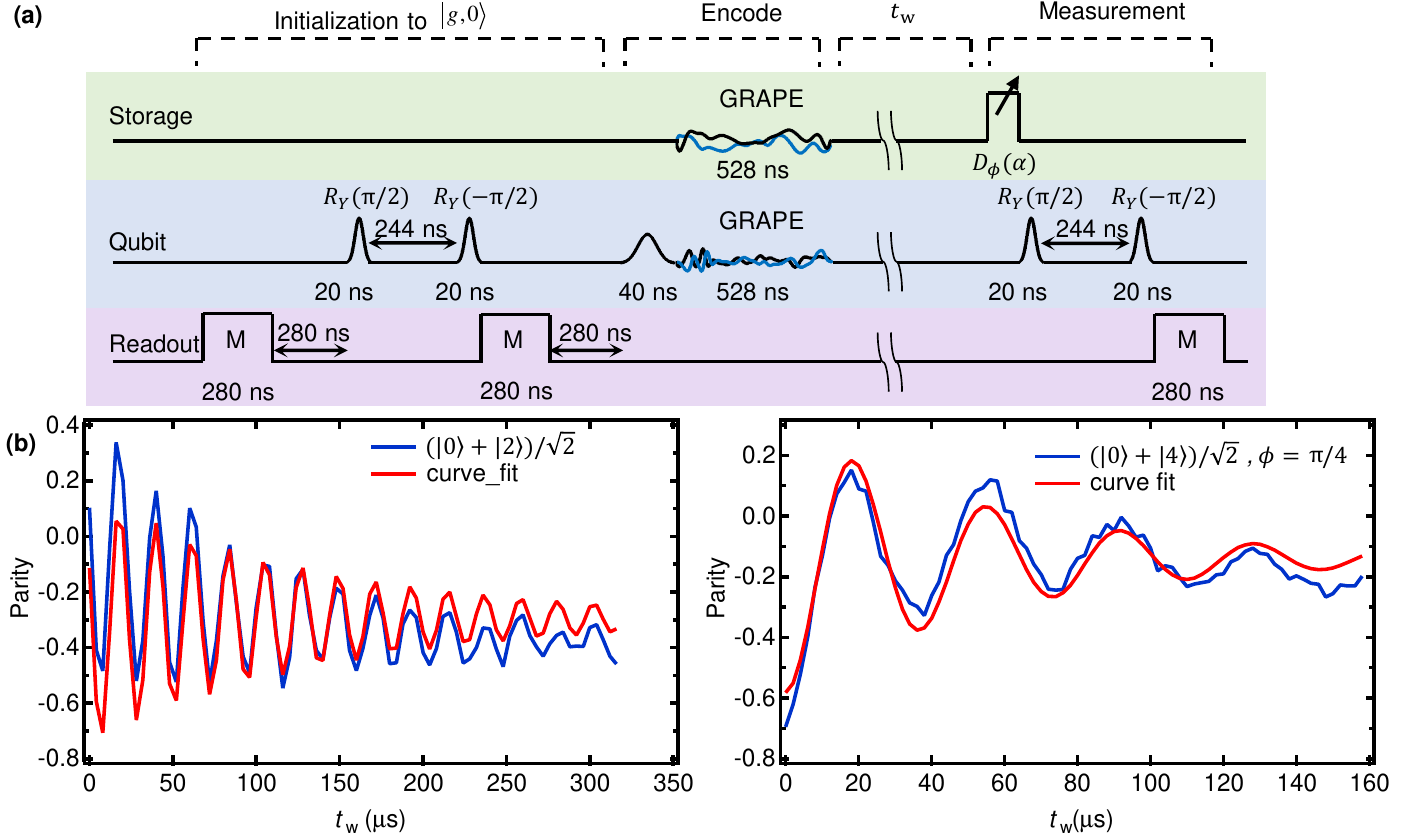} \caption{\textbf{Measurements of Kerr coefficients for the oscillator.} \textbf{(a)}
The measurement sequence. We encode the oscillator into $(\ket{0}+\ket{2})/\sqrt{2}$
and $(\ket{0}+\ket{4})/\sqrt{2}$ for measuring $K_{\mathrm{s}}$
and $K_{\mathrm{s}}^{'}$, respectively. $t_{\mathrm{w}}$ varies
with a step of $4~\mu$s for $(\ket{0}+\ket{2})/\sqrt{2}$ and $2~\mu$s
for $(\ket{0}+\ket{4})/\sqrt{2}$. Then a parity measurement is performed
following a displacement operation on the oscillator with $\alpha=0.7\sqrt{2}$
for $(\ket{0}+\ket{2})/\sqrt{2}$ and $\alpha=0.9\sqrt{2}$ for $(\ket{0}+\ket{4})/\sqrt{2}$.
The phase of the displacement operation $\phi$ varies for $(\ket{0}+\ket{2})/\sqrt{2}$
or is fixed at $\pi/4$ for $(\ket{0}+\ket{4})/\sqrt{2}$. \textbf{(b)}
Measured parity oscillations. Fitting the curves using $y=y_{0}+Ae^{-t/\tau}\mathrm{cos}(\omega t+\varphi_{0})$
gives the frequencies of the oscillation, from which we derive the
Kerr coefficients by subtracting the known detuning: the self-Kerr
$K_{\mathrm{s}}/2\pi=4.23$~kHz (in the term -$\frac{K_{\mathrm{s}}}{2}{\hat{a}_{\mathrm{s}}}^{\dagger2}{\hat{a}_{\mathrm{s}}}^{2}$)
and the correction to the self-Kerr $K_{\mathrm{s}}^{'}/2\pi=454$~Hz
(in the term $-\frac{{K_{\mathrm{s}}}^{'}}{6}{\hat{a}_{\mathrm{s}}}^{\dagger3}{\hat{a}_{\mathrm{s}}}^{3}$
).}
\label{fig:selfkerr} 
\end{figure*}

In the strong-dispersive regime, our system consisting of two cavities
and one ancilla qubit can be described by the Hamiltonian 
\begin{eqnarray}
\hat{H}/\hbar & = & \omega_{\mathrm{q}}{\hat{a}_{\mathrm{q}}}^{\dagger}{\hat{a}_{\mathrm{q}}}+\omega_{\mathrm{r}}{\hat{a}_{\mathrm{r}}}^{\dagger}\hat{a}_{\mathrm{r}}+\omega_{\mathrm{s}}{\hat{a}_{\mathrm{s}}}^{\dagger}\hat{a}_{\mathrm{s}}\nonumber \\
 & - & \chi_{\mathrm{qs}}{\hat{a}_{\mathrm{q}}}^{\dagger}{\hat{a}_{\mathrm{q}}}{\hat{a}_{\mathrm{s}}}^{\dagger}{\hat{a}_{\mathrm{s}}}-\chi_{\mathrm{qr}}{\hat{a}_{\mathrm{q}}}^{\dagger}{\hat{a}_{\mathrm{q}}}{\hat{a}_{\mathrm{r}}}^{\dagger}\hat{a}_{\mathrm{r}}-\chi_{\mathrm{sr}}{\hat{a}_{\mathrm{s}}}^{\dagger}\hat{a}_{\mathrm{s}}{\hat{a}_{\mathrm{r}}}^{\dagger}{\hat{a}_{\mathrm{r}}}\nonumber \\
 & - & \frac{K_{\mathrm{q}}}{2}{\hat{a}_{\mathrm{q}}}^{\dagger2}{\hat{a}_{\mathrm{q}}}^{2}-\frac{K_{\mathrm{r}}}{2}{\hat{a}_{\mathrm{r}}}^{\dagger2}{\hat{a}_{\mathrm{r}}}^{2}-\frac{K_{\mathrm{s}}}{2}{\hat{a}_{\mathrm{s}}}^{\dagger2}{\hat{a}_{\mathrm{s}}}^{2}-\frac{K_{\mathrm{s}}^{'}}{6}{\hat{a}_{\mathrm{s}}}^{\dagger3}{\hat{a}_{\mathrm{s}}}^{3},
\end{eqnarray}
where $\omega_{\mathrm{q,s,r}}$ are the ancilla qubit, the oscillator,
and the readout cavity frequency, respectively; $\hat{a}_{\mathrm{q,s,r}}$
are the corresponding ladder operators; $\chi_{\mathrm{qs}}$, $\chi_{\mathrm{qr}}$,
$\chi_{\mathrm{sr}}$ are the cross-Kerrs; $K_{\mathrm{{q,r,s}}}$
are the self-Kerr of corresponding quantum system; and $K_{\mathrm{s}}^{'}$
is a higher-order correction to the self-Kerr $K_{\mathrm{s}}$.
All these parameters are either measured directly or derived from
the measured ones, and are displayed in Table~\ref{Table:DeviceParameters}.
It is worth noting that although $K_{\mathrm{s}}^{'}$ is not shown
explicitly in Eq.~3 in the main text, it is important to take it
into account when calculating the control pulses based on the gradient ascent pulse engineering (GRAPE) method~\cite{Khaneja2005,DeFouquieres2011} and choosing the right waiting interval for the measurement of the uncorrected binomial code
decay time. $K_{\mathrm{s}}$ and $K_{\mathrm{s}}^{'}$ of the oscillator
are estimated by measuring the phase of $\ket{2}$ and $\ket{4}$
relative to $\ket{0}$ after an evolution of the initial state $(\ket{0}+\ket{2})/\sqrt{2}$
and $(\ket{0}+\ket{4})/\sqrt{2}$, respectively, as shown in Fig.~\ref{fig:selfkerr}.

The coherence times and thermal populations of all modes are summarized
in Table~\ref{Table:coherenttime}. The ancilla qubit has an energy
relaxation time $T_{1}=30.0$~$\mu$s and $T_{2}=40.0$~$\mu$s
corresponding to a pure dephasing time $T_{\phi}=120$~$\mu$s. The
ancilla qubit has a steady-state excited state population of ${n}_{\mathrm{th}}^{\mathrm{q}}=0.008$,
presumably due to stray infrared photons or other background noise
leaking into the system. Together with the measured $T_{1}=30.0~\mu$s,
we can derive the effective time for the ancilla being excited by thermal
environment as $1/\Gamma_{\uparrow}\approx3750~\mu$s, which is much
larger than the duration of our experiment. The oscillator has a single-photon
lifetime $\tau_{\mathrm{s}}=143$~$\mu$s ($\kappa_{\mathrm{s}}/2\pi=1.1$~kHz)
and a coherence time $T_{2}^{\mathrm{s}}=252~\mu$s corresponding
to a pure dephasing time $T_{\phi}^{\mathrm{s}}=2.1~$ms. The dephasing
time of the cavity is more than an order of magnitude larger than
the photon energy relaxation time, justifying the argument that photon
decay is the main decoherence channel for the logical qubit. Due to
the strong interaction between the oscillator and the ancilla qubit,
the pure dephasing time of the oscillator during idle when the ancilla qubit is in its ground state apparently is mainly limited
by the ancilla qubit thermal noise as $1/\Gamma_{\uparrow}$. However, for operations assisted by the ancilla qubit, the excitation of the qubit could introduce significant dephasing. Since this effect is combined with other imperfections, we do not separately estimate the dephasing of the operations but just include it in our numerically and experimentally estimated operation fidelities. Thus,
further suppressing the ancilla qubit thermal excitation and extending
the ancilla qubit energy decay time are essential to eliminate the
dephasing channel of the oscillator. Based on a similar argument,
due to the strong-dispersive interaction between the ancilla qubit
and the readout cavity, thermal population of the readout cavity can
limit the ancilla qubit pure dephasing time $T_{\phi}<1/{n}_{\mathrm{th}}^{\mathrm{r}}\kappa_{\mathrm{r}}$~\cite{Sears2012}.
Based on the measured ancilla qubit dephasing time, we can obtain
an upper limit on the thermal population of the readout cavity ${n}_{\mathrm{th}}^{\mathrm{r}}<0.0004$.
The thermal population of the oscillator ${n}_{\mathrm{th}}^{\mathrm{s}}\sim0.006$
is derived based on parity measurements in its steady state (see
below).

\begin{figure}[t]
\centering \includegraphics{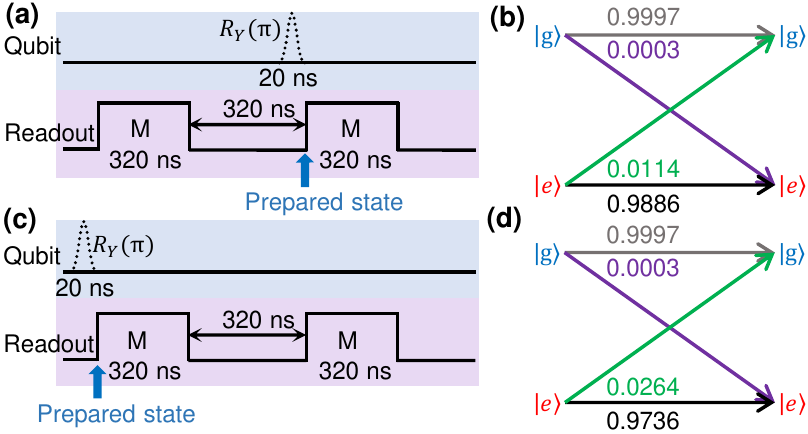} \caption{\textbf{Readout property.} \textbf{(a)} Sequences for readout fidelity determination. The
ancilla qubit is first measured and the ground state $\ket{g}$ is
post-selected, then no $\pi$ pulse or a $\pi$ pulse is performed
to prepare $\ket{g}$ or $\ket{e}$ followed immediately by a second
readout measurement. \textbf{(b)} Readout fidelity corresponding to
the sequences in \textbf{(a)}. The readout fidelity is $>0.999$ for
$\ket{g}$ and 0.989 for $\ket{e}$. The readout fidelity loss for
$\ket{e}$ comes from the ancilla qubit decay during the measurement
time. \textbf{(c)} Sequences to test QND. No $\pi$ pulse or a $\pi$
pulse is performed to prepare a state with dominant $\ket{g}$ or
$\ket{e}$ followed by the first readout. After a waiting time of
320~ns, a second readout is performed. \textbf{(d)} Readout fidelity
corresponding to sequences in \textbf{(c)}. The readout fidelity for
$\ket{g}$ when initially prepared in $\ket{g}$ by post-selection
is $>0.999$, while the fidelity for $\ket{e}$ when initially prepared
in $\ket{e}$ by post-selection is 0.974. The readout fidelity loss
for $\ket{e}$ is dominantly due to the ancilla qubit decay during
the waiting time and measurement time, demonstrating the highly QND
nature of the readout.}
\label{fig:readoutproperty} 
\end{figure}

The dispersive interaction between the ancilla qubit and the readout
cavity $\chi_{\mathrm{qr}}$, nearly matched with the decay rate $\kappa_{\mathrm{r}}$
of the readout cavity for an optimal signal-to-noise ratio, enables
a high-fidelity (with the help of a JPA) and high quantum non-demolition
(QND) single-shot readout of the ancilla qubit~\cite{Blais2004pra}.
The readout pulse has been optimized to have a width of 320~ns at
a few-photon ($<10$) level. The readout fidelity is $>0.999$ for
$\ket{g}$ and 0.989 for $\ket{e}$ (Figs.~\ref{fig:readoutproperty}a
and \ref{fig:readoutproperty}b). The readout fidelity loss for $\ket{e}$
comes from the ancilla qubit decay during the measurement time. The
readout fidelity of the ancilla qubit measured at $\ket{g}$ when
initially prepared in $\ket{g}$ by post-selection is $>0.999$,
while the fidelity for $\ket{e}$ state when initially prepared in
$\ket{e}$ by post-selection is 0.974 (Figs.~\ref{fig:readoutproperty}c
and \ref{fig:readoutproperty}d). The readout fidelity loss for $\ket{e}$
is dominantly due to the ancilla qubit decay during the following
waiting time (320~ns) and measurement time (320~ns), demonstrating
the highly QND nature of the dispersive readout.

The dispersive interaction between the ancilla qubit and the oscillator
$\chi_{\mathrm{qs}}$ allows one to do a Ramsey-type parity measurement
of the photon numbers in the oscillator~\cite{SunNature2014}, where
two unconditional $\pi/2$ rotations ($R_{\pi/2}^{Y}$) of the ancilla
qubit are separated by a pure delay of $\pi/\chi_{\mathrm{qs}}$. Since
in our QEC experiment the ancilla qubit is always prepared in the
ground state (a $\pi$ pulse is applied if the ancilla qubit is measured
in $\ket{e}$) and logical qubit has an even parity, we only focus
on the parity measurement protocol ($R_{\pi/2}^{Y}$, $\pi/\chi_{\mathrm{qs}}$,
$R_{-\pi/2}^{Y}$) in order to return the ancilla to $\ket{g}$ if the parity is even. The $\pi/2$ rotation pulses have a Gaussian envelope
truncated to 4$\sigma=20$~ns with the so-called ``derivative removal
by adiabatic gate\char`\"{} (DRAG) technique~\cite{Motzoi}.

Due to the finite bandwidth of the $\pi/2$ rotations, the parity
readout fidelities depend on the average number of photons in the
cavity, which is characterized following a similar technique in Ref.~\onlinecite{Wang2017}.
An initialization of the system based on post-selections creates a
vacuum state in the oscillator and ground state of the ancilla qubit.
Then a cavity displacement $D(\alpha=\sqrt{\bar{n}})$ and three consecutive
parity measurements are performed. Post-selections of the first two
consecutive identical parity results would give a photon state parity
with good confidence, and the third parity measurement is used for
the final parity fidelity measurement. The averaged parity measurement
fidelity results are listed in Table~\ref{Table:parity_fidelity}.
Note that for the protocol ($R_{\pi/2}^{Y}$, $\pi/\chi_{\mathrm{qs}}$,
$R_{-\pi/2}^{Y}$), the fidelity for an odd parity state is slightly
lower than an even parity state because the ancilla qubit ends up
in the excited state which has the possibility of an extra decay during the following ancilla
qubit readout. The most relevant number in our experiment is for $\bar{n}=2$,
and the corresponding averaged parity readout fidelity is $0.972$
($98.3\%$ for an even parity and $96.0\%$ for an odd parity).

\begin{figure*}[hbt]
\centering \includegraphics{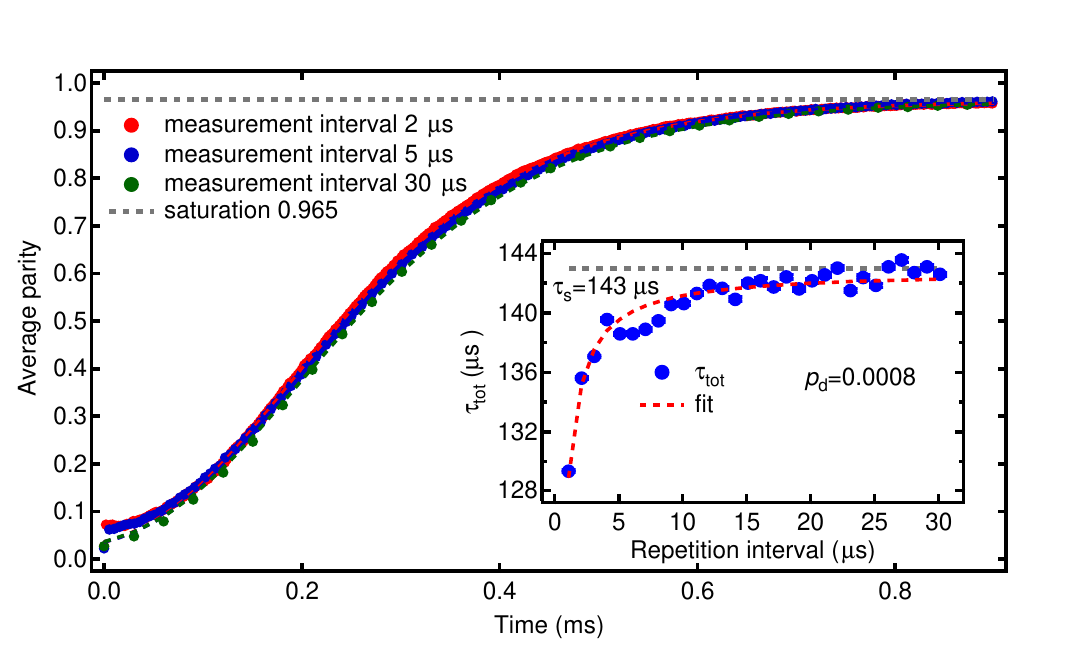} \caption{\textbf{QND property of the parity measurement.} To quantify the QND property
of the parity measurement on the oscillator, the parity decay of a
coherent state is monitored by performing repeated parity measurements
($R_{\pi/2}^{Y}$, $\pi/\chi_{\mathrm{qs}}$, $R_{-\pi/2}^{Y}$) with
different intervals, $1-30~\mu$s. This experiment is carried out
with feedback which always flips the ancilla qubit back to the ground
state after each parity measurement in order to eliminate the ancilla
qubit decay during the interval between parity measurements. The main
figure shows the averaged parity decay curves vs time with three typical
repetition intervals: $2~\mu$s, $5~\mu$s, and $30~\mu$s with an
initial displacement $\alpha=\sqrt{2}$. In the large time limit,
the cavity reaches a thermal steady state with a thermal population
${n_{th}^{s}}$ and each curve saturates at $\langle P\rangle=0.965$
(due to finite ${n_{\mathrm{th}}^{\mathrm{s}}}$ and parity measurement
fidelities). Dotted lines are fits with $\langle P\rangle=\frac{1+P0}{2}(2P_{\mathrm{e}}-1)+\frac{1-P0}{2}(1-2P_{\mathrm{o}})$,
where $P0=e^{-2|\alpha|^{2}e^{-t/\tau_{\mathrm{tot}}}/(1+2n_{\mathrm{th}}^{\mathrm{s}})}/(1+2n_{\mathrm{th}}^{\mathrm{s}})$
is the expected parity evolution with a steady-state thermal population
$n_{\mathrm{th}}^{\mathrm{s}}$ in the cavity, $\tau_{\mathrm{tot}}$
is the parity decay time under monitoring, and $P_{\mathrm{e}}$ ($P_{\mathrm{o}}$)
is the parity measurement fidelity for an even (odd) state. Note that
both $P_{\mathrm{e}}$ and $P_{\mathrm{o}}$ slightly depend on the average photon
number in the cavity and here we choose $P_{\mathrm{e}}=0.987$ and
$P_{\mathrm{o}}=0.969$ for an averaged $\bar{n}=1$. The fit also
gives ${n}_{\mathrm{th}}^{\mathrm{s}}\sim0.006$. The humps when time
is small for intervals $2~\mu$s and $5~\mu$s are mainly due to the
ancilla qubit decay during the feedback latency time (336~ns). The
insert shows the extracted $\tau_{\mathrm{tot}}$ vs the repetition
interval $\tau_{\mathrm{rep}}$. The error bar is derived from bootstrapping
of each curve (80,000 averages) in the main figure. Under parity monitoring,
$\tau_{\mathrm{tot}}$ and the natural decay rate $\tau_{\mathrm{s}}$
obey the relationship $1/\tau_{\mathrm{tot}}$=$1/\tau_{\mathrm{s}}+p_{\mathrm{d}}/\tau_{\mathrm{rep}}$,
where the $p_{\mathrm{d}}$ is the probability of inducing an extra
parity change by the parity measurement~\cite{SunNature2014}. A
fit gives $1/\tau_{\mathrm{s}}=143~\mu$s and $p_{\mathrm{d}}=0.08\%$,
indicating that a single parity measurement is 99.92\% QND.}
\label{fig:QNDP} 
\end{figure*}

The QND nature of the error syndrome measurement (that does not introduce
extra photon jump errors) is essential for QEC. To quantify the QND
property of the parity measurement on the oscillator, the parity decay
of a coherent state is monitored by performing repeated parity measurements
($R_{\pi/2}^{Y}$, $\pi/\chi_{\mathrm{qs}}$, $R_{-\pi/2}^{Y}$) with
different intervals, $1-30~\mu$s. This experiment is carried out
with real-time feedback which always flips the ancilla qubit back to the ground
state after each parity measurement in order to eliminate the ancilla
qubit decay during the interval between parity measurements. The averaged
parity decay curves vs time with an initial displacement $\alpha=\sqrt{2}$
are shown in Fig.~\ref{fig:QNDP} and are fitted with $\langle P\rangle=\frac{1+P0}{2}(2P_{\mathrm{e}}-1)+\frac{1-P0}{2}(1-2P_{\mathrm{o}})$,
where $P0=e^{-2|\alpha|^{2}e^{-t/\tau_{\mathrm{tot}}}/(1+2n_{\mathrm{th}}^{\mathrm{s}})}/(1+2n_{\mathrm{th}}^{\mathrm{s}})$
is the expected parity evolution with a steady-state thermal population
$n_{\mathrm{th}}^{\mathrm{s}}$ in the cavity, $\tau_{\mathrm{tot}}$
is the parity decay time under monitoring, and $P_{\mathrm{e}}$ ($P_{\mathrm{o}}$)
is the parity measurement fidelity for an even (odd) state. The fit
gives ${n}_{\mathrm{th}}^{\mathrm{s}}\sim0.006$. The extracted $\tau_{\mathrm{tot}}$
vs repetition interval is shown in the inset of Fig.~\ref{fig:QNDP}.
The total decay time is well modelled by a parallel combination of
the natural decay ($\tau_{\mathrm{s}}$) and a constant demolition
probability $p_{\mathrm{d}}=0.08\%$ per measurement, indicating that
a single parity measurement is $99.92\%$ QND.

\section{Experimental sequences}
In the following, we show the details of our experimental sequences.
Since both the ancilla qubit and the oscillator have non-zero but small
thermal excitations, we always initialize our system to $\ket{g,0}$
(the notation denotes $\ket{qubit,oscillator}$) by post-selection
on the ancilla qubit's ground state and a subsequent oscillator
parity measurement. This post-selection only removes about $1.4\%$
of the total data, so we do not use an adaptive control to perform
the initialization, although in principle we could.

The ancilla qubit plays two main roles: 1) serve as an ancilla for
the error syndrome detection of the logical qubit; 2) provide the
necessary non-linearity for implementing all the operations related
to the oscillator including encoding, decoding, error correcting,
as well as the universal single logical qubit operations. Through
a sequence of control pulses on the system, all these operations on
the logical qubit are realized based on the dispersive interaction
between the ancilla and the oscillator. The control pulses are synchronized
and generated by FPGAs with home-made logic. The pulse shapes are
numerically optimized with GRAPE~\cite{Khaneja2005,DeFouquieres2011} based on carefully calibrated
experimental parameters. All GRAPE pulses have a duration of 528~ns
in our experiment.

\begin{figure*}[hbt]
\centering \includegraphics{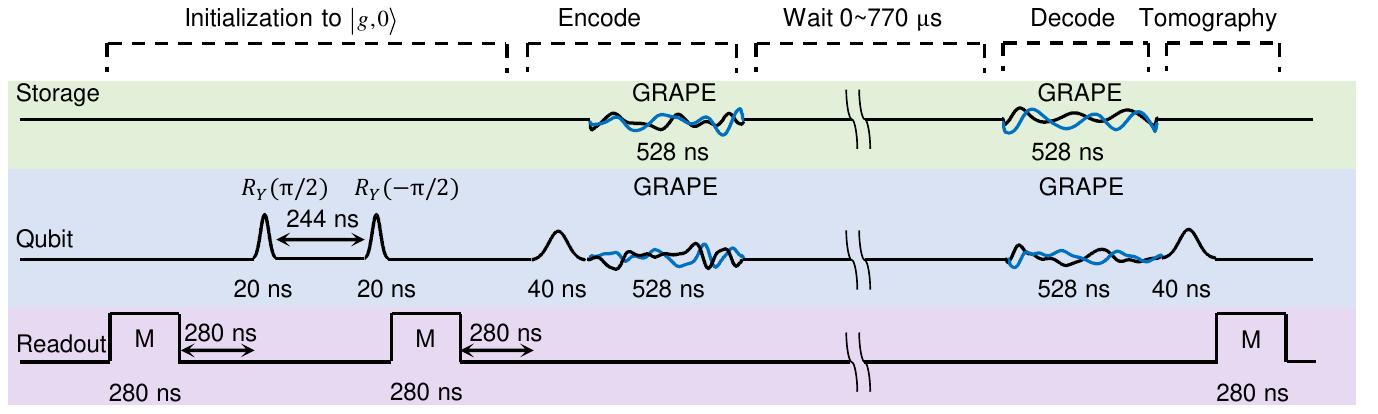} \caption{\textbf{Experimental sequence to measure the uncorrected Fock $\{\ket{0},\ket{1}\}$
encoding lifetime.} The whole experiment can be divided into five parts:
initialization of the system to $\ket{g,0}$, encoding, various waiting
time, decoding, and final state tomography of the ancilla. The encoding
and decoding are realized by GRAPE pulses. The final state tomography
of the ancilla is realized by pre-rotations of the ancilla qubit followed
by a $\sigma_{z}$ projection measurement.}
\label{fig:sequence0a1} 
\end{figure*}

\begin{figure*}[hbt]
\centering \includegraphics{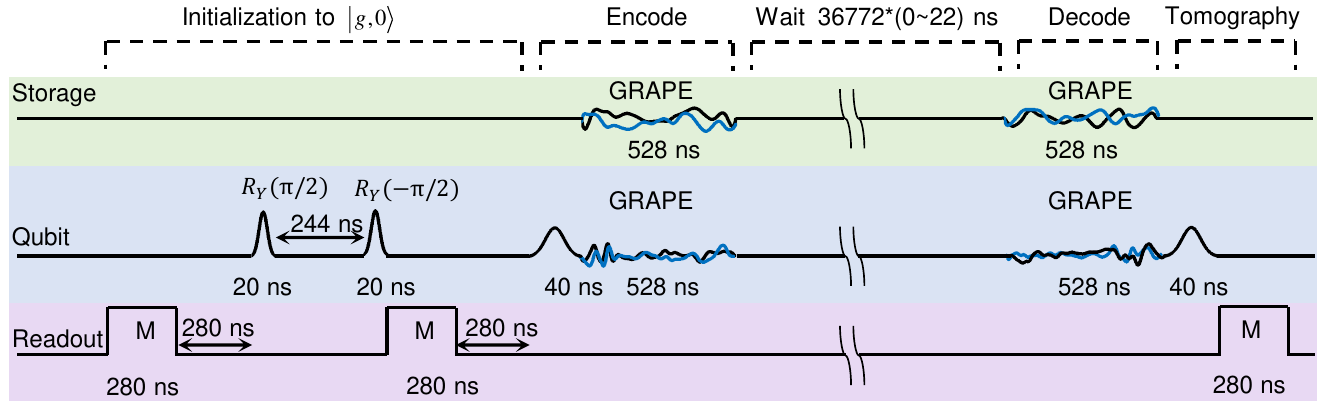} \caption{\textbf{Experimental sequence to measure the uncorrected binomial encoding lifetime.}
The GRAPE pulses for encoding and decoding are the same as those in
the QEC experiment. The waiting intervals $t_{\mathrm{w}}=N\times36,772$~ns
($N=0-22$) between encoding and decoding are chosen such that the
Kerr rotations of $\ket{4}$ are integers of $2\pi$. However, the
phase of $\ket{2}$ is $N\varphi_{0}$ that is deterministic due to
the self-Kerr of the oscillator. After decoding, this phase is transferred
to $\ket{e}$ state and is corrected in the data analysis.}
\label{fig:sequence0a2a4} 
\end{figure*}

Figure~\ref{fig:sequence0a1} shows the measurement sequence for
the process fidelity decay of the uncorrected Fock $\{\ket{0},\ket{1}\}$
encoding. The sequence for the case of uncorrected binomial encoding
(Fig.~\ref{fig:sequence0a2a4}) is similar except that the waiting
times are chosen such that the Kerr rotations of $\left|4\right\rangle $
are integer multiples of $2\pi$ to minimize the deformation of one of the
code space bases $(\ket{0}+\ket{4})/\sqrt{2}$. This is because in our experiment we have chosen a rotating frame of $\omega_s$ so that both $\ket{4}$ and $\ket{2}$ rotate relative to $\ket{0}$ due to the Kerr effects. In principle, we can choose another frame to match the rotation of $\ket{4}$ relative to $\ket{0}$ by rotating at $\omega_{4s}/4$, where $\omega_{4s}$ is the frequency difference between $\ket{0}$ and $\ket{4}$ of the cavity. Note that this is true even if the higher-order corrections to Kerr are taken into account, which is one of the nice features of the lowest-order binomial code~\cite{Michael2016}. The extra phase associated
with $\ket{2}$ is deterministic and is corrected in our final analysis
after decoding.

\begin{figure*}
\centering \includegraphics{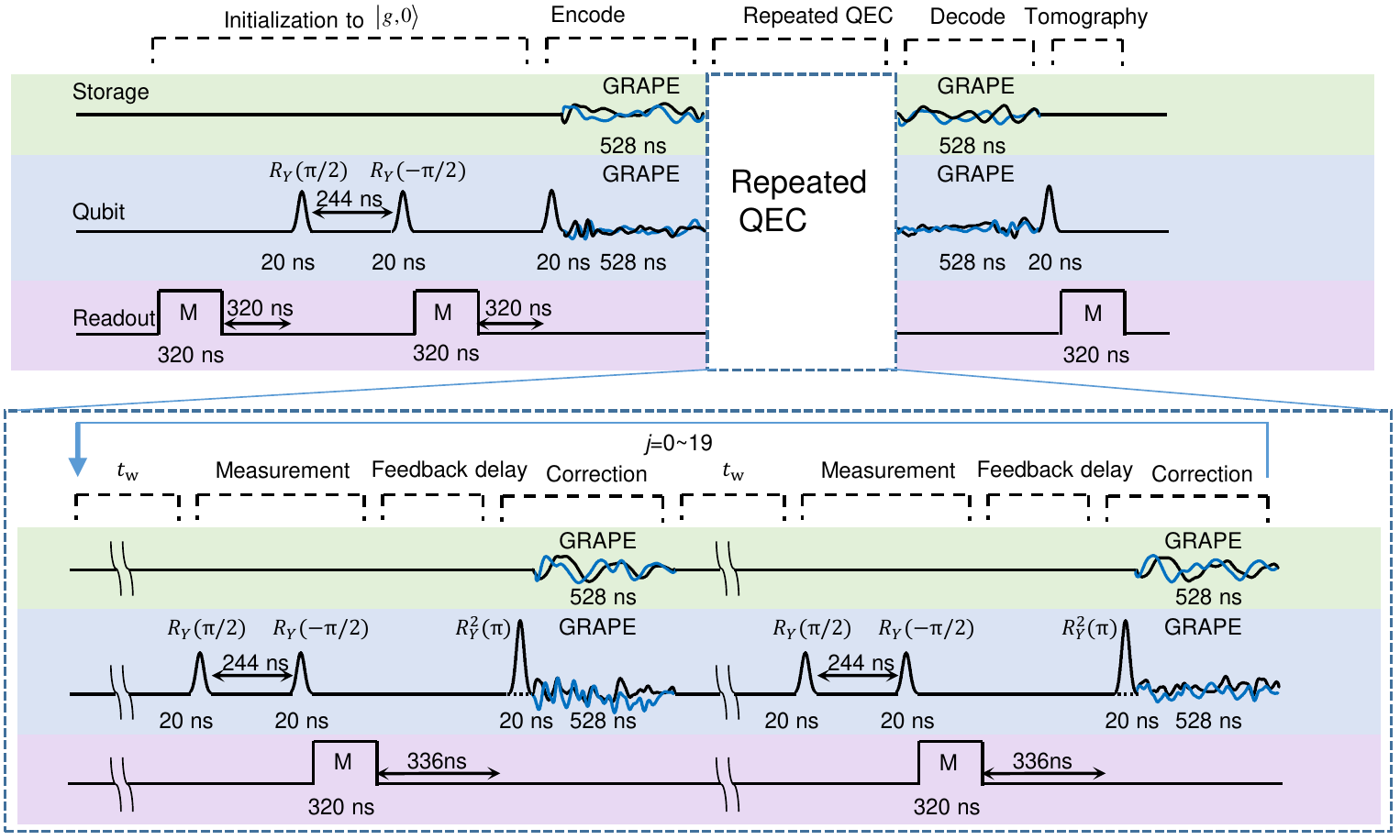} \caption{\textbf{Experimental sequence for the repetitive QEC.} The essential part is
the two-layer QEC procedure. The top layer recovers the quantum information
in the code space and is repeated 0-19 times. Two bottom-layer processes
are used to conserve parity in the deformed code space. Each bottom-layer
process consists of a waiting time $t_{\mathrm{w}}=17.895~\mu$s,
error detection, feedback delay, and correction. If no error is detected,
the $\pi$ rotation is replaced by a pure waiting time. Appropriate recovery
GRAPE pulses are performed depending on the error detection results
as shown in the protocol of Fig.~2a in the main text.}
\label{fig:sequenceerrorcorrection} 
\end{figure*}

\begin{figure*}[hbt]
\centering \includegraphics{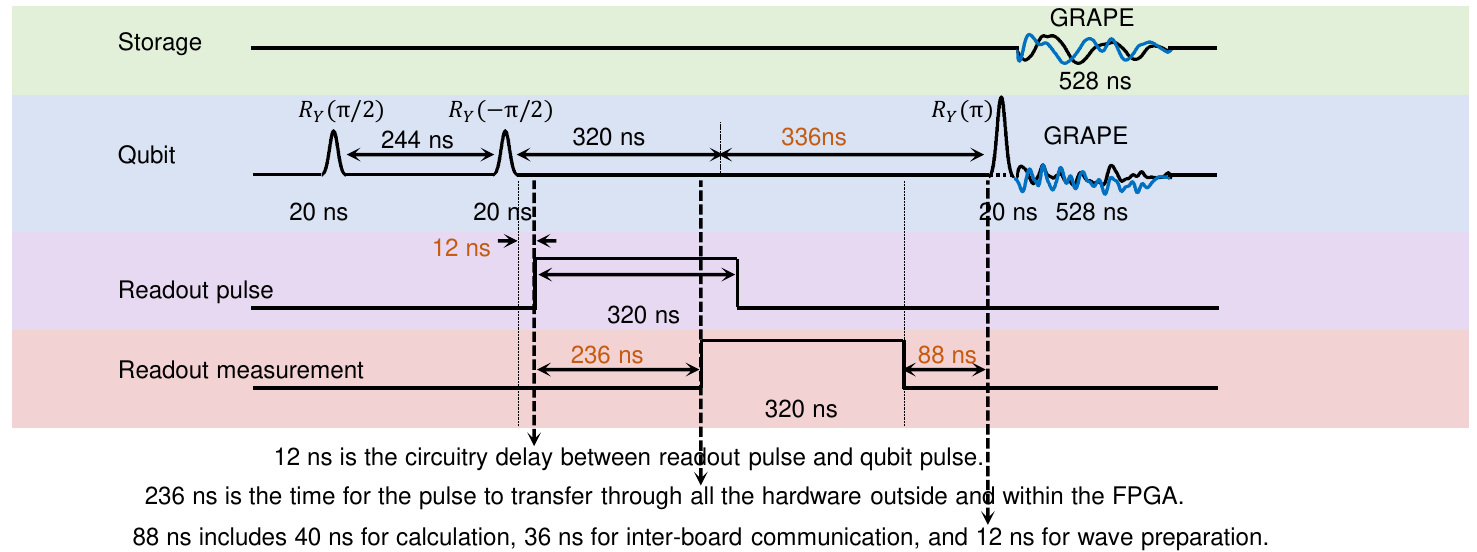} \caption{\textbf{Details of the QEC process.} Different FPGA boards are synchronized
to control the readout, the ancillary qubit, and the oscillator. The
FPGAs to control the ancilla qubit and the oscillator are the same
with no relative delay, but both have a relative delay with the one
to control the readout. The feedback delay contains three parts: 1)
the delay between the readout pulse and ancilla qubit pulse due to
the control circuitry difference; 2) the time taken by the signal
travelling through the experimental circuitry outside and within the
FPGA; 3) the time for calculation, inter-board-communication, and
response to the adaptive instruction. The whole process takes 336
ns in our system.}
\label{fig:errorcorrectiontime} 
\end{figure*}

Figures~\ref{fig:sequenceerrorcorrection} and \ref{fig:errorcorrectiontime}
show our repetitive QEC sequence and the details of the QEC process,
respectively. The latency of our adaptive control is defined as the
time interval between receiving the last point of the readout signal
and sending out the first point of control signal including the speed of light travel time through the whole
experimental circuitry and is only 336~ns, about $1\%$ of the ancilla
qubit lifetime.

\section{T Gate Fidelity Using Repeated Measurements}

\begin{figure}[hbt]
\centering \includegraphics{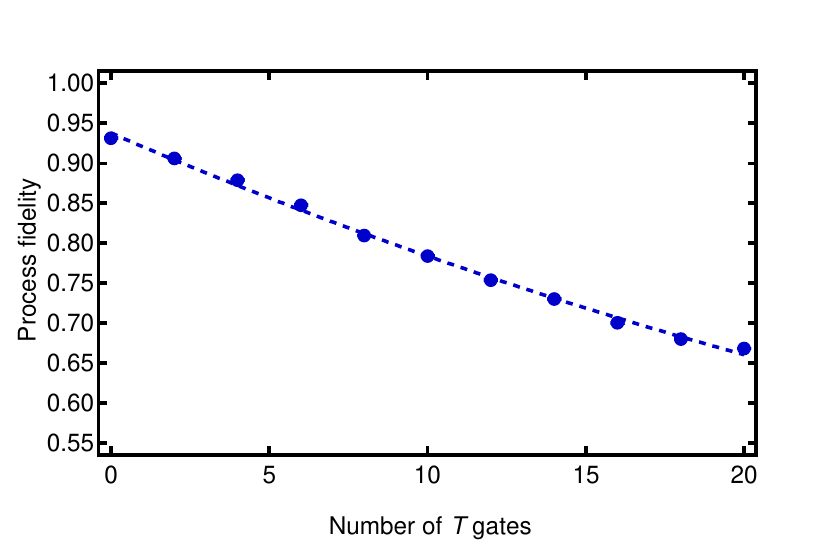} \caption{\textbf{$T$ gate process fidelity by repeated gate measurements.} The $T$
gate is repeated $0,2,4,...,20$ times and the corresponding process fidelities
are measured. A fit with $F_\chi(t)=0.25+Ae^{-n/\tau}$ gives $e^{-1/\tau}=0.974$,
implying a gate fidelity of 0.987 when converted to be consistent with the
fidelity definition in the randomized benchmarking method. The intercept at $N=0$ is the ``round trip" process fidelity of the encoding and decoding processes only, same as Fig.~3a in the main text.}
\label{fig:Tgate} 
\end{figure}

The $T$ gate does not belong to the Clifford group, so it
can not be characterized by randomized benchmarking. We instead
perform repeated gates to extract the gate fidelity of $0.987$,
as shown in Fig.~\ref{fig:Tgate}. This fidelity
has been converted to be consistent with the fidelity definition in the randomized
benchmarking method. We note that if the noise channel is a depolarization channel and the probability of having a full depolarization is $p$, then in the randomized benchmarking experiment the fidelity $F_{\mathrm{RB}}=1-p/2$. However, the $\chi$ matrix decay curve with respect to the number of repeated gates gives the normalized process fidelity (Eq.~3 defined below) $F=1-p$.

\section{Theoretical Analysis and Error Model of the QEC Performance}


In the QEC experiment, the ancilla plays multiple critical roles,
including in the encoding, decoding, error detection, error correcting
recovery operation, and the universal single logical qubit operations.
The decoherence of the ancilla introduces imperfections for the above
processes and thus eventually limits the ultimate performance of our
current experiment. As just mentioned, on one hand, the ancilla facilitates
the error detection, correction, and gate operations. Therefore, errors in these processes due to the ancilla decoherence prevent too
frequent repetitive QEC process. On the other hand, with a larger interval between
QECs, there is a higher probability of having undetectable higher-order
photon loss and gain errors and a larger dephasing effect induced
by photon losses (due to the non-commutativity of the annihilation
operation $\hat{a_{\mathrm{s}}}$ and the self-Kerr term $\frac{K_{\mathrm{s}}}{2}\hat{a_{\mathrm{s}}}^{\dagger2}\hat{a_{\mathrm{s}}}^{2}$).
These two types of error, caused by the intrinsic cavity photon loss
and the operations facilitated by the ancilla, compete with each other
and ultimately lead to an optimization of the QEC performance.

\begin{figure}
\includegraphics{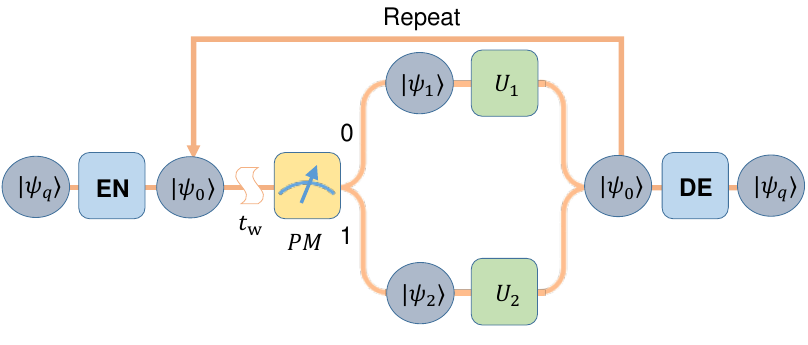} \caption{\textbf{Procedure for QEC Protocol I.} The same encoding and decoding
as in Fig.~2a in the main text are used here. However, the difference
is that the detection and correction are performed in each QEC step.
Process tomography is used to benchmark this QEC performance.}
\label{fig:ProtocolI} 
\end{figure}

In order to understand the experimental results of the QEC performance,
in this section we first investigate the two types of errors in detail,
and then establish an approximate analytical model that applies to
two experimental protocols, Protocol I and Protocol II for one and
two bottom-layer detections and corrections respectively. The model
can explain both experiments well and thus also indicates the possible
routes to further improve the performance of the binomial code QEC in
the future. The scheme and results of Protocol II have been shown
in the main text already. The experimental procedure of Protocol I
is depicted in Fig.~\ref{fig:ProtocolI}, in which the recovery gates
are performed immediately in both no-error and one-error cases after
each detection.

Both analytical models that are based on a simple error propagation treatment
and numerical simulations that are based on QuTiP in Python~\cite{Johansson2012,Johansson2013} have
been adopted to analyze each sub-process of the whole QEC protocol.
Although the analytical model provides more physical insights, the
numerical approach can be more rigorous. To compare with the experiments,
the decoherence of the ancilla and the oscillator are included in
the numerical simulations through the Lindblad master equations. In the interaction
picture, the simplified Hamiltonian used in the simulation is
\begin{equation}
H_{\mathrm{int}}=-\chi_{\mathrm{qs}}a_{\mathrm{s}}^{\dagger}a_{\mathrm{s}}\left|e\right\rangle \left\langle e\right|-\frac{K_{\mathrm{s}}}{2}a_{\mathrm{s}}^{\dagger2}a_{\mathrm{s}}^{2}-\frac{K_{\mathrm{s}}'}{6}a_{\mathrm{s}}^{\dagger3}a_{\mathrm{s}}^{3}.\label{eq:Hamiltonian_sim}
\end{equation}
Here we treat the ancilla as a two-level system and ignore the readout
cavity. The parameters in the numerical simulations are taken from the calibrated
experimental parameters, which have been listed in Tables~\ref{Table:DeviceParameters}
and \ref{Table:coherenttime}. 

Since the minimum process fidelity $\left(F_{\chi}\right)$ is 0.25,
in the following we denote the fidelity as the normalized process
fidelity with a full scale between 0 and 1: 
\begin{equation}
{F}=(F_{\chi}-0.25)/0.75.\label{eq:DefFidelity-1}
\end{equation}

\subsection{Intrinsic Errors from Oscillator Photon Loss}
The binomial code implemented in this experiment can protect quantum
information against a single photon loss error $\hat{a_{\mathrm{s}}}$. In principle, frequent error correction would suppress
the higher order errors, i.e. $\left(\tau_\mathrm{w}/\tau_\mathrm{s}\right)^{n}$ vanishes when $\tau_\mathrm{w}/\tau_\mathrm{s}\ll1$, where $\tau_\mathrm{w}$ is the time interval between QEC steps, $\tau_\mathrm{s}$ is the previously defined single-photon lifetime, and $n>1$. However, due to the imperfections in both error detection and correction (discussed below), $\tau_\mathrm{w}$ should not be too small. In this case, there are errors intrinsic to this encoding: the errors that are not contained in the error set $\{\hat{I},\hat{a_{\mathrm{s}}}\}$ would cause a
failure of the performance. The intrinsic errors include: (1) multiple photon losses for finite $\tau_\mathrm{w}/\tau_\mathrm{s}$, which could not
be detected and corrected. (2) the photon gain error due to thermal
noise injected into the oscillator. (3) the dephasing due to the combination
of photon loss jump and the self-Kerr effect. Since the detected
single photon loss may happen in any time during $\tau_{\mathrm{w}}$, the photon-number dependent Kerr phase accumulation should depend on the jump time, which induces a phase correlation between
the oscillator and the leaked photon and eventually gives rise to dephasing. (4) the non-unitary no-jump evolution $e^{-(\kappa/2) \hat{a}^\dagger\hat{a} t}$. The unitary recovery operations correct the non-unitary no-jump evolution only to first order in $\kappa t_{\mathrm{w}}$~\cite{Michael2016} as discussed in the main text.

\begin{figure}
\includegraphics{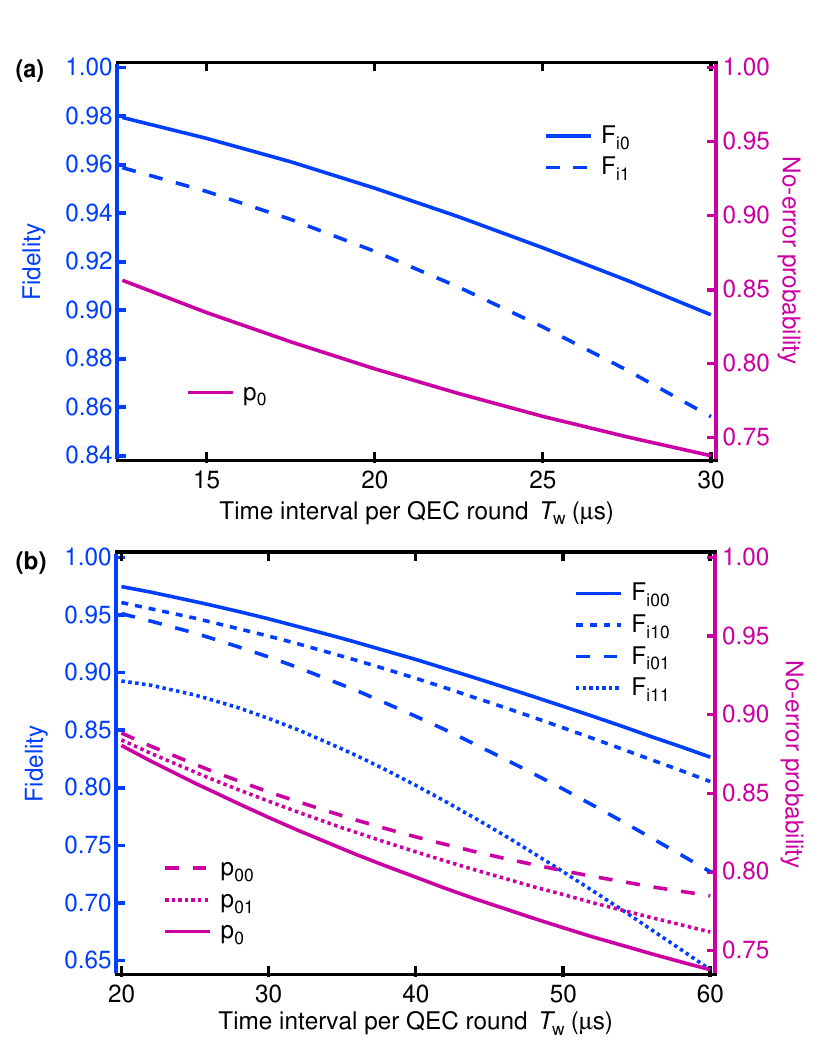} \caption{\textbf{Intrinsic errors for the binomial code.} In the numerical simulations, only the cavity decay
and Kerr effects have been included, neglecting ancilla qubit decoherence,
detection error, and correction error. \textbf{(a)} Protocol I. $F_{\mathrm{i0}}$
and $F_{\mathrm{i1}}$ are the process fidelity (blue) for detecting
no error and one error respectively. $p_{0}$ is the probability (purple)
of detecting no error. \textbf{(b)} Protocol II. $F_{\mathrm{i00}}$,
$F_{\mathrm{i01}}$, $F_{\mathrm{i10}}$, and $F_{\mathrm{i11}}$
are the process fidelities for the following cases respectively: no
error for both detections, no error for the first detection and one
error for the second one, one error for the first detection and no
error for the second one, one error for both detections. $p_{0}$
is the probability of detecting no error for the first syndrome detection,
$p_{00}$ is the probability of detecting no error for the second
syndrome detection provided no error is detected in the first syndrome
detection, $p_{10}$ is the probability of detecting no error for
the second syndrome detection provided an error is detected in the
first syndrome detection. These fidelities and probabilities for different
trajectories are different. Note that the error rates for $\ket{0_{\mathrm{L}}}$
and $\ket{1_{\mathrm{L}}}$ are slightly different and here the probabilities
in both \textbf{(a)} and \textbf{(b)} are calculated from the representative
superposition state $(\ket{0_{\mathrm{L}}}+\ket{1_{\mathrm{L}}})/\sqrt{2}$. When calculating the process fidelities, the backaction associated with the no-parity-jump evolution has been included.}
\label{fig:intrinsicloss} 
\end{figure}

To quantify the contribution of these errors, we numerically simulate the process
including only the oscillator decay and Kerr terms without the ancilla
qubit decoherence. We further construct ideal unitary encoding, decoding,
and error correction gates, and implement ideal parity projection
measurement. Figure~\ref{fig:intrinsicloss} shows the process fidelities
and the error detection probabilities as a function of the total interval
$T_{\mathrm{w}}$ per round of top-layer QEC for the two different
protocols. Note that for Protocol I $T_{\mathrm{w}}\approx\tau_{\mathrm{w}}$ and for
Protocol II $T_{\mathrm{w}}\approx2\tau_{\mathrm{w}}$ which is roughly twice as large as $t_{\mathrm{w}}$
for the bottom-layer QEC process in the main text. 

The intrinsic errors for Protocol I are summarized by the parameters
$F_{\mathrm{i0}}$ and $F_{\mathrm{i1}}$, corresponding to the the
process fidelity for detecting no error (no photon loss) or one error (single photon loss) respectively.
In addition, we also define $p_{0}$ as the probability of detecting
no error. The case for Protocol II is more complicated because of two bottom-layer QECs. As shown in Fig.~2a in the main text, there
are four different branches corresponding to four different process
fidelities $F_{\mathrm{i00}}$, $F_{\mathrm{i01}}$, $F_{\mathrm{i10}}$,
and $F_{\mathrm{i11}}$. The second subscript of $F$ represents a
detection of no error (0) and one error (1) in the first syndrome
measurement, respectively, while the third subscript represents the
result of the second syndrome detection. The probability of detecting
no error for the first syndrome detection is $p_{0}$, while $p_{00}$
and $p_{10}$ are the probability of detecting no error for the second
syndrome detection provided no error or an error is detected in the
first syndrome detection respectively. These fidelities and probabilities
are all different for different branches because of the state differences.
Note that the error rates for $\ket{0_{\mathrm{L}}}$ and $\ket{1_{\mathrm{L}}}$
are slightly different and the probabilities plotted in Figs.~\ref{fig:intrinsicloss}a
and \ref{fig:intrinsicloss}b are calculated for the representative
superposition state $(\ket{0_{\mathrm{L}}}+\ket{1_{\mathrm{L}}})/\sqrt{2}$. When calculating the process fidelities, the backaction associated with the no-parity-jump evolution has been included.

For the no-error case, the dominant intrinsic error is due to the
undetectable two photon losses. While for the one-error case, the main
intrinsic error source is the Kerr dephasing coming from the fact
that $[\hat{a_{\mathrm{s}}},\frac{K_{\mathrm{s}}}{2}\hat{a_{\mathrm{s}}}^{\dagger2}\hat{a_{\mathrm{s}}}^{2}]\neq0$
and there is no way of knowing the exact time when a photon loss occurs.
After one photon loss happens, the resulting photon state further suffers
an amplitude damping because of the imbalance of the photon number
between the error states $|3\rangle$ and $|1\rangle$. The errors
due to three photon losses and one photon gain, which cause a leakage
out of the logical space, are much smaller.

As shown in Fig.~\ref{fig:intrinsicloss} and as expected, for both protocols the probability of no photon loss increases with decreasing $T_{\mathrm{w}}$, and the intrinsic errors converge to zero in the limit $T_{\mathrm{w}}\rightarrow0$. However, a smaller
QEC interval $T_{\mathrm{w}}$ means more frequent operations of error
detection and correction, which cause extra errors mainly due to the
ancilla decoherence as will be discussed below.


\subsection{Parity Measurement Infidelity}
Projective parity measurement with a protocol ($\pi/2,\pi/\chi_{\mathrm{qs}},-\pi/2$)
is constantly implemented to detect single photon loss errors in the
QEC process. Therefore, a highly faithful parity measurement is critical
for the QEC performance. However, during the parity measurement, ancilla
qubit decoherence including both pure dephasing $\left(T_{\phi}\right)$
and energy relaxation $\left(T_{1}\right)$ could cause the photon
loss error detection to fail. The pure dephasing would give an incorrect indication
of error detection, and then the quantum information would be completely
destroyed by the following correction operation. If the ancilla qubit
decays to the ground state during the parity measurement, with $50\%$
probability the parity measurement would give a wrong indication and
the quantum information would be completely destroyed as well. With
the other $50\%$ probability, the parity measurement would still
give a correct indication. However, the part of the photon state entangled
with the ancilla qubit's excited state would rotate by an unknown
angle depending on the time when the ancilla qubit decays. This would
cause both a leakage and a dephasing of the logical state, and the
quantum information would also be strongly degenerated by the following
operations, so this case can be treated as a complete destruction
as well.


To simulate the projective measurement with a finite width (320~ns),
we simply treat the projection in the simulation occurs at the middle
point of the measurement pulse. Thus, we denote the time before and after the projection as $T_{\mathrm{BM}}=160$~ns and $T_{\mathrm{AM}}=496$~ns respectively, where the feedback latency time of 336~ns is included in $T_{\mathrm{AM}}$. Since the parity
measurement time $\sim\pi/\chi_{\mathrm{qs}}$ is much shorter than
both $T_{1}$ and $T_{\phi}$, the parity measurement process fidelity
$F_{0}$ and $F_{1}$, for an initial even and odd parity state respectively,
can be approximated by linear functions of the ancilla qubit decoherence
rates: 
\begin{equation}
{F}_{0}=C_{0}-\pi/2\chi_{\mathrm{qs}}T_{1}-\pi/2\chi_{\mathrm{qs}}T_{\phi},\label{eq:Fparityg}
\end{equation}
and 
\begin{equation}
{F}_{1}=C_{1}-\pi/2\chi_{\mathrm{qs}}T_{1}-(T_{\mathrm{AM}}+T_{\mathrm{BM}})/T_{1}-\pi/2\chi_{\mathrm{qs}}T_{\phi},\label{eq:Fparitye}
\end{equation}
where $C_{0}$ and $C_{1}$ correspond to the parity fidelity when
the ancilla qubit has no decoherence. Here ${F}_{1}$ has included
the ancilla qubit energy relaxation during and after the projection
measurement of the ancilla qubit (the second term in Eq.~\ref{eq:Fparitye},
since the ancilla qubit ends up in the excited state). Based on
the measured parity readout fidelity ${F}_{0}=0.983$ and ${F}_{1}=0.960$
with $T_{1}=30~\mu$s and $T_{\phi}=120~\mu$s, we can derive $C_{0}=0.988$
and $C_{1}=0.987$. The deviation of $C_{0}$ and $C_{1}$ from unity
could come from the non-ideal non-selective $\pi/2$ pulses in the
parity measurement due to non-perfect amplitude calibration, finite
bandwidth of the pulses, or possible calibration fluctuations. 

\subsection{Recovery Gate Infidelity}
Numerically optimized pulses using the GRAPE method are implemented in the encoding, decoding, and correction processes.
However, only the correction pulse infidelity contributes to the QEC
process fidelity decay. Because the GRAPE pulses correspond to complex
trajectories of both the ancilla qubit and the oscillator evolution,
it is not intuitive to estimate the effects from the ancilla qubit
decoherence during these pulses. Instead, we numerically simulate
the recovery GRAPE pulse process fidelity in the presence of ancilla
qubit decoherence, and extract the fidelity change as a function of
the ancilla qubit decoherence $T_{1}$ and $T_{\phi}$. Since the
GRAPE pulse duration is much shorter than both $T_{1}$ and $T_{\phi}$,
the dependence of the process fidelity on the ancilla qubit decoherence is expected to be
linear as the case for the parity fidelity above:
\begin{equation}
{F}_{U_{3}}=0.978-0.233/T_{1}-0.181/T_{\phi},\label{eq:FU3}
\end{equation}
and 
\begin{equation}
{F}_{U_{2}}={F}_{U_{4}}=0.976-0.173/T_{1}-0.188/T_{\phi},\label{eq:FU24}
\end{equation}
where $U_{2}$, $U_{3}$, and $U_{4}$ are the recovery gates defined
in Fig.~2a in the main text, and here both $T_{1}$ and $T_{\phi}$
are in units of $\mu$s. It is difficult to experimentally calibrate the fidelities
of the recovery GRAPE pulses directly, because both preparation of the initial state and measurement of the final state also require GRAPE pulses. Since the optimization procedure, pulse duration, and hardware
are all the same, we assume the fidelities of all gates implemented by GRAPE pulses are the same. The two constants (the first terms) are determined
by the extracted gate fidelity ${F}_{U}=0.970$ when $T_{1}=30~\mu$s
and $T_{\phi}=120~\mu$s in our measured QEC process fidelity decay
time, as shown in Fig.~3b in the main text.

\begin{table}[t]
\caption{Error budget for the process fidelity loss in Protocol II with $t_{\mathrm{w}}=17.9~\mu$s ($T_{\mathrm{w}}\approx36~\mu$s) as in the main text. Case~00, case~01, case~10, and case~11 are for the following four cases respectively:
no error for both detections, no error for the first detection and
one error for the second, one error for the first detection and no
error for the second, one error for both detections. Note that the
total fidelity is the product of all individual fidelities. The total
weighted average total error gives a decay time of the process fidelity
of 198~$\mu$s, in a good agreement with the experiment.}
\begin{tabular*}{0.48\textwidth}{@{\extracolsep{\fill}}@{\extracolsep{\fill}}@{\extracolsep{\fill}}ccccc}
\hline 
error source  &  &  &  & \tabularnewline
for process fidelity loss  & case~00  & case~01  & case~10  & case~11 \tabularnewline
\hline 
intrinsic  & $7.0\%$  & $8.6\%$  & $11.0\%$  & $16.6\%$ \tabularnewline
detection  & $3.4\%$  & $5.6\%$  & $5.6\%$  & $7.7\%$ \tabularnewline
recovery  & $3.1\%$  & $3.2\%$  & $6.2\%$  & $6.2\%$ \tabularnewline
ancilla thermal excitation  & $0.7\%$  & $0.7\%$  & $0.7\%$  & $0.7\%$\tabularnewline
total error  & $13.6\%$  & $17.0\%$  & $21.8\%$  & $28.3\%$\tabularnewline
weighted average total error  & $16.2\%$ &  &  & \tabularnewline
\hline 
\end{tabular*}\vspace{-6pt}
\label{T:simupara} 
\end{table}

\begin{figure}
\includegraphics{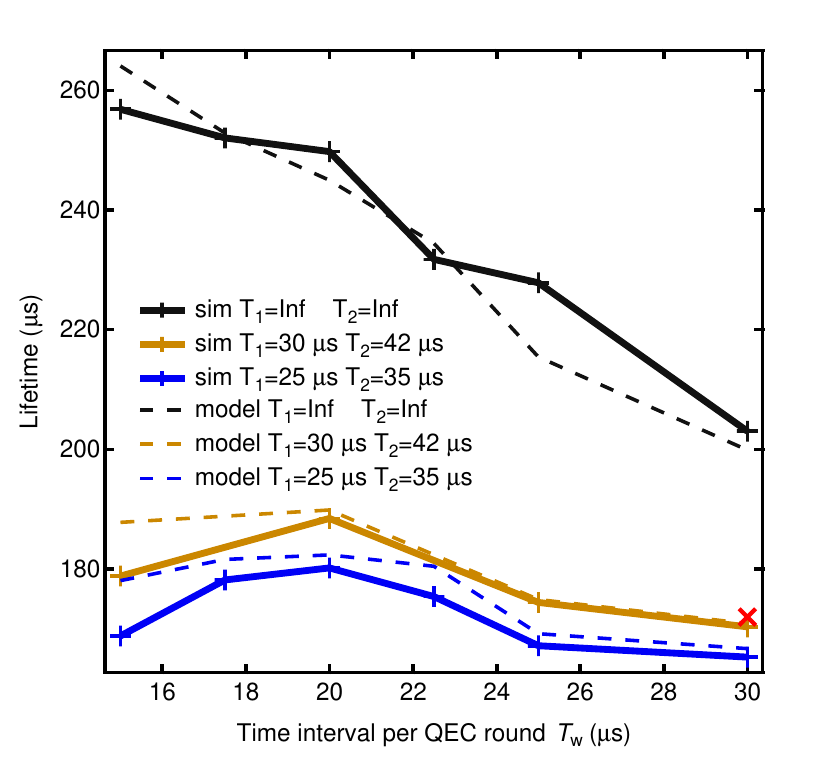} \caption{\textbf{Predictions by numerical and analytical approaches for QEC
Protocol I.} Curves with different colors present the QEC process
fidelity decay time as a function of the QEC interval $T_{\mathrm{w}}$
with different ancilla qubit coherence times. The curves are not smooth
because different optimized recovery pulses with different fidelities
are used for different QEC intervals. Nevertheless, the good agreement
between numerical simulations and the model using Eqs.~\ref{eq:errRate1}
and \ref{eq:CalLifetime} confirms the reliability of the model. The
red cross indicates our experimental result ($T_{1}=30~\mu$s and
$T_{2}=40~\mu$s), also in good agreement with the simulation.}
\label{fig:simures} 
\end{figure}

\subsection{Analytical Model and Error Propagation}
With the assistance of the numerical simulations, errors from each individual component in the whole process can be estimated. Table~\ref{T:simupara}
shows the summarized error budget for the process fidelity loss in
Protocol II with $t_{\mathrm{w}}=17.9~\mu$s ($T_{\mathrm{w}}\approx36~\mu$s) as in the main text. These numbers are obtained from numerical simulations
with the capability of choosing individual error sources separately. In the simulations, the imperfections in the experiment associated with the parity measurement and the recovery gates as mentioned above have been taken into account. The total weighted average total error gives a decay
time of the process fidelity of 198~$\mu$s, in good agreement
with the experiment.

Since the numerical simulation is time-consuming and does not provide physical
intuition, we establish an analytical model for the error propagation
during the QEC process based on a simplified probability calculation.
With all the errors for individual operations in hand, we can study
the fidelity of the whole QEC process. Here, we study Protocol I with
one bottom-layer detection and correction first.

As errors occur in a probabilistic way, the fidelity after one QEC
step in Protocol I could be described by the following equation: 
\begin{eqnarray}
{F^{(\mathrm{I})}}(T_{\mathrm{w}}) & = & [\underset{\mathrm{0\,photon\,loss}}{\underbrace{p_{0}F_{\mathrm{i0}}(T_{\mathrm{w}})F_{0}F_{U_{1}}}}+\underset{\mathrm{1\, photon\,loss}}{\underbrace{(1-p_{0})F_{\mathrm{i1}}(T_{\mathrm{w}})F_{1}F_{U_{2}}}}]\nonumber \\
 &  & \times\underset{\mathrm{no\,ancilla\,thermal\,flip}}{\underbrace{[1-n_{\mathrm{th}}^{\mathrm{q}}(1-e^{-T_{\mathrm{w}}/T_{1}})]}}.
\label{eq:errRate1}
\end{eqnarray}

The expression consists of two branches: the 
process conditioned on no photon loss and the process conditioned
on a single photon loss. Here, $F_{U_{1}}$ and $F_{U_{2}}$ are the
fidelity of the recovery gate $U_{1}$ and $U_{2}$ shown in Fig.~\ref{fig:ProtocolI},
respectively, and the last factor takes into account the small probability
of the excitation of the ancilla to $\ket{e}$ during the waiting
time $T_{\mathrm{w}}$. $F^{(\mathrm{I})}(T_{\mathrm{w}})$ is then
used to estimate the decay time $\tau$ of the QEC process according
to an exponential decay: 
\begin{equation}
\tau\approx-\frac{T_{\mathrm{w}}}{\mathrm{ln}(F(T_{\mathrm{w}}))}.\label{eq:CalLifetime}
\end{equation}

Figure~\ref{fig:simures} shows the QEC process fidelity decay time
with different ancilla qubit coherence times as a function of the
QEC interval $t_{\mathrm{w}}$. The good agreement between numerical
simulations and the analytical model using Eqs.~\ref{eq:errRate1}
and \ref{eq:CalLifetime} confirms the reliability of our
model. Our experimental result with $T_{1}=30~\mu$s and $T_{2}=40~\mu$s
($T_{\phi}=120~\mu$s) is also in a good agreement with the simulation.

The error propagation model in QEC Protocol II can be treated in a
similar way but with more probabilistic branches: 
\begin{align}
F^{(\mathrm{II})}(T_{\mathrm{w}}) = & [ \underset{\mathrm{0\,photon\,loss\,for\,both\,detections}}{\underbrace{p_{0}p_{00}\cdot F_{\mathrm{i00}}(T_{\mathrm{w}})\cdot F_{0}\cdot F_{0}\cdot F_{U_{3}}}}\nonumber \\
+ & \underset{\mathrm{0\,photon\,loss\,for\,the\,first\,;\,1\,photon\,loss\,for\,the\,second\,detection}}{\underbrace{p_{0}(1-p_{00})\cdot F_{\mathrm{i01}}(T_{\mathrm{w}})\cdot F_{0}\cdot F_{1}\cdot F_{\pi}\cdot F_{U_{4}}}}\nonumber \\
+ & \underset{\mathrm{1\,photon\,loss\,for\,the\,first\,;\,0\,photon\,loss\,for\,the\,second\,detection}}{\underbrace{(1-p_{0})p_{10}\cdot F_{\mathrm{i10}}(T_{\mathrm{w}})\cdot F_{1}\cdot F_{\pi}\cdot F_{U_{2}}\cdot F_{0}\cdot F_{U_{3}}}}\nonumber \\
+ & \underset{\mathrm{1\,photon\,loss\,for\,both\,detections}}{\underbrace{(1-p_{0})(1-p_{10})\cdot F_{\mathrm{i11}}(T_{\mathrm{w}})\cdot F_{1}\cdot F_{\pi}\cdot F_{U_{2}}\cdot F_{1}\cdot F_{\pi}\cdot F_{U_{4}}}}]\nonumber \\
\cdot[ & \underset{\mathrm{no\,ancilla\,thermal\,flip}}{\underbrace{1-n_{\mathrm{th}}^{\mathrm{q}}(1-e^{-T_{\mathrm{w}}/T_{1}})}}],
\label{eq:errRate2}
\end{align}
with all the parameters defined earlier. Compared to the no-error
case ($F_{U_{3}}$), the one-error cases ($F_{U_{2}}$ and $F_{U_{4}}$)
have an additional error source from the preceding $\pi$ pulse. The
error associated with this $\pi$ pulse can completely destroy the
encoded quantum information through the following recovery gates.
Since the $\pi$ pulse is so short, its dependence on the ancilla
qubit decoherence can be neglected. In the experiment, ${F}_{\pi}=0.984$
with the infidelity mainly coming from the finite gate bandwidth and
possible fluctuation of pulse calibration.


In order to optimize the QEC interval $T_{\mathrm{w}}$ (or $t_{\mathrm{w}}$),
in Fig.~3b in the main text we have plotted the QEC process fidelity
decay time $\tau$ by using Eqs.~\ref{eq:CalLifetime} and \ref{eq:errRate2}
as a function of the interval per QEC round ($T_{\mathrm{w}}$) while
keeping all other parameters except for the recovery gate fidelity
fixed. As expected, there is an optimized region of $T_{\mathrm{w}}$
for the longest logical qubit lifetime $\tau$. In the experiment,
a critical parameter is the fidelity of the GRAPE pulse $F_{U}$,
which is difficult to calibrate separately as mentioned earlier. Therefore,
we vary $F_{U}$ from $0.95$ to $1$ with a step of $0.005$ when
plotting Fig.~3b in the main text. The measured lifetime $\tau=200~\mu$s
implies $F_{U}=0.970$, which is exactly the one to give the calibration
for Eqs.~\ref{eq:FU3} and \ref{eq:FU24}. This recovery GRAPE
pulse fidelity is mainly limited by the ancilla qubit and oscillator
decoherence. The interval used in the experiment as expected coincides
with the optimal one from the model.


\subsection{Towards Break-Even Point}
Although the lifetime of the logical qubit achieved in our experiment
is close to that of the Fock $\{\ket{0},\ket{1}\}$ encoding, there
are still a few ways to achieve the break-even point which could
be investigated experimentally in the future.

\subsubsection{Improve the ancilla properties}
As explained by Fig.~3c in the main text, improving the
ancilla coherence properties is the most direct way to extend the QEC lifetime,
because error detection and all recovery operations utilize the ancilla
qubit. If the ancilla coherence times can be improved, our system
can work with more frequent error detection and correction, thus reducing
$\tau_{\mathrm{w}}$ and suppressing further the intrinsic errors. Our model
predicts that the break-even point could be achieved if $T_{1}$ can
be nearly doubled even with the same $T_{\phi}$.


\subsubsection{Better parameter calibration and GRAPE pulses}
The model described above has carefully analyzed the infidelities of
the parity measurement and the recovery processes. Except for the contribution
from the finite $T_{1}$ and $T_{\phi}$ of the ancilla, there are
also parameter fluctuations that lead to the fidelities $C_{0,1}\sim0.987$
for the parity measurement and $0.976$ for the recovery
gates. To minimize the parameter fluctuations, the parameters of the experimental
system should be carefully calibrated and also be stabilized during
the experiment. Then, the detection and corrections based on these parameters could
allow us to control the system more precisely.

\subsubsection{Increasing the number of bottom-layer QEC steps}
\begin{figure*}
\includegraphics{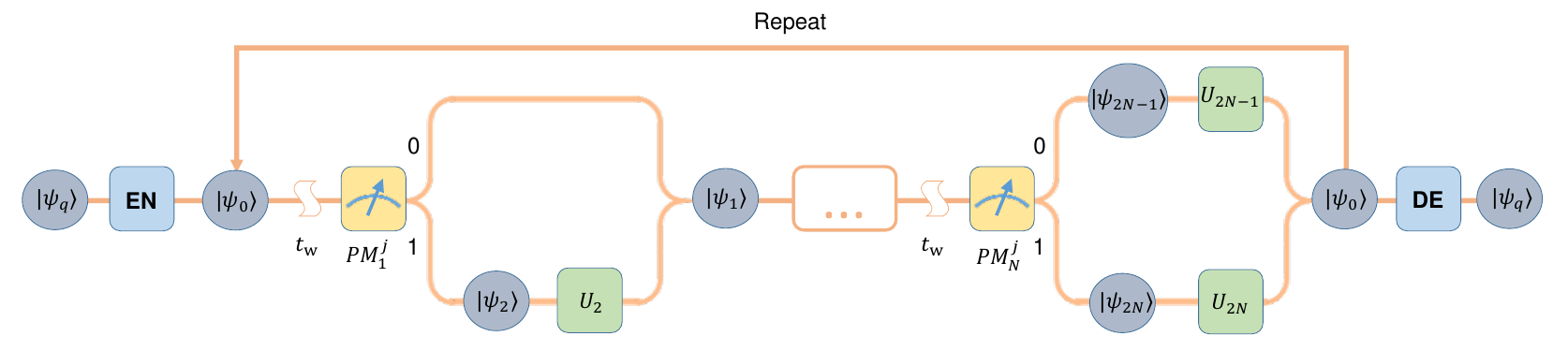} \caption{\textbf{Protocol with more bottom-layer QEC steps.} When no error
is detected, the state after the deterministic evolution should be
untouched as much as possible to minimize the errors induced by the
recovery gates. This can potentially increase the QEC protected lifetime.}
\label{fig:Protocol4} 
\end{figure*}

\begin{figure}[htb]
\includegraphics{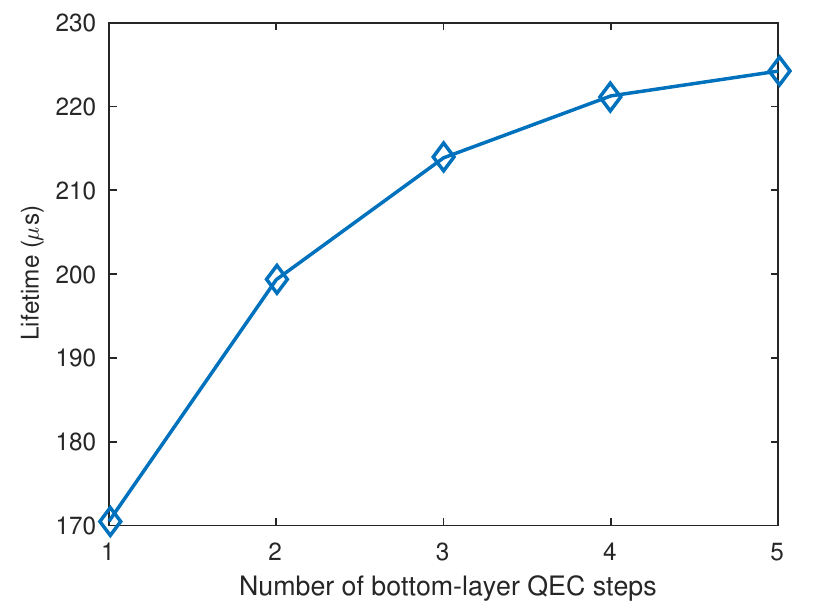} \caption{\textbf{Dependence of the logical qubit lifetime on the number of bottom-layer
QEC steps.} The lifetime is extended when the number of bottom-layer
QEC steps is increased, because the state after the deterministic
evolution is untouched as much as possible when no error is detected
and thus fewer errors are introduced.}
\label{fig:LifetimeVsNumDetection} 
\end{figure}

Since the recovery gate introduces extra errors in the QEC process,
it would be a better strategy to have the state after the deterministic
evolution untouched as much as possible when no error is detected,
i.e. to have more bottom-layer QEC steps, as shown in Fig.~\ref{fig:Protocol4}.
We extend the model of Eqs.~\ref{eq:errRate1} and \ref{eq:errRate2}
to this case while reasonably assuming the same fidelities for error
detection and recovery gates. The result is shown in Fig.~\ref{fig:LifetimeVsNumDetection}. Note that the backaction associated with the no-parity-jump evolution has been included in the model.
We estimate that if the number of bottom-layer QEC steps, which construct a single top-layer
QEC process, were increased to four, the break-even point could
be achieved with our current experimental system. However, this new
strategy demands more adaptive gate calibrations, so it is not implemented
in our current experiment but could be investigated experimentally
in the future.

